\documentclass[aps, preprint, superscriptaddress,preprintnumbers,floatfix,nofootinbib,amsmath,amssymb]{revtex4}

\usepackage{url}
\usepackage{hyperref}
\usepackage{cleveref}
\usepackage{graphicx}

\usepackage{epsfig}
\usepackage{amstext,amssymb}
\usepackage{amsmath}
\usepackage{amsfonts}

\usepackage{xspace}
\usepackage{subfigure}
\usepackage{color}
\usepackage{bm}

\usepackage{slashed} 

\newcommand{\be}{\begin{equation}}
\newcommand{\ee}{\end{equation}}
\newcommand{\bea}{\begin{eqnarray}}
\newcommand{\eea}{\end{eqnarray}}
\newcommand{\nn}{\nonumber}

\def\mlh{\hat m_\ell}

\def\mKh{\hat m_{K_1}}
\def\l1{\lambda}

\def\s1{\hat s}

\def\U1mt{U(1)_{L_\mu-L_\tau}}

\usepackage{slashed}

\begin{document}
\title{ Implications  of new physics in $ B \to K_1 \mu^+ \mu^-$ decay processes }
\author{Aishwarya Bhatta}
\email{aish.bhatta@gmail.com}

\author{Rukmani Mohanta}
\email{rmsp@uohyd.ac.in}
\affiliation{School of Physics,  University of Hyderabad, Hyderabad-500046,  India}

\begin{abstract}
In recent times, several discrepancies at the level of $(2-3)\sigma$ have been observed  in the decay processes mediated by  flavour changing neutral current (FCNC) transitions $b \to s \ell^+ \ell^-$, which may be considered as the  smoking-gun signal of New Physics (NP).  These intriguing hints of NP have attracted a lot of attention  and many attempts are made to look for the possible NP signature in other related processes, which are mediated through the same quark-level transitions. In this work, we perform a comprehensive analysis of the FCNC decays of $B$ meson to axial vector mesons $ K_1(1270)$ and $ K_1(1400)$, which are   admixture of the $1^3P_1$ and $1^1P_1$ states $K_{1A}$ and $K_{1B}$, in a model independent framework.
Using the $ B \to K_1$ form factors evaluated in the light cone sum rule approach, we investigate the rare exclusive semileptonic decays $ B \to K_1(1270) \mu^+ \mu^-$ and $ B \to K_1(1400) \mu^+ \mu^-$.  
Considering all the possible relevant operators for $b \to s \ell^+ \ell^-$ transitions, we study their effects on various observables such as branching fractions, lepton flavor universality violating ratio ($R_{K_1})$,  forward-backward asymmetries, and lepton polarization asymmetries of these processes.
These results will not only enhance the theoretical understanding of the mixing angle but also serve as a good tool for probing New Physics.

\end{abstract}

\maketitle
\flushbottom
\section{Introduction} 
Understanding the nature of  physics beyond the Standard Model (BSM) is of paramount importance today in the context of Particle Physics, Astrophysics, and Cosmology.  Although it was very much anticipated  that the LHC experiment would provide an unambiguous signature of new physics in the form  of direct observation of some new particles,  the null result so far inspires the community to look for alternative scenarios. As a consequence,   much attention has been  paid  to indirect signals, where the experimentally measured values of the observables show few sigma deviations  from their corresponding  standard model (SM) expectations. In recent times,  several such intriguing results are observed by  LHCb, Belle and BaBar experiments, in  the semileptonic decays of $B$ mesons both in the charged current $b \to c \ell \nu_\ell$ \cite{Lees:2012xj,Lees:2013uzd,Aaij:2015yra,Huschle:2015rga,Hirose:2016wfn,Aaij:2017uff,Abdesselam:2019dgh} as well as neutral current $b \to s \ell^+ \ell^-$ transitions \cite{Aaij:2013aln,Aaij:2013qta,Aaij:2014pli,Aaij:2014ora,Aaij:2015esa,Aaij:2017vbb,Aaij:2019wad,Abdesselam:2019lab,Abdesselam:2019wac}. More specifically,  hints of physics beyond the Standard Model  have been observed in the semileptonic decays of $B$ mesons in the form of lepton flavor universality violating (LFUV)  ratios. In the charged-current sector, these observables are  defined as $R_{D^{(*)}}={\rm Br}(B \to D^{(*)} \tau \overline{ \nu}_\tau)/{\rm Br}(B \to D^{(*)} \ell \overline{ \nu}_\ell)$, where $(\ell=e,\mu)$, which show nearly $3 \sigma$ deviation from their corresponding SM results, taking into account the correlation
between $R_D$ and $R_{D^*}$ \cite{Amhis:2019ckw}.  The analogous observable in the $B_c$ meson decay, i.e., $R_{J/\psi}={\rm Br}(B_c \to J/\psi \tau \bar \nu_\tau)/{\rm Br}(B_c \to J/\psi \mu  \overline{ \nu}_\mu) = 0.71 \pm 0.17 \pm 0.18$  \cite{Aaij:2017tyk} also exhibits $1.7 \sigma$ deviation from its SM value $R_{J/\psi}^{\rm SM}=  0.289 \pm 0.010$~\cite{Dutta:2017xmj}.  To  resolve these anomalies associated with the charged current transitions $b \to c \ell \nu_\ell$, it is usually assumed  the presence of new physics in the semi-tauonic mode $b \to c \tau \nu_\tau$.
In the neutral current sector, there are a plethora of observables which manifest  deviations from their SM predictions at the level of $(2-4)\sigma$.
Amongst them, the prime candidates are the  LFUV observables $R_K$ and $R_{K^*}$,  defined as 
\begin{align} \label{Eqn:RK1}
R_{K} \ & = \ \frac{{\rm Br}(B^+ \to K^{+} \mu^+ \mu^-)}{{\rm Br}(B^+ \to K^{+} e^+ e^-)} \, , \qquad 
R_{K^{*}} \  = \ \frac{{\rm Br}(\overline B \to \overline K^* \mu^+ \mu^-)}{{\rm Br}(\overline B \to \overline K^* e^+ e^-)} \, .
\end{align}
In 2014, the measurement on the LFUV ratio $R_K=0.745^{+0.090}_{-0.074}\pm 0.036$, in the low $q^2\in [1,6]~{\rm GeV}^2$ region by the LHCb experiment~\cite{Aaij:2014ora}
 attracted huge attention, as it manifested a discrepancy of $2.6\sigma$ from its SM prediction 
~\cite{Bobeth:2007dw} (see also~\cite{Bordone:2016gaq}) 
\bea \label{Eqn:RK-SM1}
R_K^{\rm SM} \ = \ 1.0003\pm 0.0001\,.
\eea
The updated LHCb measurement of $R_K$ in the $q^2\in [1.1,6]~{\rm GeV}^2$ region by combining the Run 1 data with $2~{\rm fb}^{-1}$ of Run 2 data~\cite{Aaij:2019wad}
\bea \label{Eqn:RK-Exp-new1}
R_K^{\rm LHCb} \ = \ 0.846^{+0.060+0.016}_{-0.054-0.014}\,  ,
\eea
also exhibits a discrepancy at the level of  $2.5\sigma$. 

Recently, the LHCb Collaboration  reported the updated result on $R_K$ in the dilepton mass-squared region $1.1<q^2<6.0~{\rm GeV}^2$, based on the data collected at the center-of-mass energy of 7, 8 and 13 TeV corresponding to an integrated luminosity of $9~{\rm fb}^{-1}$ \cite{Aaij:2021vac} as
\bea \label{RK-Exp-new1}
R_K^{\rm LHCb} \ = \ 0.846^{+0.044}_{-0.041}\,  ,
\eea
which shows $3.1 \sigma$ deviation with the SM prediction.

In addition, the LHCb Collaboration has also measured the $R_{K^{*}}$ ratio in two bins of low-$q^2$ region~\cite{Aaij:2017vbb}
\bea
R_{K^*}^{\rm LHCb}& \ = \ & \begin{cases}0.660^{+0.110}_{-0.070}\pm 0.024 \qquad q^2\in [0.045, 1.1]~{\rm GeV}^2 \, , \\ 
0.685^{+0.113}_{-0.069}\pm 0.047 \qquad q^2\in [1.1,6.0]~{\rm GeV}^2 \, ,
\end{cases}
\eea 
which also depict $2.2\sigma$ and $2.4\sigma$ deviations from their corresponding SM results~\cite{Capdevila:2017bsm}
\bea
R_{K^*}^{\rm SM} \ = \  \begin{cases} 0.92\pm 0.02 \qquad q^2\in [0.045, 1.1]~{\rm GeV}^2 \, , \\  
1.00\pm 0.01\qquad q^2\in [1.1,6.0]~{\rm GeV}^2 \, .
\end{cases}
\eea
 These discrepancies associated with the flavor changing neutral current (FCNC) transitiona $b \to s \ell^+ \ell^-$ are generally attributed to the presence of new physics (NP) in $b \to s \mu \mu$ decay channel. In addition to these LHCb results, the Belle experiment has recently announced new measurements on $R_K$~\cite{Abdesselam:2019lab} and  $R_{K^*}$~\cite{Abdesselam:2019wac} in several other bins, which are though consistent with SM, but have large uncertainties.

There are also quite a few other deviations from the SM expectations in the measurement involving $b \to s \mu \mu$ transition, such as the branching fractions of $B_s \to \mu^+ \mu^-,$ $B \to K^{(*)} \mu^+\mu^-$, $B_s \to \phi \mu^+ \mu^-$, the angular observable $P_{4,5}'$ in $B \to K^* \mu^+ \mu^-$, etc \cite{Tanabashi:2018oca}.  Additionally, LHCb Collaboration measured the lepton flavor universality observable in $\Lambda_b \to pK \ell^+ \ell^-$ channel, using 7, 8 and 13 TeV data corresponding to integrated  luminosity $4.7~{\rm fb}^{-1}$ in the $0.1<q^2 <6~{\rm GeV^2}$ bin \cite{Aaij:2019bzx}
\bea
R_{pK}^{-1}=\frac{{\rm Br} (\Lambda_b \to pK e^+ e^-)}{{\rm Br}( \Lambda_b \to pK~J/\psi (\to  e^+ e^-))}\Big/\frac{{\rm Br}( \Lambda_b \to pK \mu^+ \mu^-)}{{\rm Br} (\Lambda_b \to pK~J/\psi (\to  \mu^+ \mu^-))}
=1.17^{+0.18}_{-0.16}\pm 0.01,
\eea
which is compatible with unity, i.e., the SM prediction, within $1 \sigma$ deviation.

Hence, it is natural to address all these anomalies associated with the semileptonic FCNC transitions $ b \to s \ell^+ \ell^-$  by assuming the presence of new physics only in the muon sector.  It is thus quite reasonable to expect that if new physics is indeed responsible for the above mentioned anomalies, it might also leave its footprints in the other related decay modes mediated by $b \to s \mu^+ \mu^-$ transition. In this context, we would like to analyze the decay channels $B \to (K_1(1270)/K_1(1400)) \mu^+ \mu^-$, where  $K_1(1270)$ and $K_1(1400)$ are axial vector mesons, which are an admixture of $1^3P_1$ and $1^1P_1$ states $K_{1A}$ and $K_{1B}$ respectively,
\bea
&&|K_1(1270) \rangle =|K_{1A} \rangle \sin \theta +|K_{1B} \rangle \cos \theta, \nn\\
&&|K_1(1400) \rangle =|K_{1A} \rangle \cos \theta -|K_{1B} \rangle \sin \theta ,
\eea
where $\theta$ is the mixing angle, which is not yet determined precisely.
 Its value has been estimated to be $-(34 \pm 13)^\circ $ from the decay of $B \to K_1(1270) \gamma$ and $\tau \to K_1(1270) \nu_\tau$ \cite{Hatanaka:2008xj}. However,  it is experimentally challenging to separate the $K_1(1270)$ and $K_1(1400)$ states as these are broad resonances and have the common decay channel $K_1 \to K \pi \pi$. The $K_1(1270)$ state decays predominantly through intermediate $K \rho$ state, while $K_1(1400)$ decays almost exclusively via $K^* \pi$ channel. Therefore, separating these two channels requires dedicated amplitude analysis.
An unbinned maximum-likelihood Dalitz plot method can be used simultaneously fit the data in the three dimensional invariant mass-squared plane: $M^2(K\pi \pi)$, $M^2(K \pi)$ and $M^2 (\pi \pi)$ as done for the case of nonleptonic decays $B \to J/\psi  K_1(\to K \pi \pi)$ and $B \to  \psi' K_1(\to K \pi \pi)$ by Belle Collaboration \cite{Guler:2010if}.
  In the recent past, the $B \to (K_1(1270)/K_1(1400)) \ell^+ \ell^-$, decay modes have been the subject of many  theoretical discussions,  both in the SM 
 \cite{Hatanaka:2008gu, Li:2009rc,Paracha:2007yx,Bashiry:2009wq} as well as in various new physics scenarios, such as supersymmetric model \cite{Bashiry:2009wh}, extra dimension  \cite{Ahmed:2008ti, Saddique:2008xj},  fourth generation model \cite{Ahmed:2011vr}, nonuniversal $Z'$ model \cite{ Li:2011nf,Huang:2018rys}, two Higgs doublet model \cite{Falahati:2014yba}  etc., and also in the model independent approach \cite{Ahmed:2010tt}. The study of these semileptonic decays provide a complementary framework  to corroborate the results of the observed anomalies associated with $b \to s\mu^+ \mu^-$ transitions, as a number of observables associated with these modes, such as  branching fractions, forward-backward asymmetry, lepton polarization asymmetry,  are quite sensitive to new physics. In this context, we would like to investigate these decay processes in a model independent framework, where the possible new physics effects are quantified by introducing additional new operators to the SM effective Hamiltonian. 
 
 It should be further emphasized that  the differential branching ratio  of $ B^+ \to K^+\pi^+ \pi^-\mu^+ \mu^-$ process has been reported in the LHCb paper using the 7 TeV and 8 TeV data set corresponding to an integrated luminosity of $3.0 ~{\rm fb}^{-1}$ \cite{Aaij:2014kwa} as
\begin{eqnarray}
{\rm Br}(B^+ \to K^+\pi^+ \pi^-\mu^+ \mu^-)= \Big( 4.36^{+0.29}_{-0.27}({\rm stat}) \pm 0.21 ({\rm syst}) \pm 0.18 ({\rm norm}) \Big) \times 10^{-7}.
\end{eqnarray}
Since the branching fraction of the rare decay $B^+ \to K_1(1270)^+ \mu^+ \mu^-$ is expected to contribute significantly, it is strongly argued to perform the analysis for $ B^+ \to K^+\pi^+ \pi^-\mu^+ \mu^-$ process with 13 TeV data set as well as to look for $ B^+ \to K^+\pi^+ \pi^-e^+ e^-$ process so that the lepton flavour universality violation parameter 
\begin{eqnarray}
R_{K \pi \pi}=\frac{B^+ \to K^+\pi^+ \pi^-\mu^+ \mu^-}{B^+ \to K^+\pi^+ \pi^- e^+ e^-}
\end{eqnarray}
can also be tested independently in another  semileptonic flavour changing neutral current process $b \to s \ell^+ \ell^-$ process, preferably in the low $q^2$ bin, i.e.,  $q^2\in[1.1,6]~{\rm GeV}^2$.

 The layout of the paper is as follows. In section II,  we discuss the generalized effective Hamiltonian describing the  semileptonic transition $b \to s \ell^+ \ell^-$, both in the SM and in the context of  NP.  We then proceed to constrain the NP parameters performing a two-dimensional fit to the existing $b \to s \mu \mu$ observables, which show more than $1\sigma$ deviation from their corresponding SM predictions and relatively free from hadronic uncertainties. The discussion on differential decay distribution and other relevant observables is presented in Section III. The implications of new physics on various decay observables of $B \to (K_{1}(1270/K_1(1400)) \mu^+ \mu^-$ processes are presented in section IV followed by our conclusions and outlook in Section V.

\section{Theoretical Framework}
The SM effective Hamiltonian responsible for $b \to s \ell^+ \ell^-$ transition can be expressed as 
\bea
&&{\cal H}_{\rm eff}^{\rm SM}=-\frac{\alpha G_F }{\sqrt 2 \pi} V_{tb}V_{ts}^* \Big[2 \frac{C_7^{\rm eff}}{q^2} \left [\bar s \sigma^{\mu \nu} q_\nu (m_s P_L +m_b P_R) b \right ]  (\bar \ell \gamma_\mu \ell) \nn\\
&&~~~~~~+ C_9^{\rm eff} (\bar s \gamma^\mu P_L b) (\bar \ell \gamma_\mu \ell)+ C_{10} (\bar s \gamma^\mu P_L b) (\bar \ell \gamma_\mu \gamma_5 \ell) \Big],\label{Ham-SM}
\eea
where $\alpha$ is the fine structure constant, $G_F$ is the Fermi coupling, $V_{tb}, V_{ts}$ are the CKM matrix elements, and $P_{L,R}=(1\mp\gamma_5)/2$ are the chiral projection operators, $C_7^{\rm eff}, C_9^{\rm eff}$ and $C_{10}$ are  the Wilson coefficients,  evaluated at the $m_b$ scale. It should be noted  that the coefficient $C_9^{\rm eff}$ contains both short-distance contributions from the 4-quark operators, away from the charmonium resonance domain, which are known to be calculated precisely in the perturbation theory and  long distance part  associated with real $c \bar c$ intermediate states, i.e.,
it can be expressed as: $C_9^{\rm eff}(m_b, q^2)= C_9(m_b)+Y_{\rm pert}(q^2)+Y_{\rm LD}(m_b,q^2)$. The explicit forms of $Y_{\rm pert}$ and  $Y_{\rm LD}$ are widely discussed in the literature \cite{Buras:1994dj, Lim:1988yu, Deshpande:1988bd,ODonnell:1991cdx, ODonnell:1992ppr,Kruger:1996cv} and their values are taken from \cite{Falahati:2014yba}.  The values of the Wilson coefficients $C_{1,\cdots,6}$  at $m_b$ scale calculated in Next-to-next-to leading-logarithmic (NNLL) order by matching the full and effective theories at the electroweak scale and subsequently evolved down to the $b$ quark scale using renormalization group equations \cite{Bobeth:2003at,Huber:2005ig, Altmannshofer:2008dz},  while the values of $C_7^{\rm eff}$, $C_9$ and $C_{10}$ are taken from \cite{Bhom:2020lmk}, which are presented in Table-\ref{Wilson}. 
\begin{table}[ht]
\begin{center}
\vspace*{0.1 true in}
\begin{tabular}{c c  c c c c c c c  }
\hline
\hline
 $C_1$ & $C_2$ & $C_3$ & $C_4$ & $C_5$ & $C_6$ & $C_7^{eff}$ &  $C_9$ & $C_{10}$ \\
\hline
$-0.257 $ &~ $1.009 $ ~& $-0.005 $ ~&~ $-0.078 $ ~&~ $0.000 $~ &~ $0.001 $ ~& $-0.292 $ 
~& ~ $4.08 $ ~&$-4.31 $\\
\hline\hline
\end{tabular}
\end{center}
\caption{Values of the SM Wilson coefficients evaluated at the $m_b$ scale.}\label{Wilson}
\end{table} 

Keeping in mind that, the new physics solutions, which can explain the observed anomalies in $b \to s \mu^+ \mu^-$ transition are  only in the form of vector and axial-vector operators,  we consider only these additional operators to the SM Hamiltonian for both  chiral quark currents. Thus, the total effective Hamiltonian describing the $b \to s \mu^+ \mu^-$ transition processes can be represented as
\bea
{\cal H}_{\rm eff}^{\rm tot}={\cal H}_{\rm eff}^{\rm SM}+{\cal H}_{\rm eff}^{\rm NP}\;,
\eea
where ${\cal H}_{\rm eff}^{\rm NP}$ denotes the new physics effective Hamiltonian, which can be expressed as
\bea
&&{\cal H}_{\rm eff}^{\rm NP}=-\frac{\alpha G_F }{\sqrt 2 \pi} V_{tb}V_{ts}^* \Big[C_9^{\rm NP} (\bar s \gamma^\mu P_L b) (\bar \ell \gamma_\mu \ell)+ C_{10}^{\rm NP} (\bar s \gamma^\mu P_L b) (\bar \ell \gamma_\mu \gamma_5 \ell) \nn\\
&&~~~~~~~~~+ C_9^{'\rm NP} (\bar s \gamma^\mu P_R b) (\bar \ell \gamma_\mu \ell)+ C_{10}^{'\rm NP} (\bar s \gamma^\mu P_R b) (\bar \ell \gamma_\mu \gamma_5 \ell) \Big],\label{Ham-NP}
\eea
where $C_{9,10}^{\rm NP}$ and $C_{9,10}^{'\rm NP}$ are the new Wilson coefficients, and their values can be obtained from the global fit to the observed $b \to s \mu^+ \mu^-$ data. A commonly acceptable presumption that emerged from the global fits performed by various groups, see e.g. \cite{Alok:2019ufo}, by considering one NP coefficient at a time is either (I) $C_9^{\rm NP}=-1.09 \pm 0.18$ or  (II)$~C_9^{\rm NP}=-C_{10}^{\rm NP} = -0.53 \pm 0.09 $ with pull values 6.24 and 6.40 respectively. 
Recently, a combined global fit is performed  in \cite{Bhom:2020lmk}, to constrain the $C_{7,9,10}^{\rm NP}$ Wilson coefficients, considering them as real, and the best-fit results obtained are given as $(C_7^{\rm NP}, C_9^{\rm NP}, C_{10}^{\rm NP})=(0.013, -1.03,0.08)$,  which are pretty well consistent with the scenario I, with only one NP coefficient at a time.

In this work, we perform a two-dimensional global fit, by taking two new operators at a time with the following   possible combinations: $(C_9^{\rm NP}, C_9^{'\rm NP})$,  $(C_{10}^{\rm NP}, C_{10}^{'\rm NP})$ and $(C_9^{\rm NP}, C_{10}^{\rm NP})$. In our fit, we include only those observables associated with $b \to s \mu^+ \mu^-$ anomalies, which are relatively free from hadronic uncertainties and are listed below:
\subsubsection{$R_K$ and $R_{K^*}$}
 
The recently updated   lepton flavour universality   violating (LFUV) ratios  $R_K$ \cite{Aaij:2021vac} and $R_{K^*}$  \cite{Aaij:2017vbb}, by LHCb measurement in the low $q^2$ bins: 
\bea \label{Eqn:RK-Exp-new}
R_K^{\rm LHCb} \ = \ 0.846^{+0.044}_{-0.041}\,  ,\qquad q^2\in [1.1, 6]~{\rm GeV}^2 \, ,
\eea
\bea
R_{K^*}^{\rm LHCb}& \ = \ & \begin{cases}0.660^{+0.110}_{-0.070}\pm 0.024 \qquad q^2\in [0.045, 1.1]~{\rm GeV}^2 \, , \\ 
0.685^{+0.113}_{-0.069}\pm 0.047 \qquad q^2\in [1.1,6.0]~{\rm GeV}^2 \, .
\end{cases}
\eea 
Besides the LHCb results, the Belle experiment has recently announced new measurements on $R_K$~\cite{Abdesselam:2019lab} and 
$R_{K^*}$~\cite{Abdesselam:2019wac} in several other bins:
\bea
R_K^{\rm Belle} \ & = \ \begin{cases} 
1.01 ^{+0.28}_{-0.25}\pm 0.02 \qquad q^2\in [0.1, 4.0]~{\rm GeV}^2 \, , \\  
0.85 ^{+0.30}_{-0.24}\pm 0.01 \qquad q^2\in [4.0,8.12]~{\rm GeV}^2 \, , \\  
1.03 ^{+0.28}_{-0.24}\pm 0.01 \qquad q^2\in [1.0,6.0]~{\rm GeV}^2 \, , \\  
1.97 ^{+1.03}_{-0.89}\pm 0.02 \qquad q^2\in [10.2,12.8]~{\rm GeV}^2 \, , \\  
1.16 ^{+0.30}_{-0.27}\pm 0.01 \qquad q^2> 14.18~{\rm GeV}^2 \, , \end{cases} \\
R_{K^*}^{\rm Belle} \ & = \ \begin{cases} 0.52^{+0.36}_{-0.26}\pm 0.05 \qquad q^2\in [0.045, 1.1]~{\rm GeV}^2 \, , \\   
0.96^{+0.45}_{-0.29}\pm 0.11 \qquad q^2\in [1.1, 6]~{\rm GeV}^2 \, , \\  
0.90^{+0.27}_{-0.21}\pm 0.10 \qquad q^2\in [0.1, 8.0]~{\rm GeV}^2 \, , \\  
1.18^{+0.52}_{-0.32}\pm 0.10 \qquad q^2\in [15, 19]~{\rm GeV}^2 \, . 
\end{cases}
\eea
As  the  Belle results have relatively larger uncertainties,  we do not include them in our fit.
\subsubsection{$B_s \to \mu^+ \mu^-$} 
The  combined ATLAS, CMS and LHCb results on the branching franction of  $B_s \to \mu^+ \mu^-$ process  is ~\cite{ATLAS:2020acx}: 
\begin{align}
{\rm Br}(B_s^0\to \mu^+\mu^-) \ = \ \left(2.69^{+0.37}_{-0.35}\right)\times 10^{-9} \, ,
\end{align}
which shows  $2.4\sigma$ discrepancy with   the SM prediction~\cite{Bobeth:2013uxa} 
\begin{align}
{\rm Br}(B_s^0\to \mu^+\mu^-)^{\rm SM} \ = \ \left(3.65\pm 0.23\right)\times 10^{-9} \, .
\end{align}

\subsubsection{Angular observables of  $B\to K^*\mu\mu$ and $B_s\to \phi \mu\mu$ processes} 

\begin{itemize}
 
\item  
 The angular observables of $ B^0 \to K^{*0}  \mu^+ \mu^-$ decay process, such as the form factor independent (FFI) observables:   ($P_{1,2,3}, P_{4,5,6,8}^\prime)$,  longitudinal polarization asymmetry $(F_L)$, and the forward-backward asymmetry  $(A_{FB})$   in the following  $q^2$ bins: $(0.1\to 0.98, ~1.1\to 2,~2\to3, ~3\to 4, ~4\to 5, ~5\to 6 , 1\to 6)$  taken from \cite{Aaij:2015oid}.

\item For  $B_s \to\phi \mu^+ \mu^-$ process, we consider the logitudinal polarization asymmetry $(F_L)$ and CP averaged angular observables ($S_{3,4,7}, A_{5,6,8,9}$) of  in three $q^2$ bins: $0.1\to 2,~ 2\to 5$, and $1\to 6$ \cite{Aaij:2015esa}.
\end{itemize} 

\begin{figure}
\includegraphics[scale=0.55]{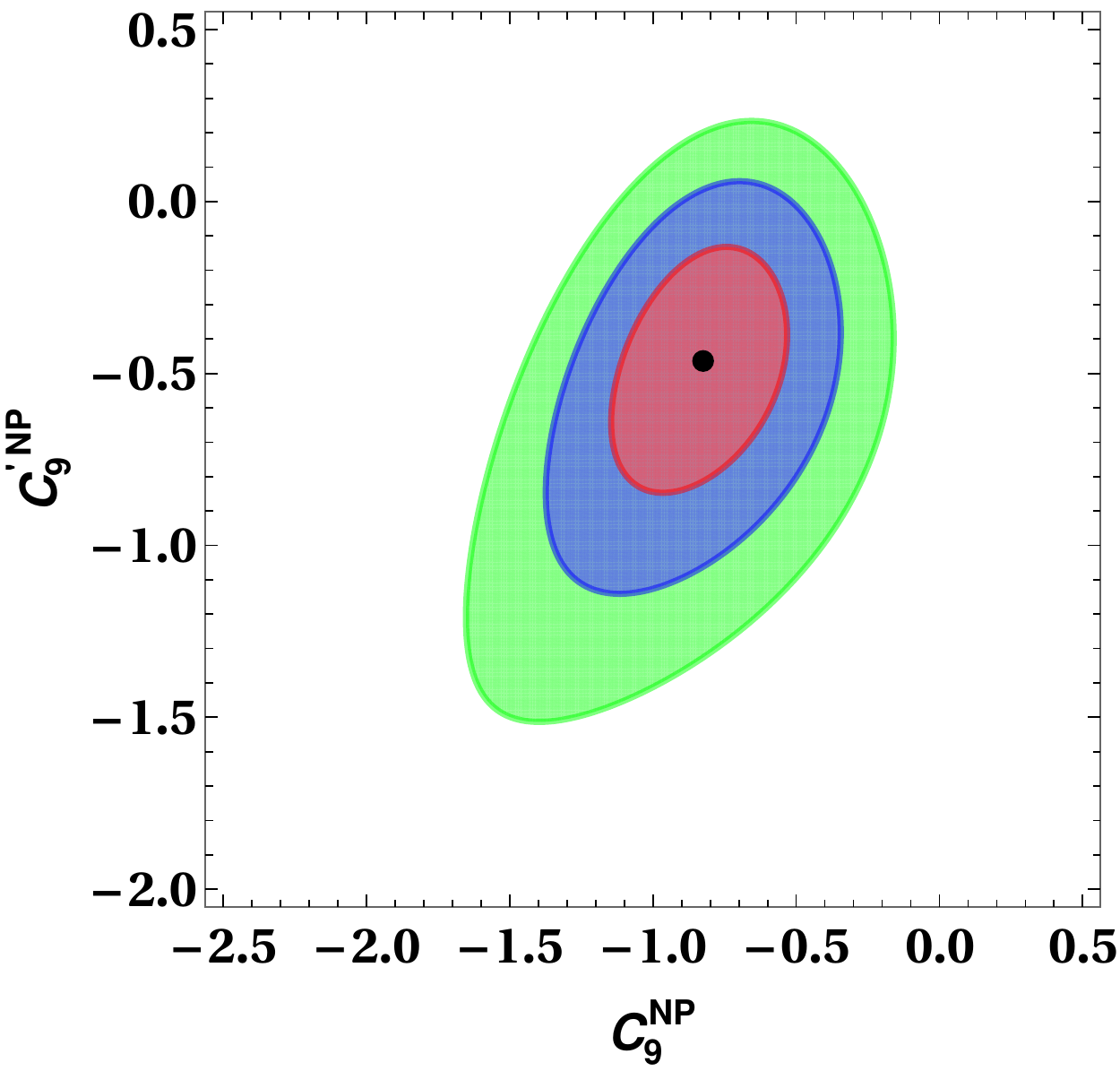}
\hspace*{0.3 true cm}
\includegraphics[scale=0.55]{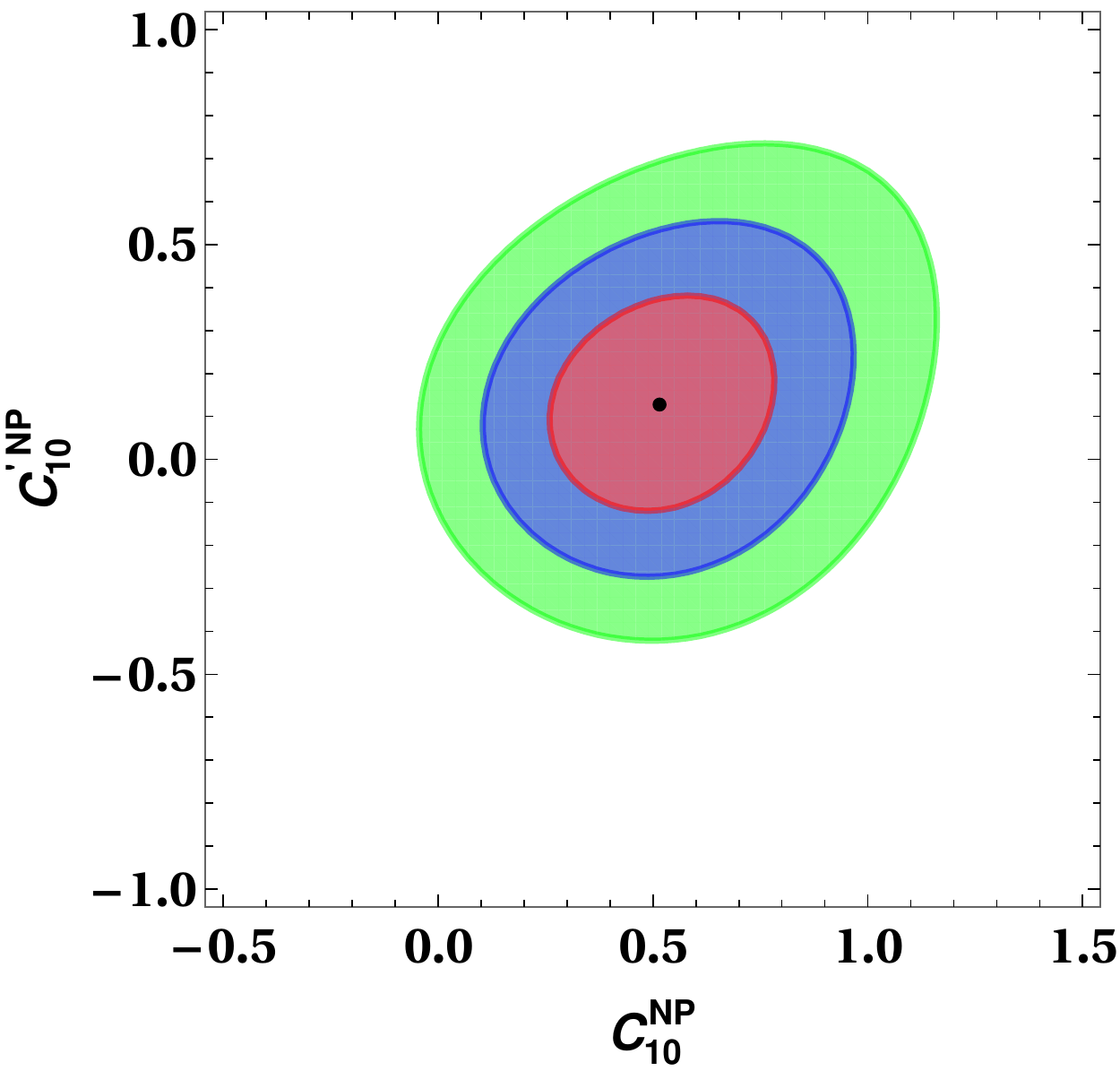}
\includegraphics[scale=0.55]{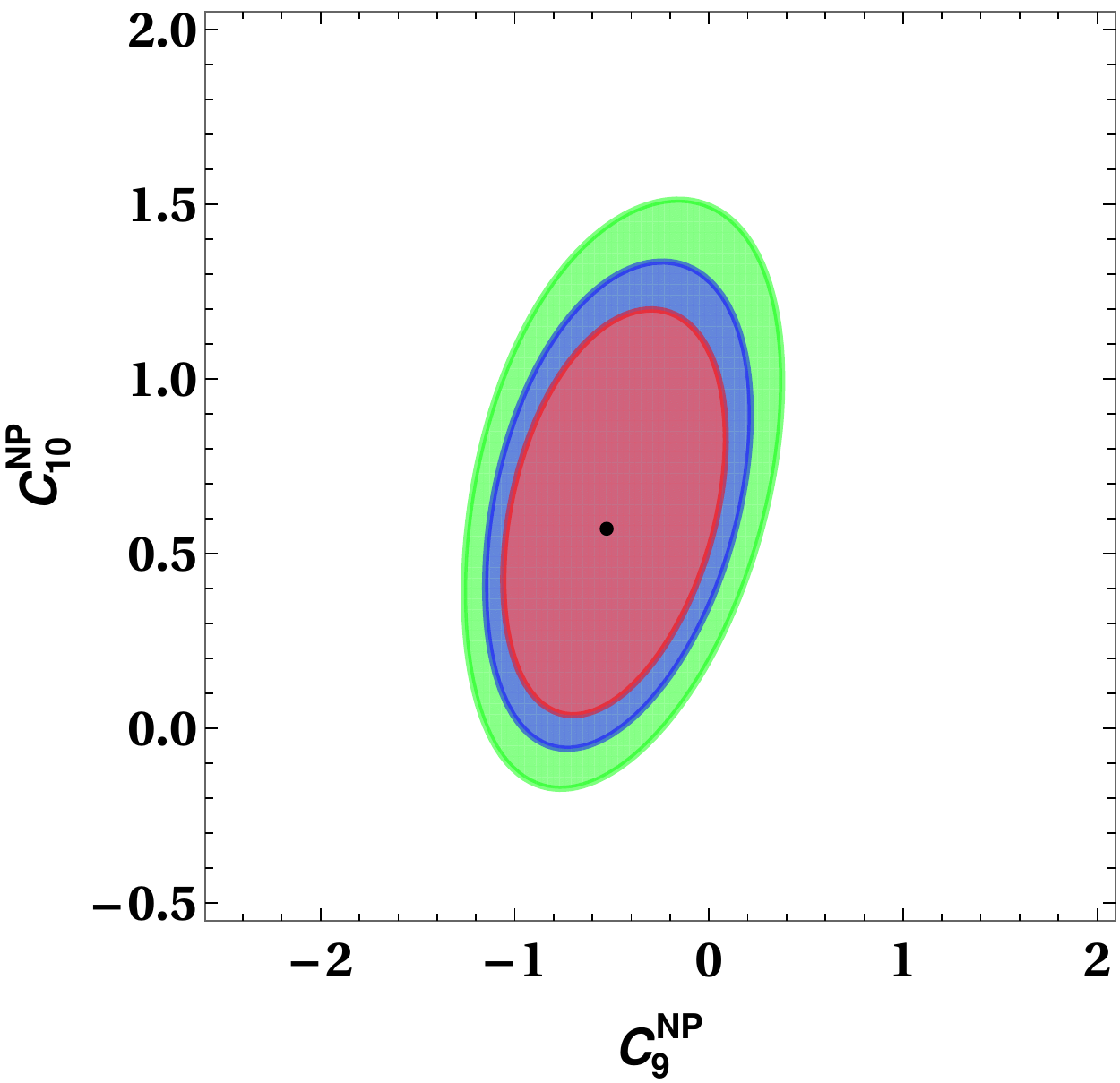}
\caption{Allowed  parameter space in  $C_9^{\rm NP}-C_9^{'\rm NP}$ plane (top-left panel), $C_{10}^{\rm NP}-C_{10}^{'\rm NP}$ plane (top-right panel) and $C_{9}^{\rm NP}-C_{10}^{\rm NP}$ plane (bottom panel). Different colors represent the $1\sigma$,  $2\sigma$ and $3\sigma$
contours and the black points represent the best-fit values. }\label{Fig:constraint}
\end{figure}

The theoretical expressions for different observables of $B \to V \ell^+ \ell^-$  processes where $V$ denotes the vector meson, are used from \cite{Altmannshofer:2008dz} and the form factors are calculated using the light cone sum rule approach \cite{Straub:2015ica}. Using these observables, the new Wilson coefficients are constrained by assuming the presence of two new real coefficients at a time. We consider three possible scenarios, i.e., the simultaneous presence of  ($C_9^{\rm NP}, C_9^{'\rm NP}$), ($C_{10}^{\rm NP},~C_{10}^{'\rm NP} $) and ($C_9^{\rm NP}, C_{10}^{\rm NP}$) new physics coefficients and perform a $\chi^2$ analysis. The expression for $\chi^2$ is 
delineated as
\bea
\chi^2(C_i^{\rm NP})= \sum_i  \frac{\Big ({\cal O}_i^{\rm th}(C_i^{\rm NP}) -{\cal O}_i^{\rm exp} \Big )^2}{(\Delta {\cal O}_i)^2},
\eea
where ${\cal O}_i^{\rm th}(C_i^{\rm NP})$ are the theoretical expectations for the observables used in our fit,  ${\cal O}_i^{\rm exp}$ represent the measured central values of the observables   and $(\Delta {\cal O}_i)^2=(\Delta {\cal O}_i^{\rm exp})^2+(\Delta {\cal O}_i^{\rm th})^2$ encompasses the $1\sigma$ uncertainties from theory and experiment.  In Fig.\ref{Fig:constraint}, we present the allowed parameter space of the new Wilson coefficients  in $C_9^{\rm NP}-C_9^{'\rm NP}$ (top-left panel),    $C_{10}^{\rm NP}-C_{10}^{'\rm NP}$ (top-right panel) and $C_9^{\rm NP}-C_{10}^{\rm NP}$ (bottom panel)  planes, where the red, blue and green colors represent the $1\sigma$, $2 \sigma$ and $3 \sigma$ contours and the black dots characterize the best-fit values. The best-fit values of the new coefficients along with the corresponding $\chi^2_{\rm min}/{\rm d.o.f}$ and the ${\rm pull}=\sqrt{\chi^2_{\rm SM}-\chi^2_{\rm best-fit}}$, for these three scenarios are presented in Table \ref{best-fit}.

\begin{table}[h]
\begin{center}
\caption{The best-fit values of new coefficients, $\chi^2_{\rm min}/{\rm d.o.f}$ and pull values  for different scenarios.  }\label{best-fit}
\begin{tabular}{|c | c | c| c|}
\hline
~New Coefficients~ &~Best-fit Values~&~ $\chi^2_{\rm min}/{\rm d.o.f}$~~&~~Pull~~\\
\hline

$(C_9^{\rm NP},~C_9^{'\rm NP})$ & ~$(-0.829, -0.463)$~ & 1.04 & 4.8 \\
 \hline
$(C_{10}^{\rm NP},~C_{10}^{'\rm NP})$ & $(0.513, ~0.125)$ & 1.3 & 3.0  \\
 \hline
$(C_9^{\rm NP},~C_{10}^{\rm NP})$ & $(-0.526, ~0.573)$ & 1.02 & 5.4 \\
 \hline
\end{tabular}
\end{center}
\end{table}
From Table \ref{best-fit},  it should be noted that   for 
$C_{10}^{\rm NP}-C_{10}^{'\rm NP}$ case, the $\chi^2_{\rm min}/{\rm d.o.f}$  is greater than 1, with a lower pull value,  this scenario is  not very  robust, as also inferred in \cite{Aebischer:2019mlg}. While for $C_{9}^{\rm NP}-C_{9}^{'\rm NP}$ and $(C_9^{\rm NP}, C_{10}^{\rm NP})$ cases, the $\chi^2_{\rm min}/{\rm d.o.f} \simeq 1$, with a larger pull, hence these scenarios are  acceptable.   Therefore, in our analysis, we will consider the impact of three different classes of NP scenarios: the first
scenario includes NP contributions only in operators which are non-zero in the SM,
 and the values of NP coefficients are taken from \cite{Bhom:2020lmk}  as $(C_7^{\rm NP}, C_9^{\rm NP}, C_{10}^{\rm NP})=(0.013, -1.03,0.08)$ (NP1),    in the second case we will consider the presence of $C_9^{\rm NP}-C_9^{'\rm NP}$  and use the extracted  best-fit values of the NP coefficients:  $(-0.829,-0.463)$ (NP2) and for the third case, we consider  the new physics due to  $(C_9^{\rm NP}, C_{10}^{\rm NP})$ Wilson coefficients as $(-0.526,0.573)$ (NP3) on various observables. Since the effect due to the NP3 coefficients are  similar to NP1 case, we have not shown explicitly the corresponding results in the  plots and provided only the corresponding numerical results. 
\section{Differential decay distribution and other relevant Observables}

In this section, we discuss the differential decay distribution and other relevant angular observables like forward-backward asymmetries and lepton polarization asymmetries for the $B \to \left(K_1(1270)/K_1(1400)\right) \mu^+ \mu^-$ processes. As mentioned before, 
the physical states $K_1(1270)$ and $K_1(1400) $ are related to the flavour states $K_{1A}$ and $K_{1B}$ through the relation
\bea
\begin{pmatrix}
|\bar K_1(1270) \rangle \\
|\bar K_1(1270) \rangle
\end{pmatrix} = M  \begin{pmatrix}
|\bar K_{1A} \rangle \\
|\bar K_{1B} \rangle
\end{pmatrix} ,~~~{\rm where}~~~  
M= \begin{pmatrix}
\sin \theta & \cos \theta \\
\cos \theta & -\sin \theta
\end{pmatrix},
\eea
is the mixing matrix with mixing angle  $\theta = -(34 \pm 13)^\circ$ \cite{Hatanaka:2008xj}. 

Now using the effective Hamiltonian given in Eqns. (\ref{Ham-SM}) and (\ref{Ham-NP}), the matrix elements for $B \to K_1 \ell ^+ \ell^-$ process can be obtained  using the relation $
{\cal M} = \langle K_1 \ell^+ \ell^-|{\cal H}_{\rm eff}^{\rm SM} + {\cal H}_{\rm eff}^{\rm NP}| B\rangle $, which requires the knowledge  of the   $\overline B \to {\overline K}_1$  transition form factors.  The required form factors for both vector and axial vector current mediated transitions are defined as 
\bea
&&\langle \overline K_1(p_{K_1}, \varepsilon)|\bar s \gamma_\mu (1\pm \gamma_5) b | \overline{B}(p_B) \rangle = \pm i \frac{2}{m_B +m_{K_1}}\epsilon_{\mu \nu \rho \sigma} \varepsilon^{* \nu} p_B^\rho p_{K_1}^\sigma A^{K_1}(q^2)\nn\\
&&~~~~~~~~- \Big [(m_B +m_{K_1}) \varepsilon_\mu^{*} V_1^{K_1}(q^2)-(p_B+p_{K_1})_\mu (\varepsilon^* \cdot p_B) \frac{V_2^{K_1}(q^2)}{m_B +m_{K_1}} \Big ]\nn\\
&&~~~~~~~~+ 2 m_{K_1} \frac{\varepsilon^* \cdot p_B}{q^2} q_\mu \Big [V_3^{K_1}(q^2)-V_0^{K_1}(q^2) \Big ],
\eea
where $\varepsilon$ is the polarization vector of $K_1$, $V_i(q^2)$'s are the vector form factors and $A(q^2)$ is the axial-vector form factor, which  depend on the square of momentum transfer $q^2$. Analogously, the tensor form factors are expressed as
\bea
&&\langle \overline K_1(p_{K_1}, \varepsilon)|\bar s \sigma_{\mu \nu} q^\nu (1\pm \gamma_5) b| \overline{B}(p_B) \rangle =\pm  2 T_1^{K_1}(q^2) \epsilon_{\mu \nu \rho \sigma} \varepsilon^{* \nu} p_B^\rho p_{K_1}^\sigma \nn\\
&&~~~~~~~~- i T_2^{K_1}(q^2) \Big [(m_B ^2-m_{K_1}^2) \varepsilon_\mu^{*}- (\varepsilon^* \cdot q)(p_B+p_{K_1})_\mu \Big ]\nn\\
&&~~~~~~~~-i T_3^{K_1} (q^2)( \varepsilon^* \cdot q) \Big [  q_\mu -\frac{q^2}{m_B^2 -m_{K_1}^2}(p_{K_1}+p_B)_\mu \Big ],
\eea
with $T_i(q^2)$ as the relevant tensorial  form factors.

Thus, the matrix elements of $B \to K_1(1270)/ K_1(1400)$ processes can be parmetrized in terms of $B\to K_{1A}/K_{1B}$ form factors  as \cite{Li:2011nf}
\bea
\begin{pmatrix}
\langle \bar K_1(1270)|\bar s \gamma_\mu (1 \pm \gamma_5)b|\bar B \rangle \\
\langle \bar K_1(1400)|\bar s \gamma_\mu (1\pm \gamma_5)b|\bar B \rangle 
\end{pmatrix}=M \begin{pmatrix}
\langle \bar K_{1A} |\bar s \gamma_\mu (1 \pm \gamma_5)b|\bar B \rangle \\
\langle \bar K_{1B}|\bar s \gamma_\mu (1\pm \gamma_5)b|\bar B \rangle 
\end{pmatrix},
\eea
and analogously for tensor form factors. 
More explicitly the various form factors are related as
\begin{eqnarray*}
\begin{pmatrix}
A^{K_1(1270)}/(m_B +m_{K_1(1270)})\\
A^{K_1(1400)}/(m_B +m_{K_1(1400)})
\end{pmatrix}=M \begin{pmatrix}
A^{K_{1A}}/(m_B +m_{K_{1A}})\\
A^{K_{1B}}/(m_B +m_{K_{1B}})
\end{pmatrix} ,\nn\\
\end{eqnarray*}
\begin{eqnarray*}
\begin{pmatrix}
(m_B +m_{K_1(1270)}) V_1^{K_1(1270)}\\
(m_B +m_{K_1(1400)}) V_1^{K_1(1400)}
\end{pmatrix}=M \begin{pmatrix}
(m_B +m_{K_{1A}}) V_1^{K_{1A}}\\
(m_B +m_{K_{1B}}) V_1^{K_{1B}}
\end{pmatrix},
\end{eqnarray*}
\begin{eqnarray*}
\begin{pmatrix}
V_2^{K_1(1270)}/(m_B +m_{K_1(1270)})\\
V_2^{K_1(1400)}/(m_B +m_{K_1(1400)})
\end{pmatrix}=M \begin{pmatrix}
V_2^{K_{1A}}/(m_B +m_{K_{1A}})\\
V_2^{K_{1B}}/(m_B +m_{K_{1B}})
\end{pmatrix},
\end{eqnarray*}
\begin{eqnarray*}
\begin{pmatrix}
m_{K_1(1270)} V_0^{K_1(1270)}\\
m_{K_1(1400)} V_0^{K_1(1400)}
\end{pmatrix}=M \begin{pmatrix}
m_{K_{1A}} V_0^{K_{1A}}\\
m_{K_{1B}} V_0^{K_{1B}}
\end{pmatrix},~~~~~\begin{pmatrix}
 T_{1,3}^{K_1(1270)}\\
 T_{1,3}^{K_1(1400)}
\end{pmatrix}=M \begin{pmatrix}
 T_{1,3}^{K_{1A}}\\
 T_{1,3}^{K_{1B}}
\end{pmatrix}, 
\end{eqnarray*}
\begin{eqnarray*}
\begin{pmatrix}
(m_B ^2-m_{K_1(1270)}^2) T_2^{K_1(1270)}\\
(m_B^2-m_{K_1(1400)}^2) T_2^{K_1(1400)}
\end{pmatrix}=M \begin{pmatrix}
(m_B ^2-m_{K_{1A}}^2) T_2^{K_{1A}}\\
(m_B^2 -m_{K_{1B}}^2) T_2^{K_{1B}}
\end{pmatrix}.
\end{eqnarray*}
Additionally, the form factors satisfy the the following relations, which can be obtained using the equation of motion
\bea
&&V_3^{K_1}(0) = V_0^{K_1}(0)\;,~~~~~~~~~T_1^{K_1}(0) = T_2^{K_1}(0)\;,\nn\\
&&V_3^{K_1}(q^2)= \frac{m_B +m_{K_1}}{2 m_{K_1}} V_1^{K_1}(q^2) - \frac{m_B -m_{K_1}}{2 m_{K_1}} V_2^{K_1}(q^2)\;.
\eea
The form factors are calculated in the light cone sum rule (LCSR) approach \cite{Yang:2008xw}, and their $q^2$ dependence  in  the whole kinematical region
 is parametrized in the three parameter form  as
\begin{eqnarray} \label{eq:FFpara}
 F(q^2)=\,{F(0)\over 1-a(q^2/m_{B}^2)+b(q^2/m_{B}^2)^2}\;.
\end{eqnarray}
The values of different parameters involved  in (\ref{eq:FFpara}) are taken from \cite{Hatanaka:2008gu} and are provided in Table \ref{tab:FFinLF}. Now we list below the various observables associated with $B \to K_1 \ell^+ \ell^-$ processes.
 
\begin{table}[htb]
\caption{Transition form factors for $B\to K_{1A},K_{1B}$ processes, obtained in the LCSR approach \cite{Hatanaka:2008gu}, along with the fitted parameters for $q^2$ dependence as shown in Eq. (\ref{eq:FFpara}).} \label{tab:FFinLF}
\begin{ruledtabular}
\begin{tabular}{clll|clll}
      ~~~~$F$~~~~~~
    & ~~~~~$F(0)$~~~~~
    & ~~~$a$~~~
    & ~~~$b$~~
    & ~~~~$F$~~~~~~
    & ~~~~~$F(0)$~~~~~
    & ~~~$a$~~~
    & ~~~$b$~~
 \\
    \hline
$V_1^{BK_{1A}}$
    & $0.34\pm0.07$
    & $0.635$
    & $0.211$
&$V_1^{BK_{1B}}$
    & $-0.29^{+0.08}_{-0.05}$
    & $0.729$
    & $0.074$
    \\
$V_2^{BK_{1A}}$
    & $0.41\pm 0.08$
    & $1.51$
    & $1.18~~$
&$V_2^{BK_{1B}}$
    & $-0.17^{+0.05}_{-0.03}$
    & $0.919$
    & $0.855$
    \\
$V_0^{BK_{1A}}$
    & $0.22\pm0.04$
    & $2.40$
    & $1.78~~$
&$V_0^{BK_{1B}}$
    & $-0.45^{+0.12}_{-0.08}$
    & $1.34$
    & $0.690$
    \\
$A^{BK_{1A}}$
    & $0.45\pm0.09$
    & $1.60$
    & $0.974$
&$A^{BK_{1B}}$
    & $-0.37^{+0.10}_{-0.06}$
    & $1.72$
    & $0.912$
    \\
$T_1^{BK_{1A}}$
    & $0.31^{+0.09}_{-0.05}$
    & $2.01$
    & $1.50$
&$T_1^{BK_{1B}}$
    & $-0.25^{+0.06}_{-0.07}$
    & $1.59$
    & $0.790$
    \\
$T_2^{BK_{1A}}$
    & $0.31^{+0.09}_{-0.05}$
    & $0.629$
    & $0.387$
&$T_2^{BK_{1B}}$
    & $-0.25^{+0.06}_{-0.07}$
    & $0.378$
    & $-0.755$
    \\
$T_3^{BK_{1A}}$
    & $0.28^{+0.08}_{-0.05}$
    & $1.36$
    & $0.720$
&$T_3^{BK_{1B}}$
    & $-0.11\pm 0.02$
    & $-1.61$
    & $10.2$
    \end{tabular}
\end{ruledtabular}
\end{table}

\subsection{Differential decay rate }
The differential decay width with respect to the dilepton invariant mass ($q^2 \equiv s$)  for the process  $\bar B \to \bar K_1 \ell^+ \ell^-$ is given as
\bea
\frac{d {\Gamma}(\bar B \to \bar K_1 \ell^+ \ell^-)}{d \s1}= \frac{G_F^2 \alpha^2 m_B^5 \tau_B}{2^{12} \pi^5~}|V_{tb}V_{ts}^*|^2 ~v \sqrt{\lambda}~ \Delta (\s1),\label{decay-rate}
\eea
where $v=\sqrt{1-4 \mlh^2/\s1}$, $\lambda=1+\mKh^2+\s1^2- 2\s1 -2\mKh -2 \mKh \s1$, with $\mKh=m_{K_1}^2/m_B^2$, $\mlh=m_\ell/m_B$ and $\s1=q^2/m_B^2$. The expression for $\Delta(\s1)$ is given as \cite{Falahati:2014yba}
\bea
\Delta(\s1) &=&- \frac{4}{3} |{\cal F}_1|^2(2 \mlh^2 +\s1) \l1 
+\frac{1}{3 \mKh} |{\cal F}_2|^2\Big (-3-3 \mKh^2 +6 \mKh(1-8 \mlh^2-3 \s1)      +6 \s1-3 \s1^2+v^2 \l1 \Big )\nn\\
&-&\frac{1}{3 \mKh} |{\cal F}_3|^2 \l1 \Big (3+3 \mKh^2 -6 \s1+3 \s1^2-6 \mKh(1+ \s1)-v^2 \l1 \Big )\nn\\
&+& |{\cal F}_5|^2\left ( 4 \mlh^2 \l1  -\frac{\s1}{3}(3+ 3 \mKh^2 -6 \s1 +3\s1^2 -6 \mKh(1+\s1)+v^2 \l1)\right )\nn\\
&+& \frac{1}{3 \mKh}|{\cal F}_6|^2\left (-3- 3 \mKh^2 +6 \mKh(1+16 \mlh^2 -3 \s1) +6 \s1 -3 \s1^2+v^2 \l1 \right )\nn\\
&-& \frac{1}{3 \mKh}|{\cal F}_7|^2 \lambda  \Big ( 3+3 \mKh^2 +12 \mlh^2 (2+2 \mKh -\s1) -6 \s1 +3 \s1^2 -6 \mKh (1+\s1) -v^2 \l1 \Big )\;,\nn\\
&-& \frac{4}{ \mKh}|{\cal F}_8|^2   \mlh^2 \s1 \l1+\frac{8}{ \mKh}{\rm Re}[{\cal F}_6 {\cal F}_8^*]\mlh^2 \l1+\frac{8}{ \mKh}{\rm Re}[{\cal F}_7 {\cal F}_8^*]\mlh^2 \l1(-1+\mKh)\nn\\
&+& \frac{2}{3 \mKh}{\rm Re}[{\cal F}_6  {\cal F}_7^*](12 \mlh^2 \l1 -(-1 +\mKh +\s1)(3 +3 \mKh^2 -6 \s1 +3 \s1^2 -6 \mKh (1+\s1) -v^2 \l1))\nn\\
&-& \frac{2}{3 \mKh}{\rm Re}[{\cal F}_2  {\cal F}_3^*](-1+\mKh+\s1)(3+3 \mKh^2-6 \s1+3 \s1^2-6 \mKh(1+\s1)-v^2 \l1 )\;.
\eea
The functions ${\cal F}_{1,2, \cdots, 8}$ are related to the form factors and the Wilson coefficients and are expressed as
\begin{eqnarray}
{\cal F}_1^{K_1}(\s1)&=& \frac{2}{1+\sqrt{\mKh}}(C_9^{\rm eff}+C_9^{\rm NP}+C_9^{'\rm NP}) A^{K_1}(\s1)+ \frac{4 \hat m_b}{\s1} C_7^{\rm eff} T_1^{K_1}(\s1)\;,\nn\\
{\cal F}_2^{K_1}(\s1)&=& {1+\sqrt{\mKh}}\Big[(C_9^{\rm eff}+C_9^{\rm NP}-C_9^{'\rm NP}) A^{K_1}(\s1)+ \frac{2 \hat m_b}{\s1} (1-\sqrt{\mKh})C_7^{\rm eff} T_2^{K_1}(\s1)\Big]\;,\nn\\
{\cal F}_3^{K_1}(\s1)&=& \frac{1}{1-{\mKh}}\Big[(1-\sqrt{\mKh}) (C_9^{\rm eff}+C_9^{\rm NP}-C_9^{'\rm NP}) V_2^{K_1}(\s1)\nn\\
&&~~~~~+ \frac{2 \hat m_b}{\s1} C_7^{\rm eff}\left ( T_3^{K_1}(\s1) + \frac{1- \mKh}{\s1}T_2^{K_1}(\s1)\right )\Big]\;,\nn\\
{\cal F}_4^{K_1}(\s1)&=& \frac{1}{\s1}\Big[(C_9^{\rm eff} +C_9^{\rm NP}-C_9^{'\rm NP})\Big( (1+\sqrt{\mKh})  V_1^{K_1}(\s1)- (1-\sqrt{\mKh})V_2^{K_1}(\s1) - 2 \sqrt{\mKh} V_0^{K_1}(\s1) \Big) \nn\\
&&~~~~~- 2 \hat m_b  C_7^{\rm eff}  T_3^{K_1}(\s1) \Big]\;,\nn\\
{\cal F}_5^{K_1}(\s1)&=& \frac{2}{1+\sqrt{\mKh}}( C_{10}+C_{10}^{\rm NP}+C_{10}^{'\rm NP}) A^{K_1}(\s1)\;,\nn\\
{\cal F}_6^{K_1}(\s1)&=& \left(1+\sqrt{\mKh}\right ) (C_{10}+C_{10}^{\rm NP}-C_{10}^{'\rm NP})  V_1^{K_1}(\s1)\;,\nn\\
{\cal F}_7^{K_1}(\s1)&=& \frac{1}{1+\sqrt{\mKh}}( C_{10}+C_{10}^{\rm NP}-C_{10}^{'\rm NP})  V_2^{K_1}(\s1)\;,\nn\\
{\cal F}_8^{K_1}(\s1)&=& \frac{1}{\s1}(C_{10}+C_{10}^{\rm NP}-C_{10}^{'\rm NP})  \Big[ (1+\sqrt{\mKh}) V_1^{K_1}(\s1)-(1-\sqrt{\mKh}) V_2^{K_1} -2 \sqrt{\mKh} V_0^{K_1}(\s1)\Big]. \hspace{0.5 true cm} 
\eea

\subsection{LFU violating observable}
Analogous to $R_{K^{(*)}}$, the lepton flavor universality violating observable in $B \to K_1 \ell^+ \ell^-$ processes can be defined as
\bea
R_{K_1}(q^2)= \frac{d{\rm  Br}(B \to K_1 \mu^+ \mu^-)/d q^2}{d{\rm Br}(B \to K_1 e^+ e^-)/d q^2}\;.
\eea
\subsection{$R_\mu$ observable}
The observable $R_\mu$ is  defined as
\bea
R_{\mu}(q^2)= \frac{d{\rm  Br}(B \to K_1(1400) \mu^+ \mu^-)/d q^2}{d{\rm Br}(B \to K_1(1270) \mu^+ \mu^-)/d q^2}\;.
\eea
Since the $K_1$ mesons depend on the mixing angle $\theta$, $R_\mu$ can be used for its determination.

\subsection{Forward-backward asymmetries}
The  unpolarized forward-backward asymmetry,  defined as
\bea
A_{\rm FB}(\s1)=\left (\int_{-1}^0  d \cos \theta_\ell \frac{d^2 \Gamma}{ d \s1 d \cos \theta_\ell}- \int_{0}^1 d \cos \theta_\ell \frac{d^2 \Gamma}{d \s1~ d \cos \theta_\ell} \right )\Big/\frac{ d\Gamma}{d\s1}
\;,
\eea
where $\theta_\ell$ represents the angle between the initial $B$ meson and final lepton  $\ell^-$ in the C.o.M. frame of the outgoing lepton pair.  In terms of the angular amplitudes, it can  be expressed  as
\bea
A_{\rm FB}(\s1)=\frac{2 }{\Delta} v \sqrt{\lambda}\s1 \Big[ 2 {\rm Re}({\cal F}_1 {\cal F}_6^*) + {\rm Re}({\cal F}_2 {\cal F}_5^*) \Big].
\eea
Next, we focus on the  differential forward-backward asymmetries, that are  associated with the polarized leptons. In this regard,  first we  define two sets of orthogonal vectors belonging to the polarization of $\ell^-$ and $\ell^+$, which are denoted as $S_i$ and $W_i$, with $i=L,N$ and $T$, corresponding to longitudinal, normal and transverse spin projections:
\bea
&&S_L^\mu \equiv (0, {\bf e}_L)=\left (0, \frac{{\bf p}_{\ell^-}}{|{\bf p}_{\ell^-}|}  \right ),\nn\\
&& S_N^\mu \equiv (0, {\bf e}_N)=\left (0, \frac{{\bf p}_{K_1}\times {\bf p}_{\ell^-}}{|{\bf p}_{K_1}\times {\bf p}_{\ell^-|}}  \right ),\nn\\
&& S_T^\mu \equiv (0, {\bf e}_T)=\left (0, {\bf e}_N \times{{\bf e}_L}  \right ),\nn\\
&&W_L^\mu \equiv (0, {\bf w}_L)=\left (0, \frac{{\bf p}_{\ell^+}}{|{\bf p}_{\ell^+}|}  \right ),\nn\\
&& W_N^\mu \equiv (0, {\bf w}_N)=\left (0, \frac{{\bf p}_{K_1}\times {\bf p}_{\ell^+}}{|{\bf p}_{K_1}\times {\bf p}_{\ell^+}|}  \right ),\nn\\
&& W_T^\mu \equiv (0, {\bf w}_T)=\left (0, {\bf w}_N \times{{\bf w}_L}  \right ),
\eea 
where ${\bf p}_{\ell^\mp}$,  and ${\bf p}_{K_1}$ represent the three-momenta of the outgoing particles  $\ell^\mp$,  and $K_1$ respectively. It should be emphasized that the polarization vectors $S_i^\mu~(W_i^\mu)$ are defined in the  $\ell^-(\ell^+)$ rest frame.  Thus, while  Lorentz boost is applied to bring these  vectors from the rest frame of $\ell^-$ and $\ell^+$ to the C.o.M. frame of $\ell^- \ell^+$ system, only longitudinal component  gets boosted, while the other two components remain unchanged. Hence,  the longitudinal polarization four vectors have the form
\bea
S_L^\mu= \left (\frac{ |{\bf p}_{\ell^-}|}{m_\ell}, \frac{E_\ell ~{\bf p}_{\ell^-}}{m_\ell ~|{\bf p}_{\ell^-}|} \right ),~~~~~W_L^\mu= \left (\frac{|{\bf p}_{\ell^+}|}{m_\ell}, \frac{E_\ell~ {\bf p}_{\ell^+}}{m_\ell~ |{\bf p}_{\ell^+}|} \right ).
\eea
The polarized forward-backward asymmetry is defined as \cite{Falahati:2014yba}
\bea
&&A_{\rm FB}(\s1)=\left (\frac{ d\Gamma}{d\s1}\right )^{-1}   \Big \{\int_{0}^1  d \cos \theta_\ell   -  \int_{-1}^0 d \cos \theta_\ell\Big \}\Big\{  \Big[
\frac{d^2 \Gamma({\bf s^-}={\bf i},{\bf s^+}={\bf j})}{ d \s1 d \cos \theta_\ell}-\frac{d^2 \Gamma({\bf s^-}={\bf i},{\bf s^+}=-{\bf j})}{ d \s1 d \cos \theta_\ell} \Big] \nn\\
&&~~~~~~~~~~~~~~~~~~~~~~~- \Big[
\frac{d^2 \Gamma({\bf s^-}=-{\bf i},{\bf s^+}={\bf j})}{ d \s1 d \cos \theta_\ell}-\frac{d^2 \Gamma({\bf s^-}=-{\bf i},{\bf s^+}=-{\bf j})}{ d \s1 d \cos \theta_\ell} \Big]\Big \} \nn\\
&&~~~~~~~~~~=A_{\rm FB}({\bf s^-}={\bf i},{\bf s^+}={\bf j})-   A_{\rm FB}({\bf s^-}={\bf i},{\bf s^+}=-{\bf j})
-A_{\rm FB}({\bf s^-}=-{\bf i},{\bf s^+}={\bf j})\nn\\
&&~~~~~~~~~~~~~~~~~~~~~~~ +A_{\rm FB}({\bf s^-}=-{\bf i},{\bf s^+}=-{\bf j})
\;,
\eea
where ${\bf s}^\mp$ are the spin projections of $\ell^\mp$ and ${\bf i,j}=L,N,T$, are the unit vectors. Thus, the expressions for double polarized forward-backward asymmetries are given as:
\bea
&&A_{\rm FB}^{LL}= \frac{2 }{\Delta} v \sqrt{\lambda}\s1 \Big[ 2 {\rm Re}({\cal F}_1 {\cal F}_6^*) + {\rm Re}({\cal F}_2 {\cal F}_5^*) \Big],\nn\\
&&A_{\rm FB}^{LN}= \frac{4 v \lambda}{3 \mKh \sqrt{\s1} \Delta}{\rm Im}\Big[\mlh  \lambda ({\cal F}_3{ \cal F}_7^*)+ \mlh   ({\cal F}_3{ \cal F}_6^*+{\cal F}_2{ \cal F}_7^*)(-1+ \mKh +\s1)\nn\\
&&~~~~~~~~~~~~+ \mlh  ({\cal F}_2{ \cal F}_6^*) - \mlh \s1 \mKh  ({\cal F}_1{ \cal F}_5^*)\Big]\;,
\nn\\
&&A_{\rm FB}^{LT}= \frac{4 \lambda}{3 \mKh \sqrt{\s1}\Delta}\Big[ \mlh \lambda |{\cal F}_3|^2 + 2 \mlh {\rm Re}[{\cal F}_2 {\cal F}_3^*](-1+\mKh +\s1)  +\mlh |{\cal F}_2|^2 -\mlh \s1 \mKh |{\cal F}_1|^2\Big]\;,\nn\\
&& A_{\rm FB}^{NT}=\frac{2 \sqrt{\lambda}}{\mKh \sqrt{\s1}\Delta} {\rm Im}\Big [-2 \mlh^2 \lambda ({\cal F}_3 {\cal F}_7^*)(1-\mKh) + 2 \mlh^2 \lambda ({\cal F}_3 {\cal F}_6^*) \nn\\
&&~~~~~~~- 2 \mlh^2  ({\cal F}_2 {\cal F}_6^*)(1-\mKh -\s1)  +2 \mlh^2  ({\cal F}_2 {\cal F}_7^*)(1-\mKh)(1-\mKh -\s1)\nn\\
&&~~~~~~- 2 \mlh^2\s1 \lambda ({\cal F}_3 {\cal F}_8^*) + 2 \mlh^2 \s1 ({\cal F}_2 {\cal F}_8^*) (1-\mKh -\s1)  \Big ],
 \eea
 along with the relations
\bea
A_{\rm FB}^{LN}=A_{\rm FB}^{NL},~~~~A_{\rm FB}^{LT}=A_{FB}^{TL},~~~A_{\rm FB}^{TN} = -A_{\rm FB}^{NT}.
\eea
It should be noted that $A_{\rm FB}^{LL}$ has the same form as the unpolarized forward-backward asymmetry $A_{\rm FB}$.
\subsection{Lepton polarization Asymmetries} 
Next, we pay our attention to  the single-lepton polarization asymmetry parameters in $B \to K_1 \ell^+ \ell^- $, defined as
\bea
P_i=\frac{d \Gamma({\bf s}^\pm={\bf i})/d \s1- d \Gamma({\bf s}^\pm=-{\bf i})/d \s1 }{d \Gamma({\bf s}^\pm={\bf i})/d \s1+d \Gamma({\bf s}^\pm=-{\bf i})/d \s1}\;,
\eea
where ${\bf i}$ denotes the unit vector along longitudinal ($L$), normal ($ N$) and transverse ($T$)  polarization directions of the lepton and ${\bf s}^\pm$ denote the spin direction of $\ell^\pm$. The  polarized and unpolarized invariant dilepton mass spectra for the $B \to K_1 \ell^+ \ell^-$ processes are  related as
\bea
\frac{d \Gamma({\bf s}^\pm)}{ds}= \frac{1}{2}\left (\frac{d \Gamma}{ds} \right )\Big[1+\left (P_L {\bf e}_L+P_N {\bf e}_N +P_T {\bf e}_T \right )\cdot {\bf s}^\pm \Big ].
\eea
Thus. by using the decay rate (\ref{decay-rate}), one can obtain the  expressions for the single polarization asymmetries as \cite{Bashiry:2009wq}:
\bea
P_L &=& \frac{1}{3 \mKh \Delta}\Big[ 2 {\rm Re}[{\cal F}_2 {\cal F}_7^*] v(\mKh+\s1-1)(3 \mKh^2 -6(\s1+1)\mKh+3(\s1-1)^2 -\l1)\nn\\
&+&  2 {\rm Re}[{\cal F}_3 {\cal F}_6^*]v(\mKh+\s1-1)(3 \mKh^2 -6(\s1+1)\mKh+3(\s1-1)^2 -\l1)\nn\\
&+&  2 {\rm Re}[{\cal F}_3 {\cal F}_7^*]v \l1 (3 \mKh^2 -6(\s1+1)\mKh+3(\s1-1)^2 -\l1)\nn\\
&+& 2 \mKh  {\rm Re}[{\cal F}_1 {\cal F}_5^*]v\s1 (3 \mKh^2-6(\s1+1)\mKh+3(\s1-1)^2+\l1)\nn\\
&-& {2}  {\rm Re}[{\cal F}_2 {\cal F}_6^*]v (\l1 -3(\mKh^2+(6 \s1-2)\mKh +(\s1-1)^2))\Big]\;.
\eea
\bea
P_T  &=& \frac{\pi \mlh \sqrt{\l1}}{\Delta} \Big [\frac{{\rm Re}[{\cal F}_3 {\cal F}_8^*] \sqrt{\s1} \l1}{\mKh} - \frac{{\rm Re}[{\cal F}_3 {\cal F}_6^*]  \l1}{\mKh \sqrt{\s1}} - \frac{{\rm Re}[{\cal F}_3 {\cal F}_7^*] (\mKh -1) \l1}{\mKh \sqrt{\s1}}\nn\\
&+& \frac{{\rm Re}[{\cal F}_2 {\cal F}_8^*]  \sqrt{\s1} (\mKh +\s1 -1)}{\mKh}
-  \frac{{\rm Re}[{\cal F}_2 {\cal F}_6^*]  (\mKh +\s1 -1)}{\mKh  \sqrt{\s1}}
\nn\\
&-& \frac{{\rm Re}[{\cal F}_2 {\cal F}_7^*]  (\mKh-1)(\mKh +\s1 -1)}{\mKh  \sqrt{\s1}}+ 4{\rm Re}[{\cal F}_1 {\cal F}_2^*] \sqrt{\s1}
\Big].
\eea
\bea
P_N &=&-\frac{ \pi \mlh \sqrt{\l1}}{\Delta} \Big [\frac{{\rm Im}[{\cal F}_7 {\cal F}_8^*] \sqrt{\s1} \l1}{\mKh} +\frac{{\rm Re}[{\cal F}_6 {\cal F}_7^*]  \sqrt{\s1} (-3 \mKh +\s1-1)}{\mKh }\nn\\
&+& \frac{{\rm Im}[{\cal F}_6 {\cal F}_8^*]  \sqrt{\s1} (\mKh +\s1 -1)}{\mKh}
-  2{\rm Im}[{\cal F}_1 {\cal F}_6^*]\sqrt{\s1} -  2{\rm Im}[{\cal F}_1 {\cal F}_5^*]\sqrt{\s1} \Big].
\eea
The averaged asymmetries can be obtained by using the formula
\bea
\langle P_i \rangle =\displaystyle{\frac {\int_{4 m_\ell^2}^{s_{\rm max}^2} P_i ~\frac{d \Gamma}{ds} ds}{\int_{4 m_\ell^2}^{ s_{\rm max}^2 }\frac{d \Gamma}{ds} ds}}\;,
\eea
where $s_{\rm max}=(m_B-m_{K_1})^2$.
\begin{figure}
\includegraphics[scale=0.5]{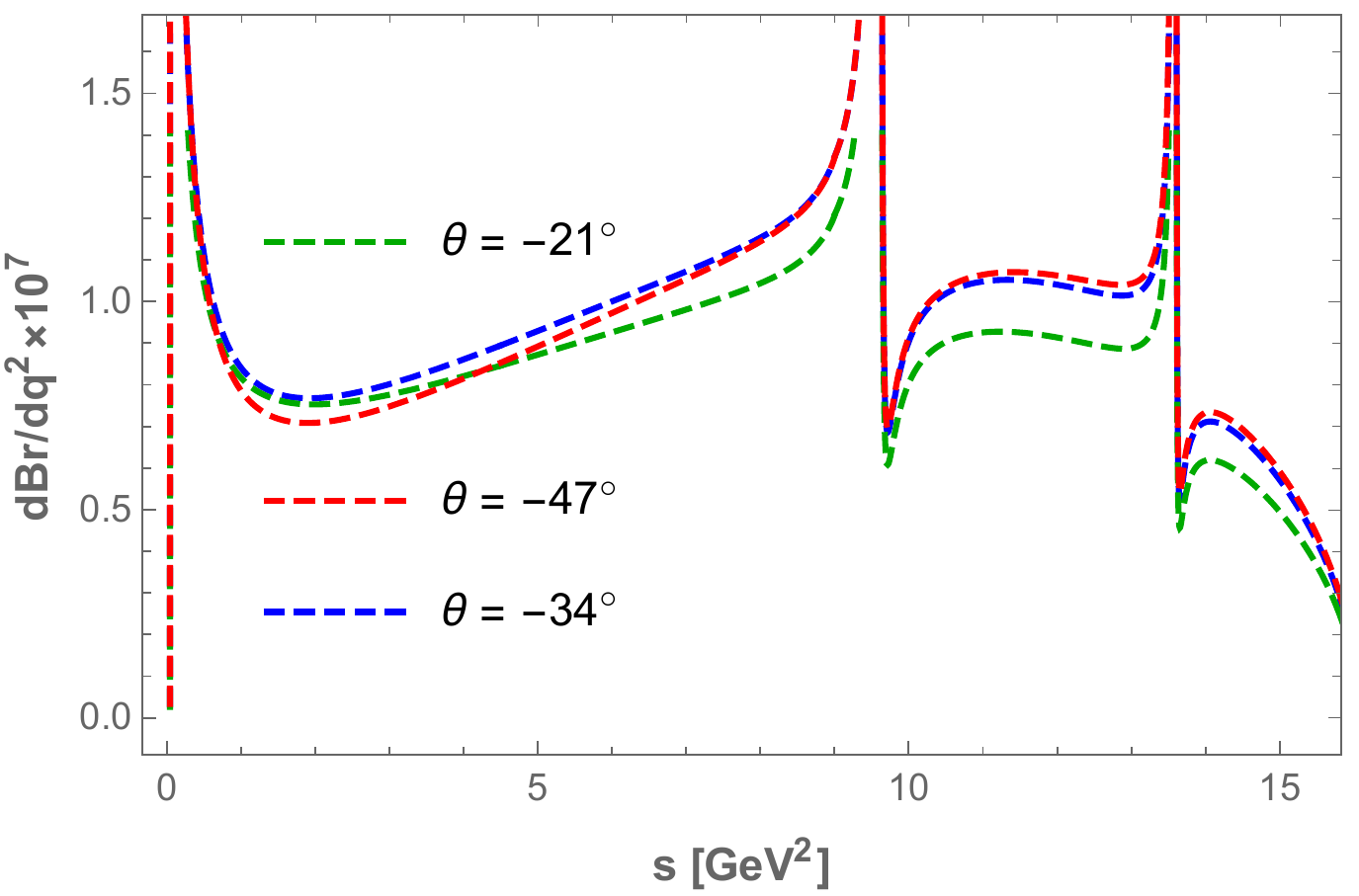}
\quad
\includegraphics[scale=0.5]{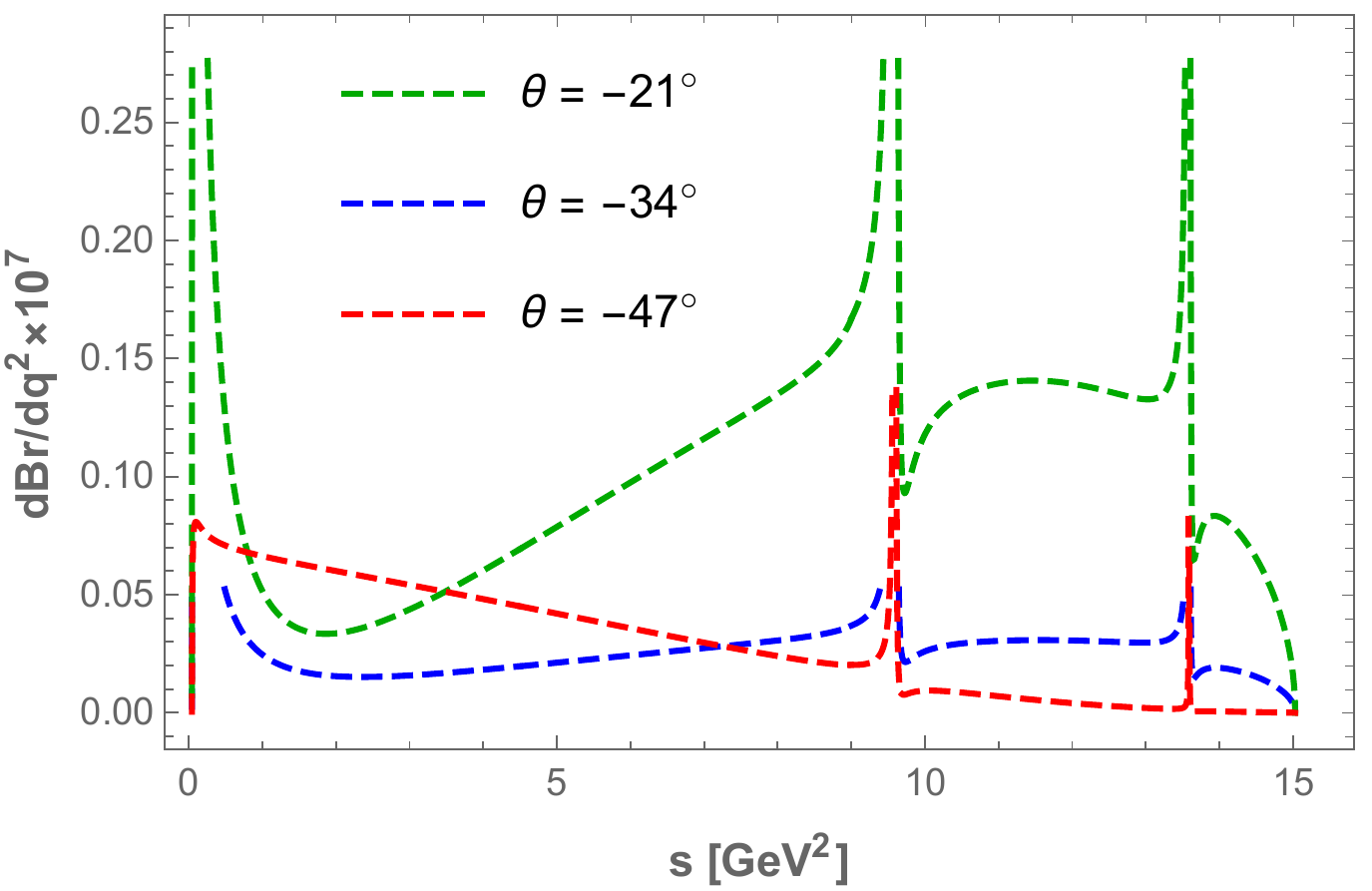}
\quad
\includegraphics[scale=0.5]{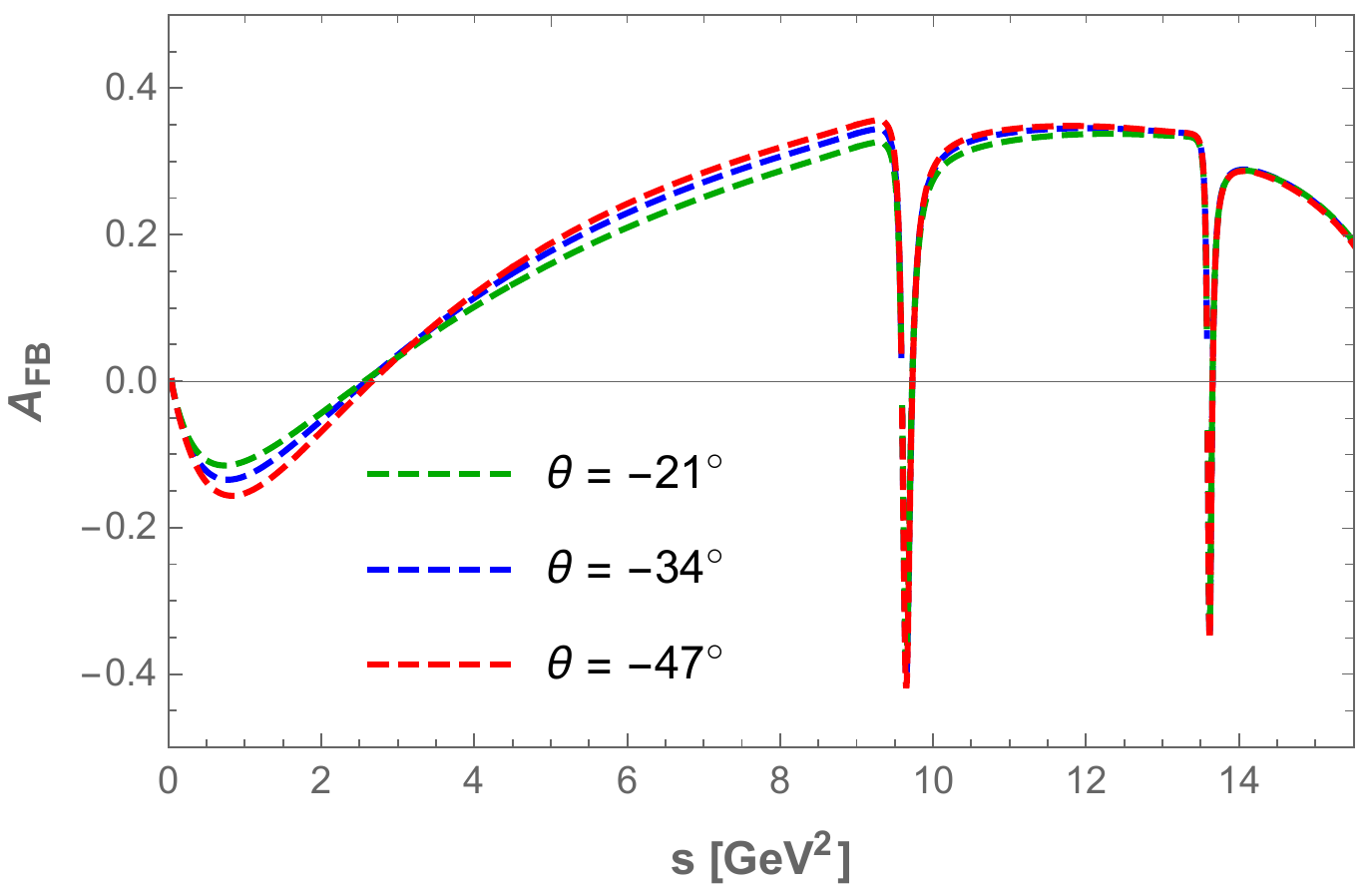}
\quad
\includegraphics[scale=0.5]{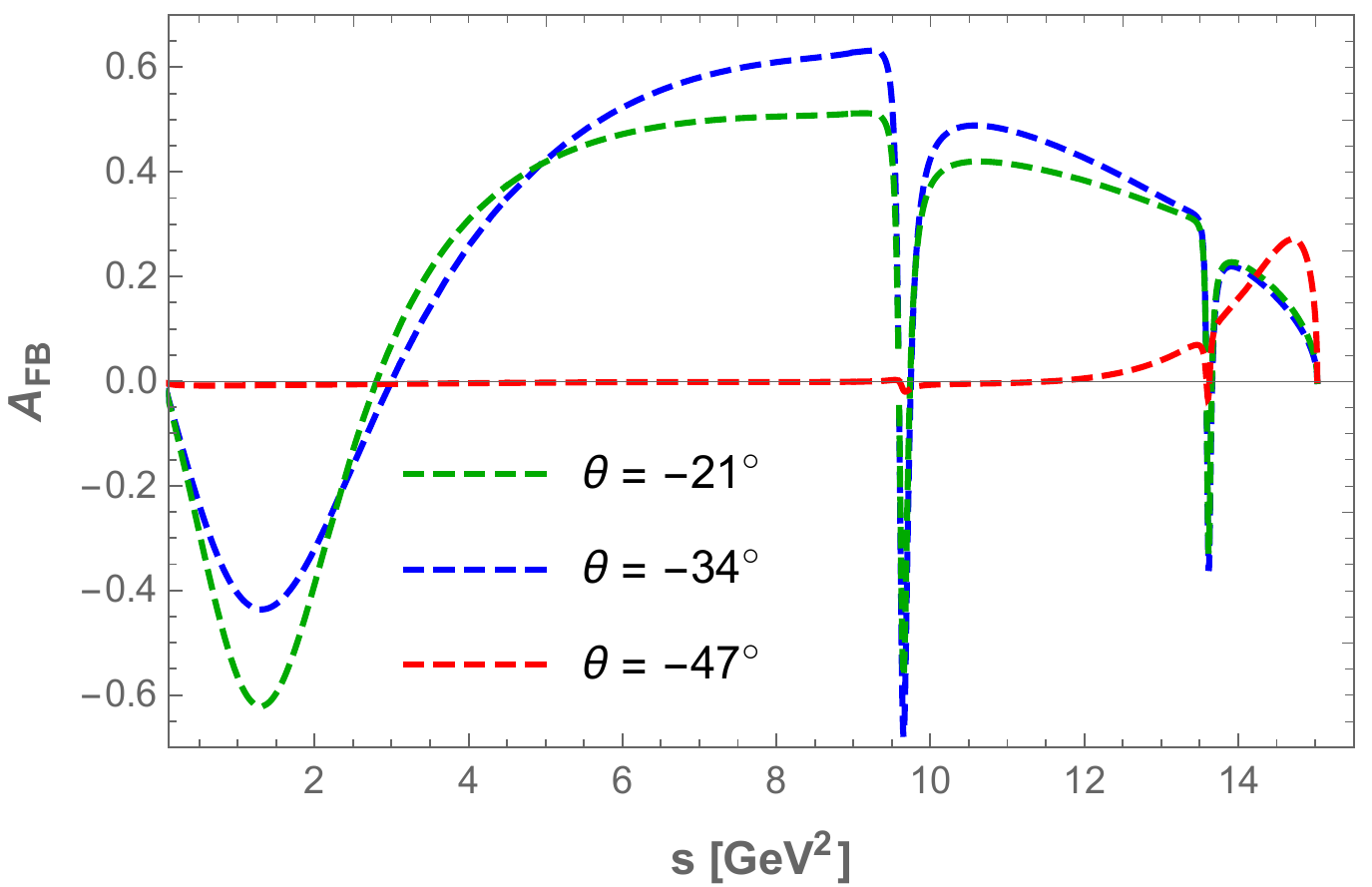}
\quad
\includegraphics[scale=0.5]{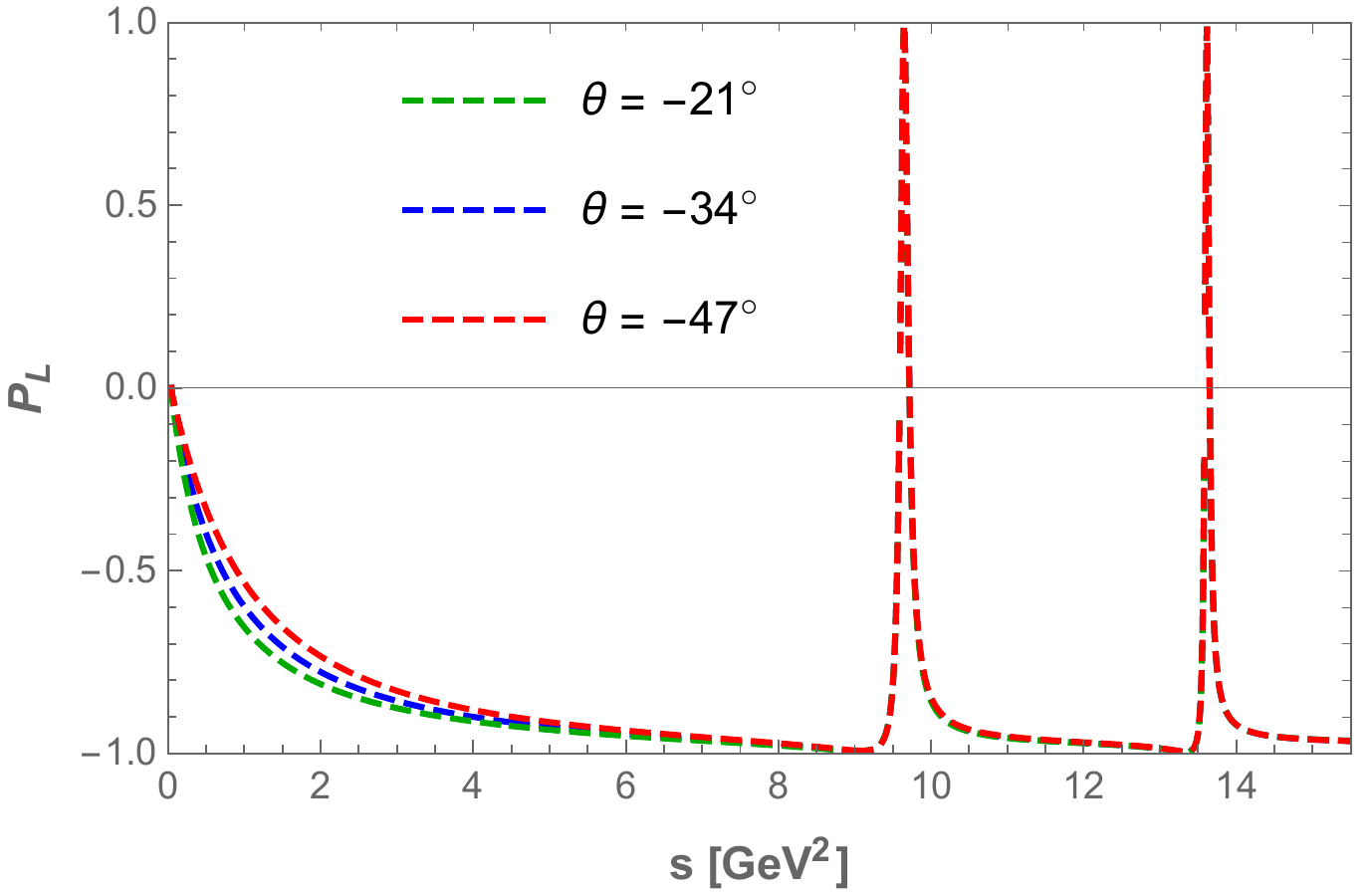}
\quad
\includegraphics[scale=0.5]{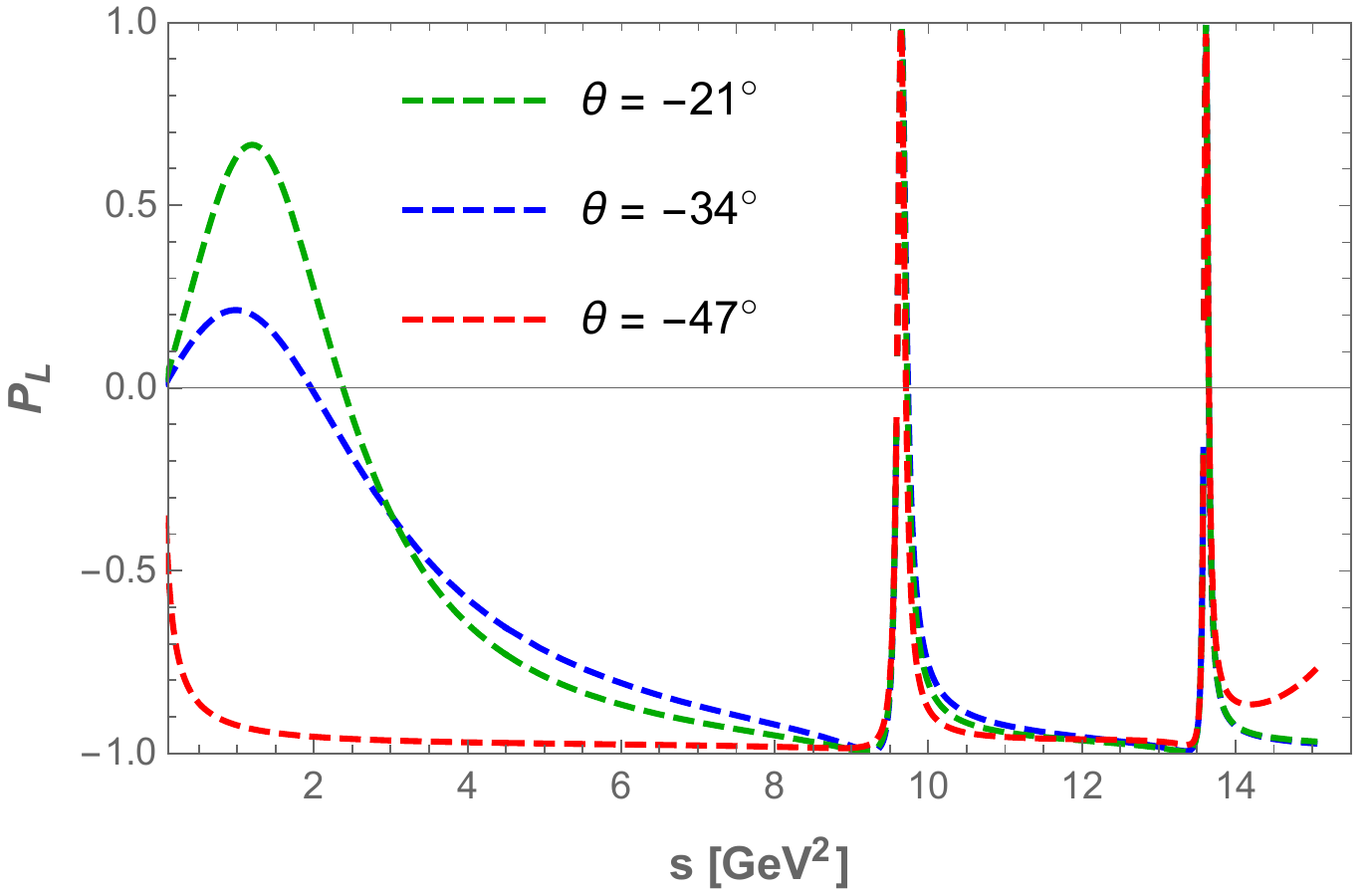}
\caption{Variation of differential branching ratio, forward-backward asymmetry and longitudinal polarization fraction with $s$ for different values of the mixing angle $\theta$. The plots in the left  panel are for $B \to K_1(1270) \mu^+ \mu^-$ and  those in the right panel are for $B \to K_1(1400) \mu^+ \mu^-$ process}\label{Fig:diff-theta}
\end{figure} 
\begin{figure}
\includegraphics[scale=0.5]{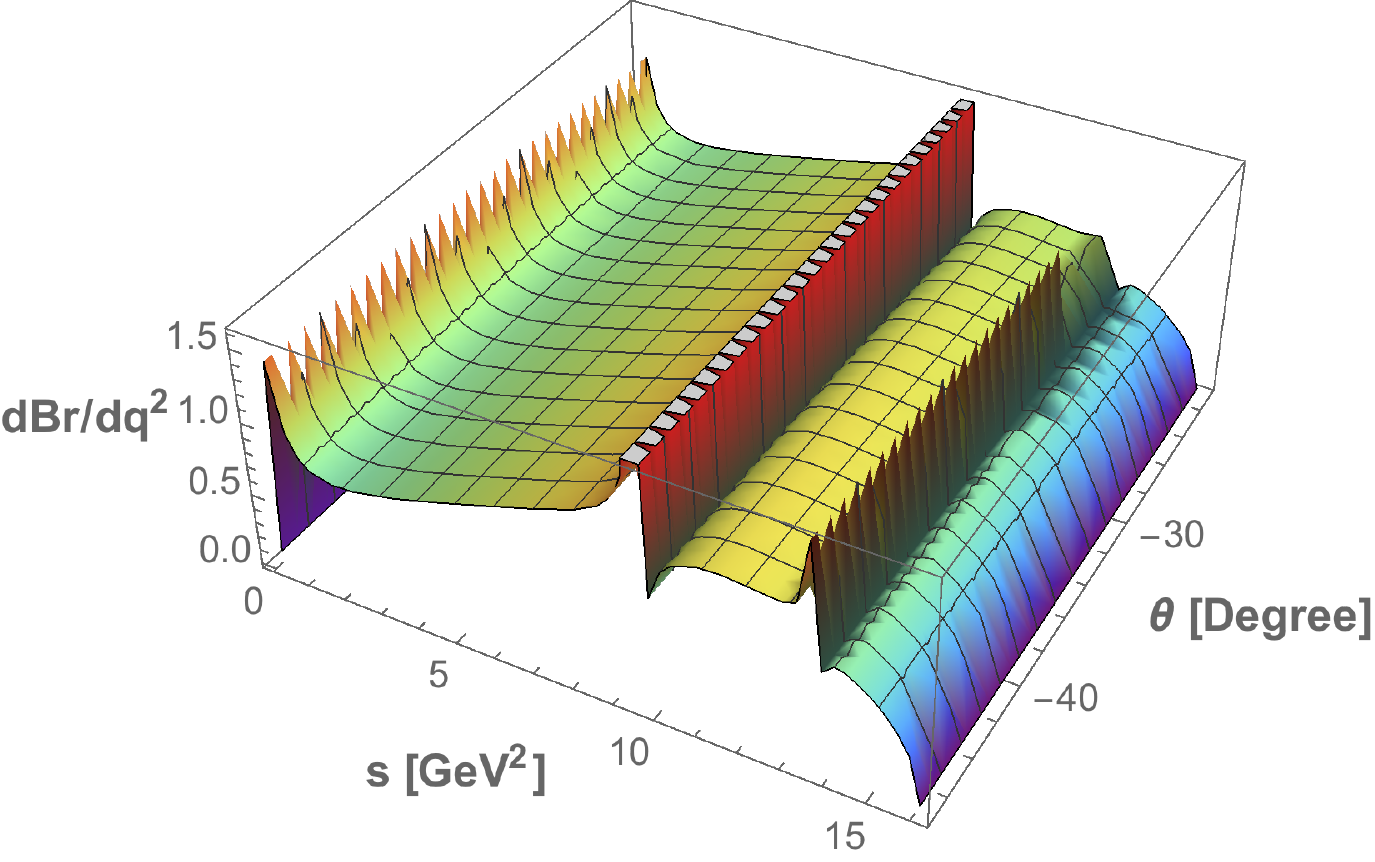}
\quad
\includegraphics[scale=0.5]{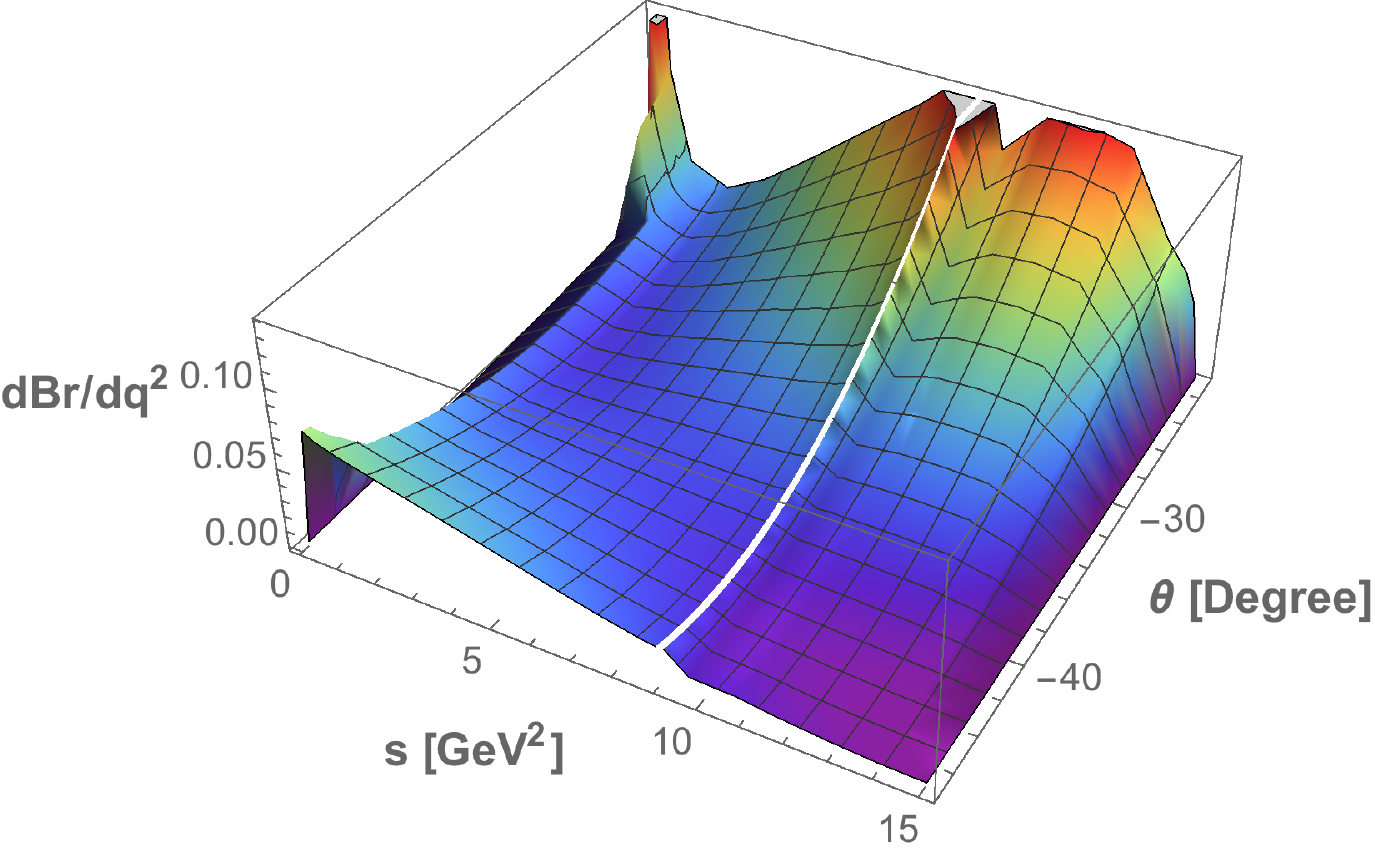}
\quad
\includegraphics[scale=0.5]{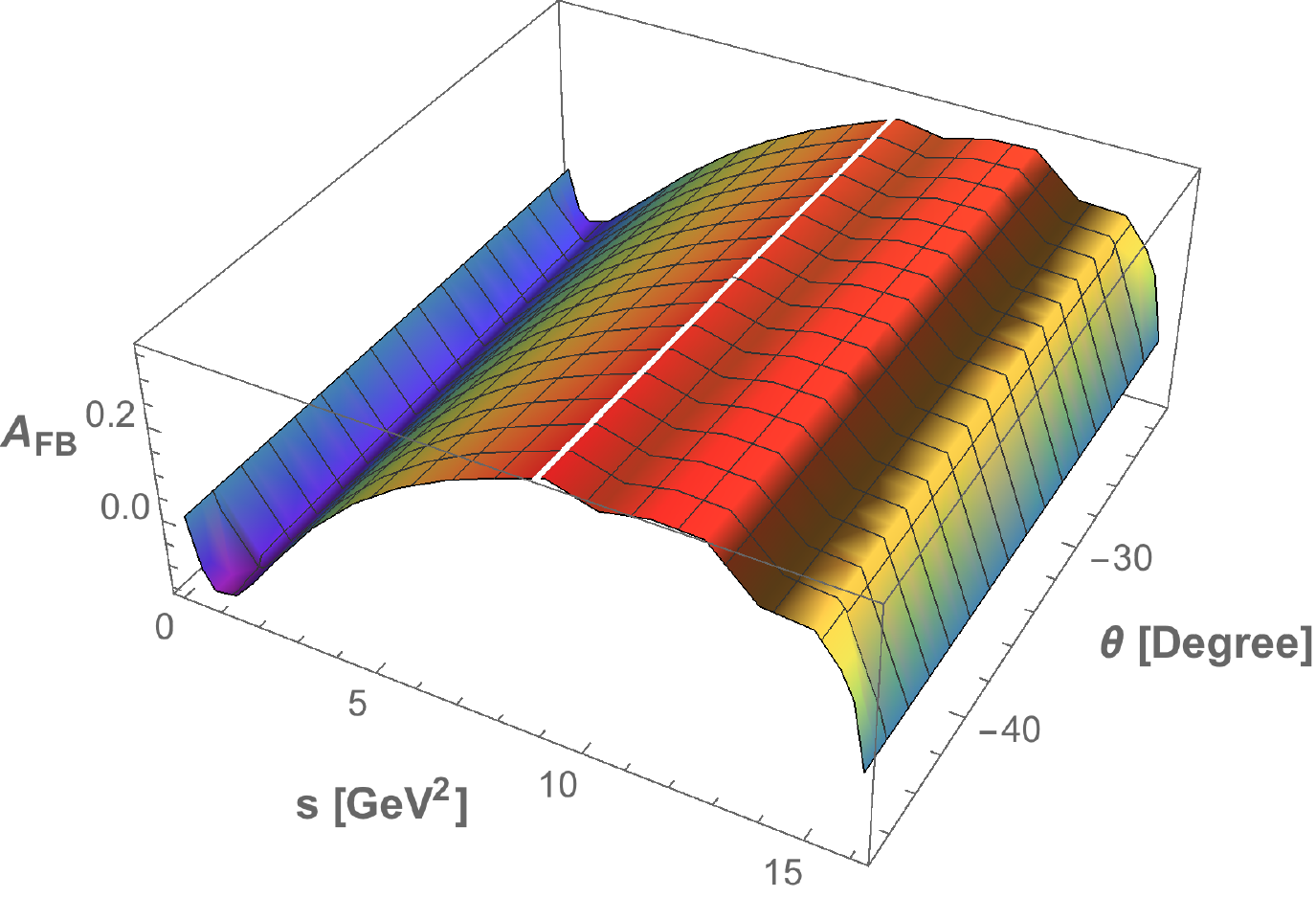}
\quad
\includegraphics[scale=0.5]{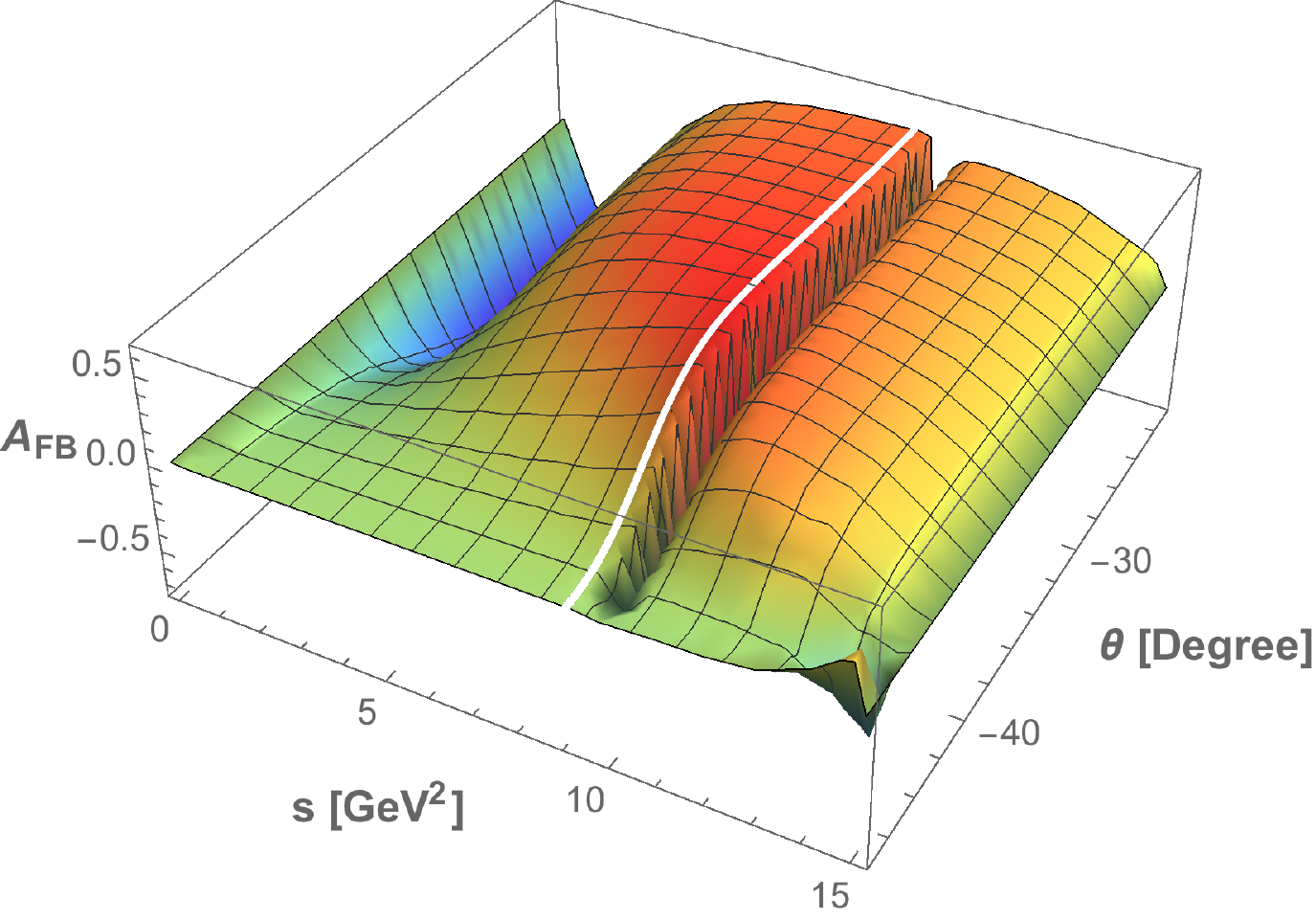}
\quad
\includegraphics[scale=0.5]{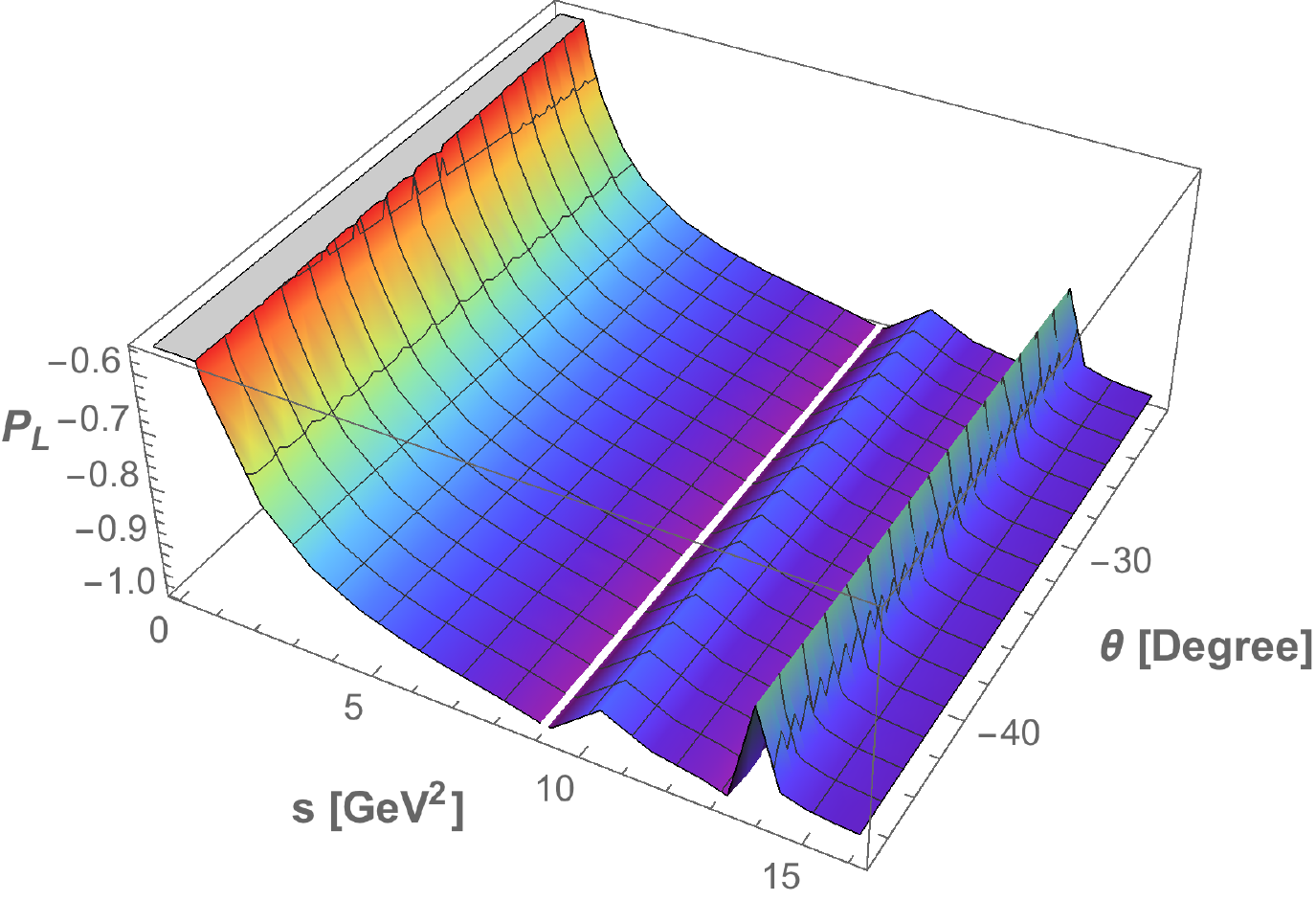}
\quad
\includegraphics[scale=0.5]{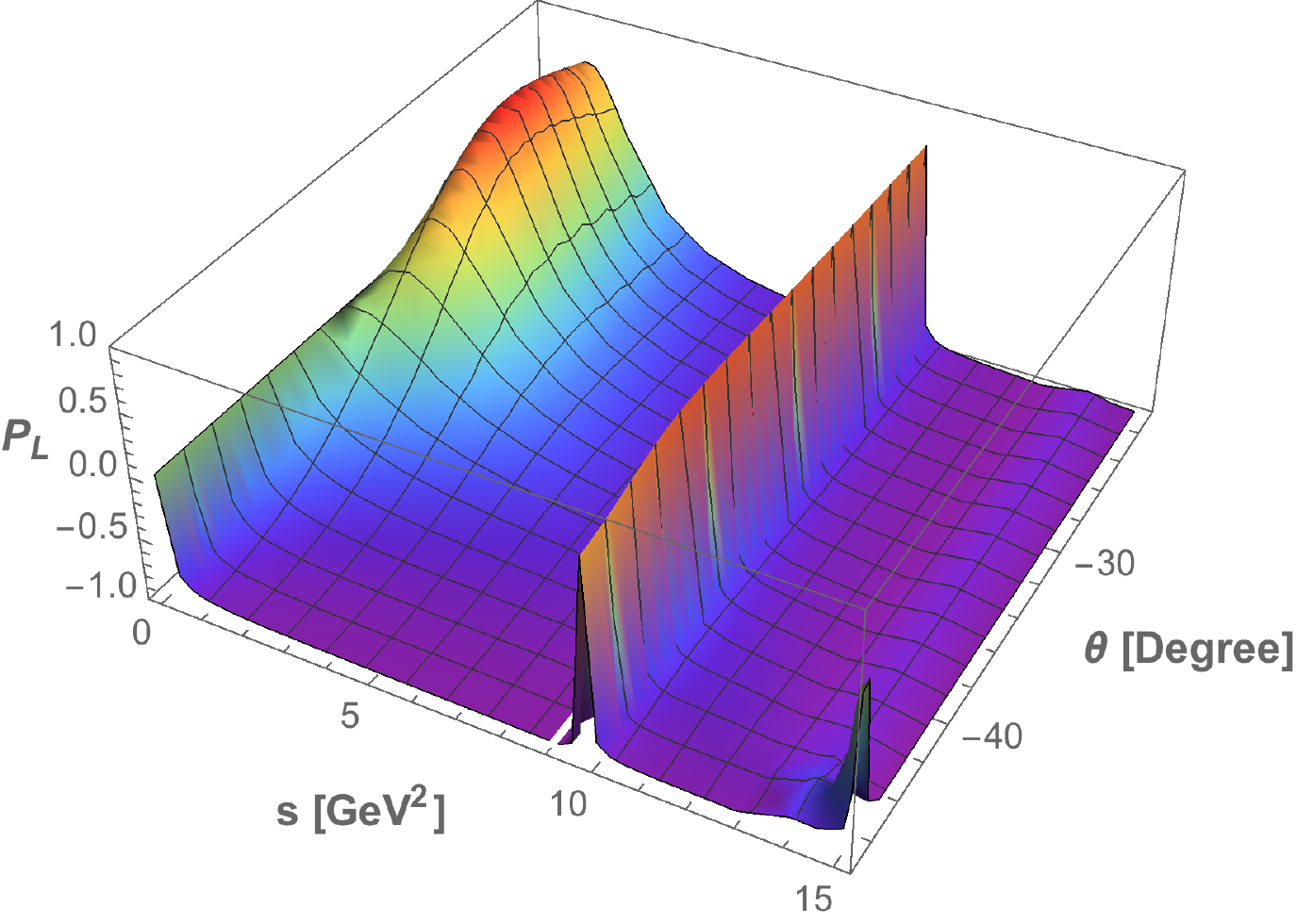}
\caption{Three-dimensional representation of differential branching ratio (in units of $10^{-7})$, forward-backward asymmetry and longitudinal polarization  with $s$ and the mixing angle $\theta$. The plots in the left  panel are for $B \to K_1(1270) \mu^+ \mu^-$ and  those in the right panel are for $B \to K_1(1400) \mu^+ \mu^-$ process.}\label{Fig:3D}
\end{figure} 
\begin{figure}
\includegraphics[scale=0.5]{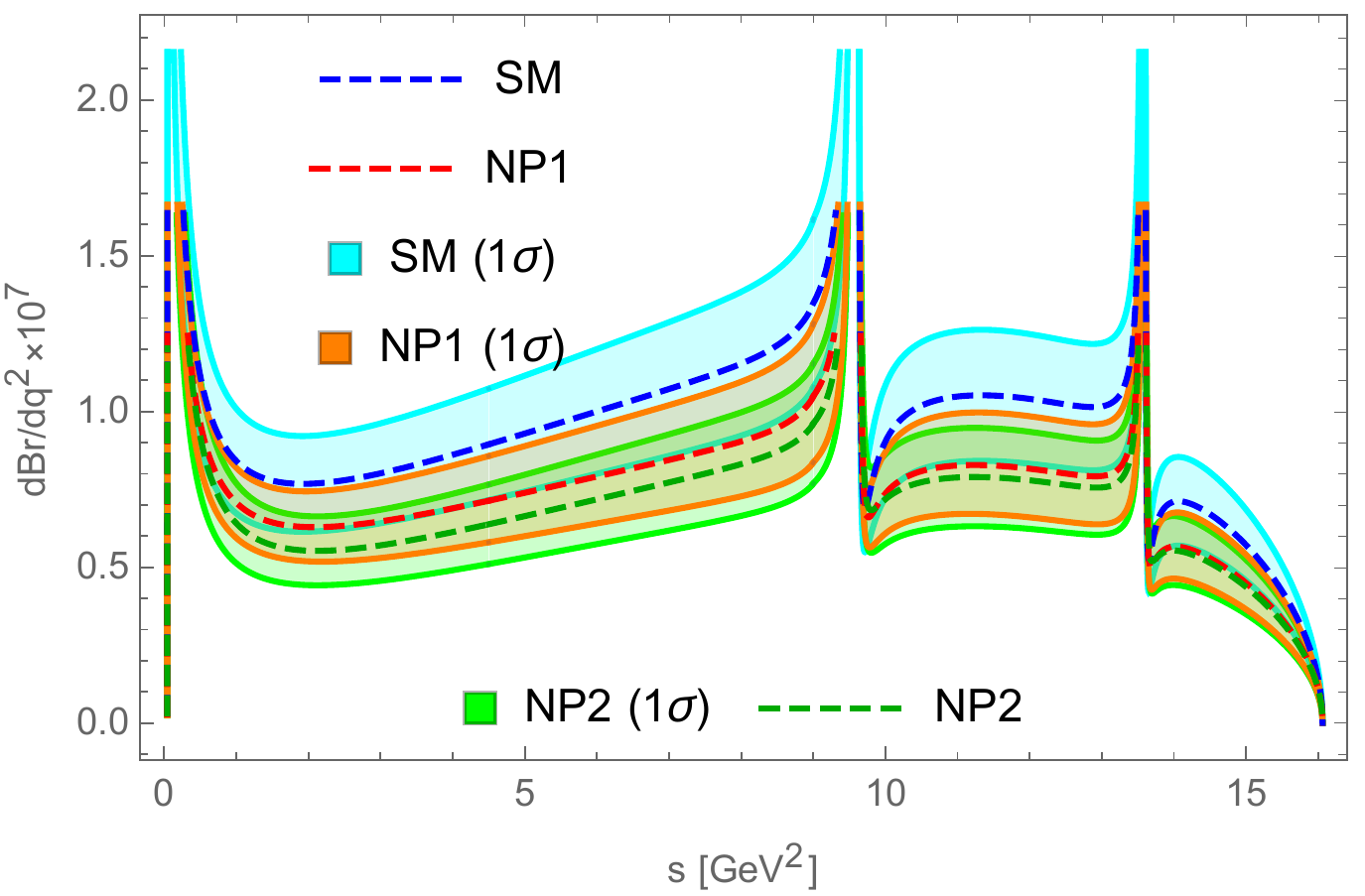}
\quad
\includegraphics[scale=0.5]{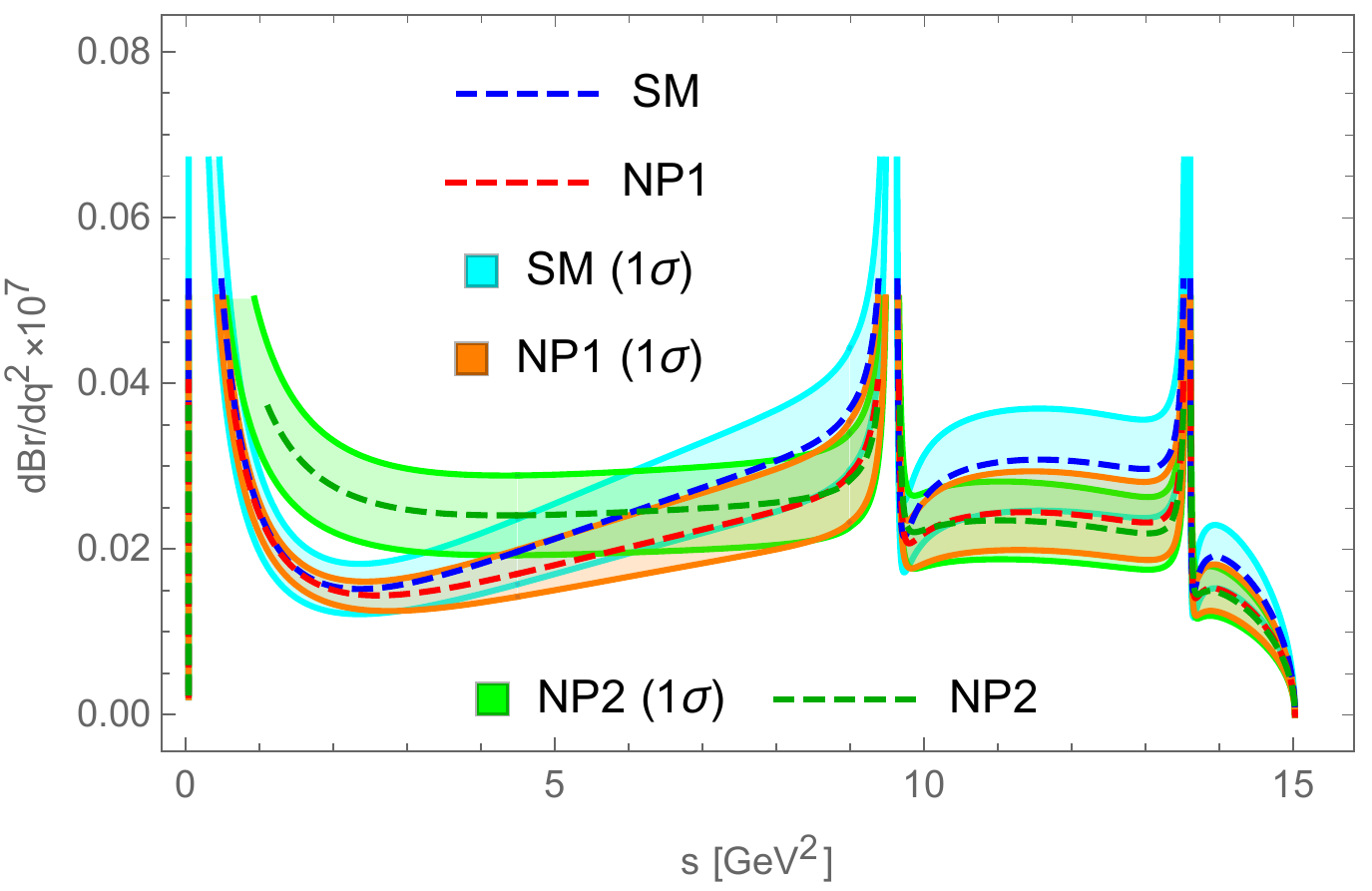}
\quad
\includegraphics[scale=0.5]{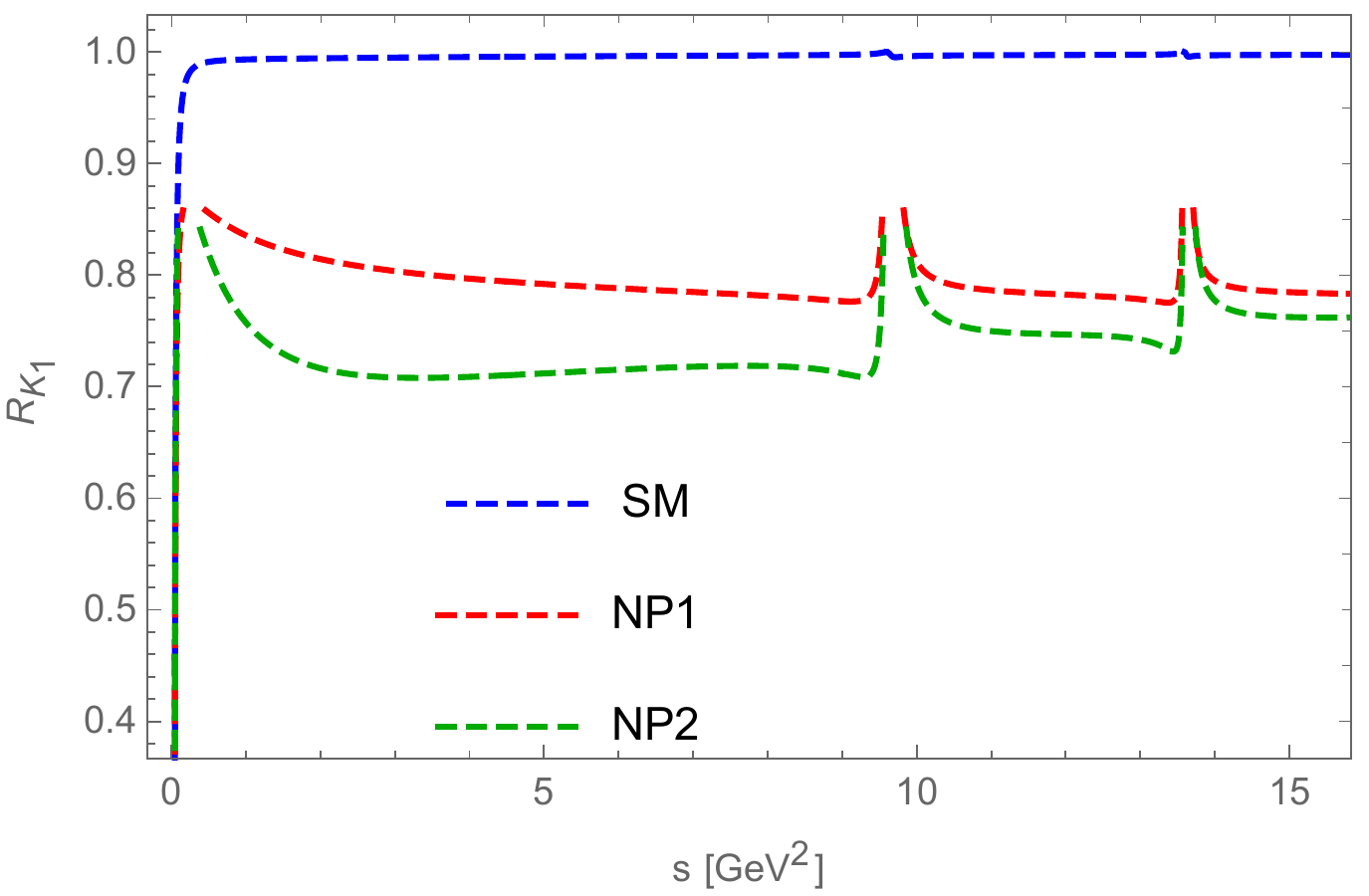}
\quad
\includegraphics[scale=0.5]{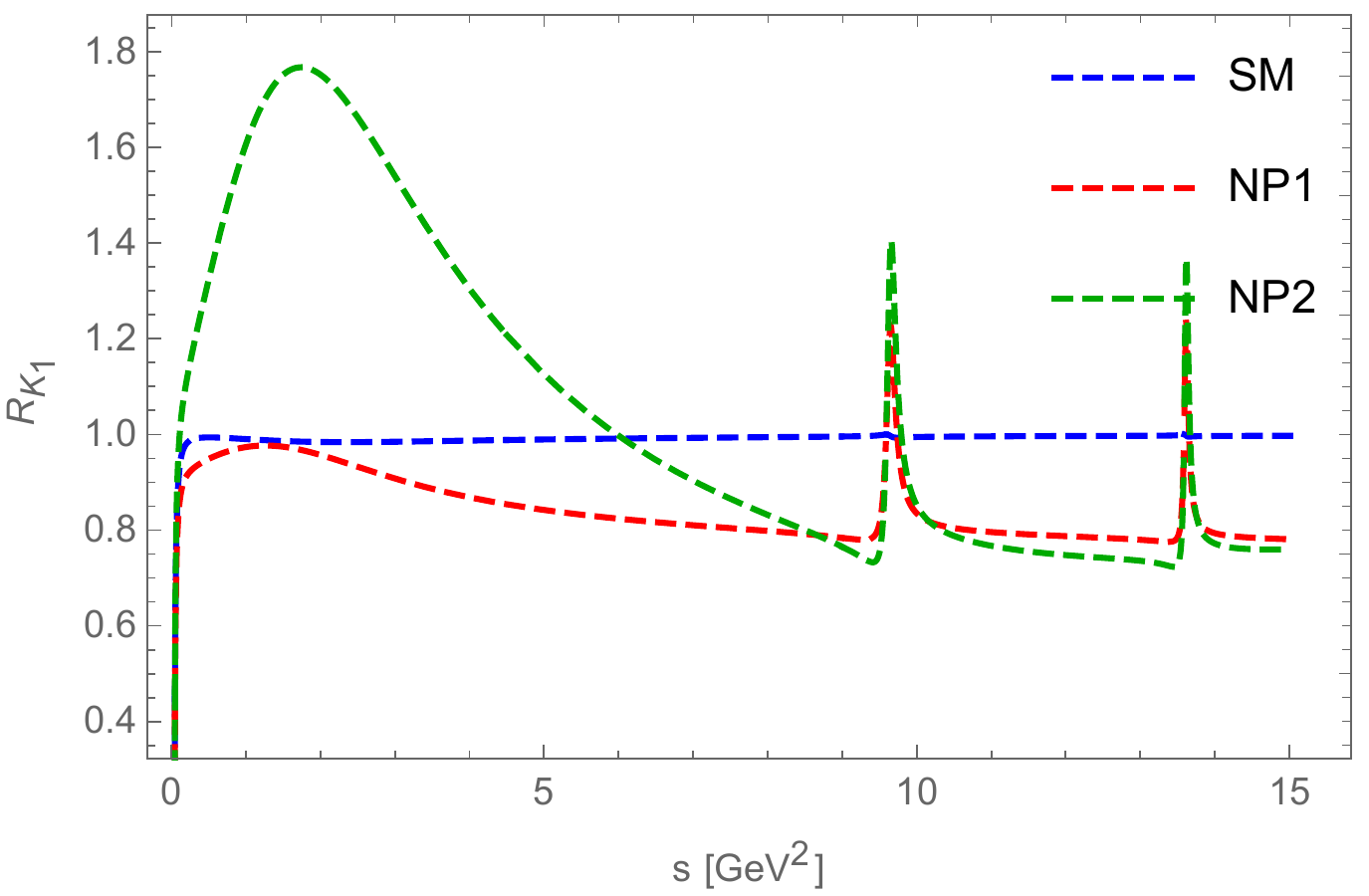}
\quad
\includegraphics[scale=0.5]{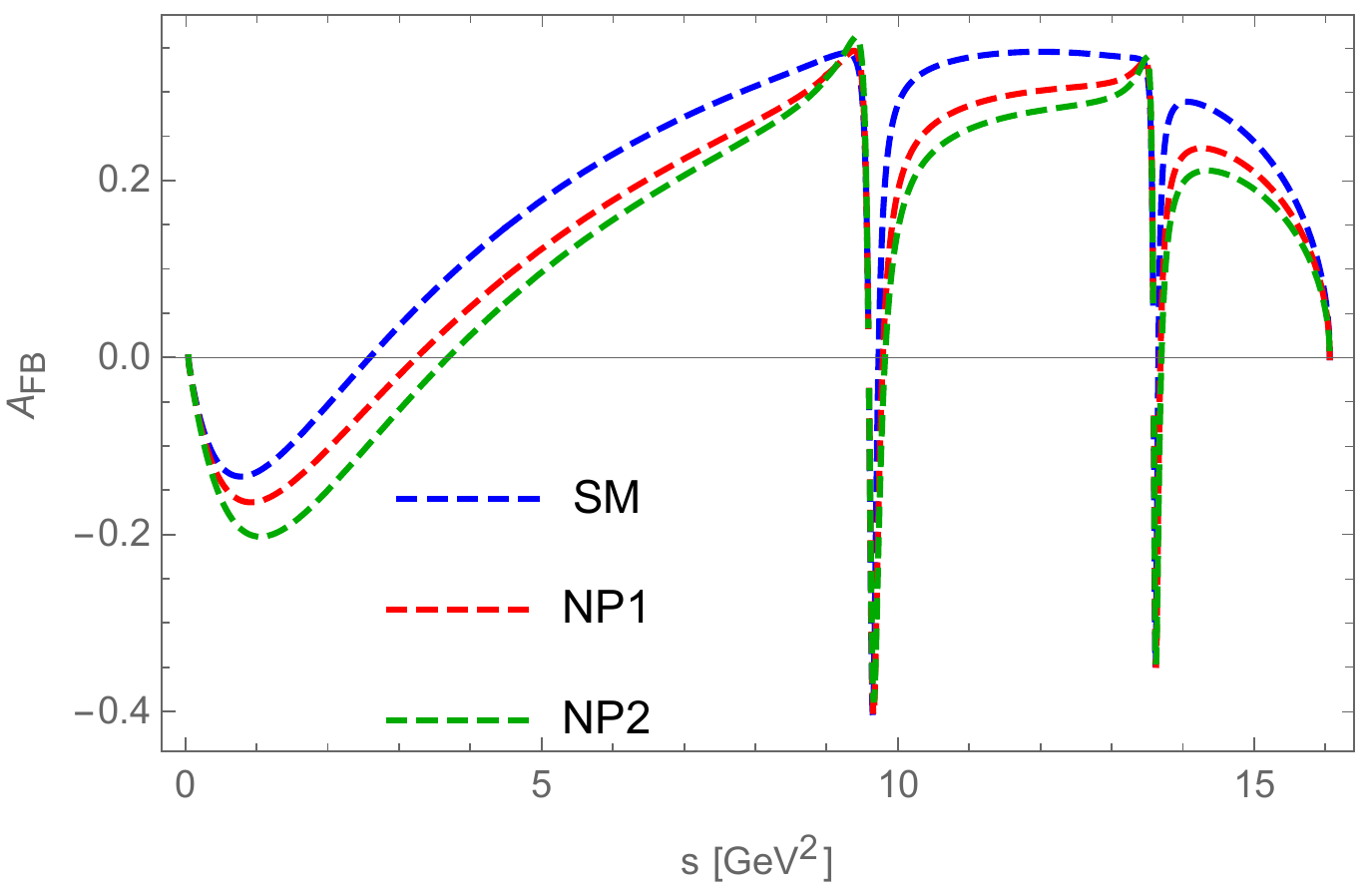}
\quad
\includegraphics[scale=0.5]{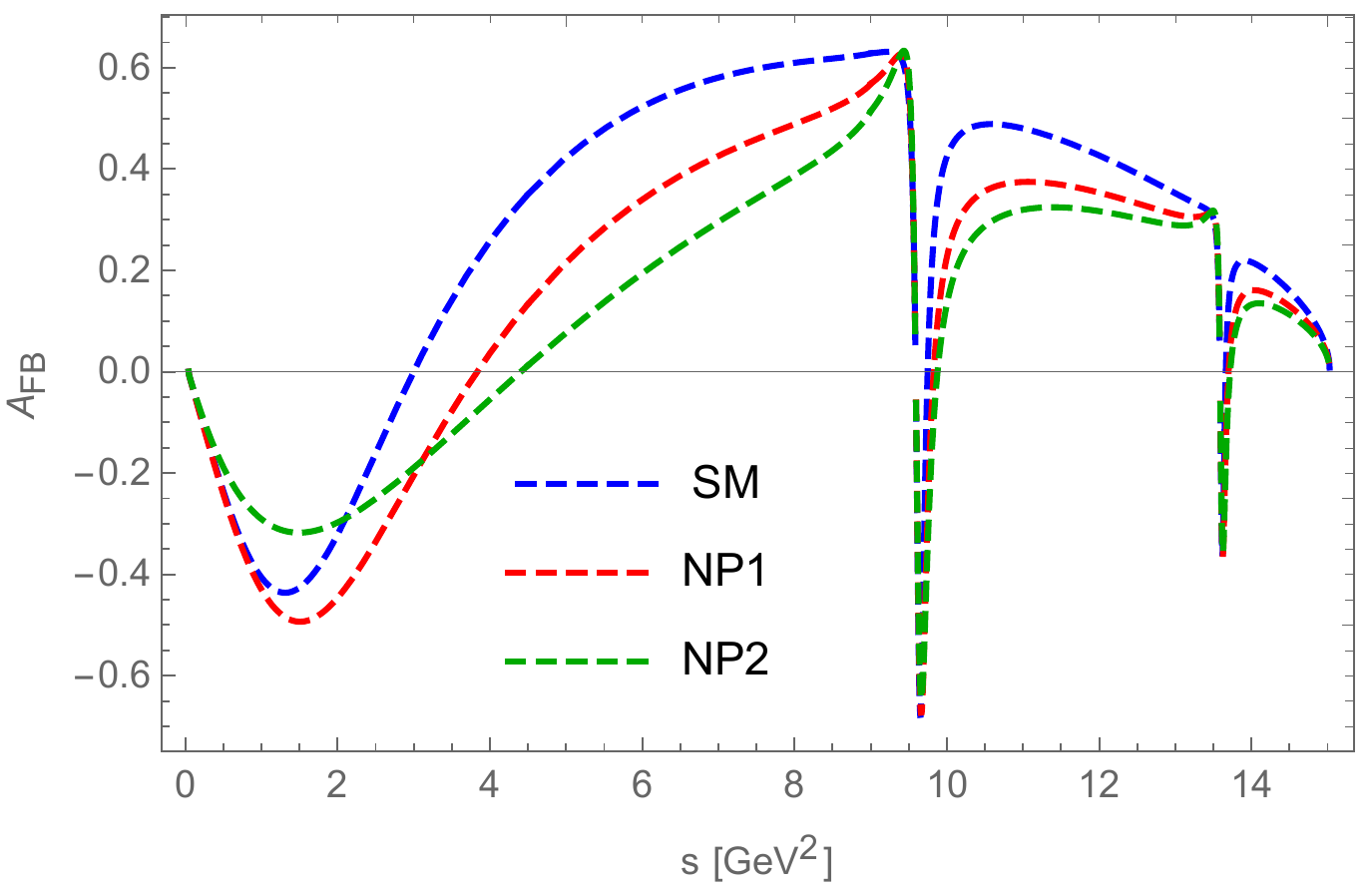}
\caption{The $s$ variation of the differential branching franction, lepton non-universality observable and the forward-backward asymmetry in the SM as well as the NP scenarios. The plots in left panel correspond to $B \to K_1(1270) \mu^+ \mu^-$ process whereas the right panel plots are for $B \to K_1(1400) \mu^+ \mu^-$ process.  }\label{Fig:BR-NP}
\end{figure} 

\section{Results and Discussion}
After gathering the required information about all the relevant observables, we now proceed for numerical estimation. The particle masses, the $B$ meson  lifetime and the values of CKM  matrix elements are taken from \cite{Tanabashi:2018oca}. The $B \to K_1$ form factors used in this analysis are taken from \cite{Yang:2008xw}, which are calculated in light cone sum rule (LCSR) approach.  The $q^2 $ dependence of the form factors are parametrized in double pole form (\ref{eq:FFpara}) and the necessary parameters are listed in Table \ref{tab:FFinLF}.  Since the mixing angle $\theta$ is not known precisely, to see its impact on various observables, we first show the $q^2$ variation of SM differential branching fraction, forward-backward asymmetry and the longitudinal lepton polarization asymmetry of $B \to K_1 (1270) \mu^+\mu^-$ process for three different $\theta$ values from its allowed range, i.e., the central value ($\theta=-34^\circ$), and the one-sigma limiting values ($\theta=(-34\pm 13)^\circ=-21^\circ$, and $-47^\circ$) in the left panel of Fig. \ref{Fig:diff-theta}, and the corresponding plots for $B \to K_1 (1400) \mu^+\mu^-$ process are shown in the right panel. From these plots, it should be noted that the observables of $B \to K_1(1270) \mu^+ \mu^-$ process are almost insensitive to mixing angle $\theta$ whereas for $B \to K_1(1400) \mu^+ \mu^-$ process, they are strongly dependent on $\theta$, due to the cancellation of the contributions from $B \to K_{1A}$ and $B \to K_{1B}$ form factors. As expected, these observables are found to have their minimal values for $\theta=-47^\circ $, which is very close to the maximal mixing. For completeness, we show the $q^2$ variation of these observables  for the  one-sigma allowed range of the mixing angle $\theta$  in Fig. \ref{Fig:3D}. 
Therefore, the measurement of various observables of $B \to K_1(1400) \mu^+ \mu^-$ process will shed light on the determination of the mixing angle.

\begin{figure}
\includegraphics[scale=0.5]{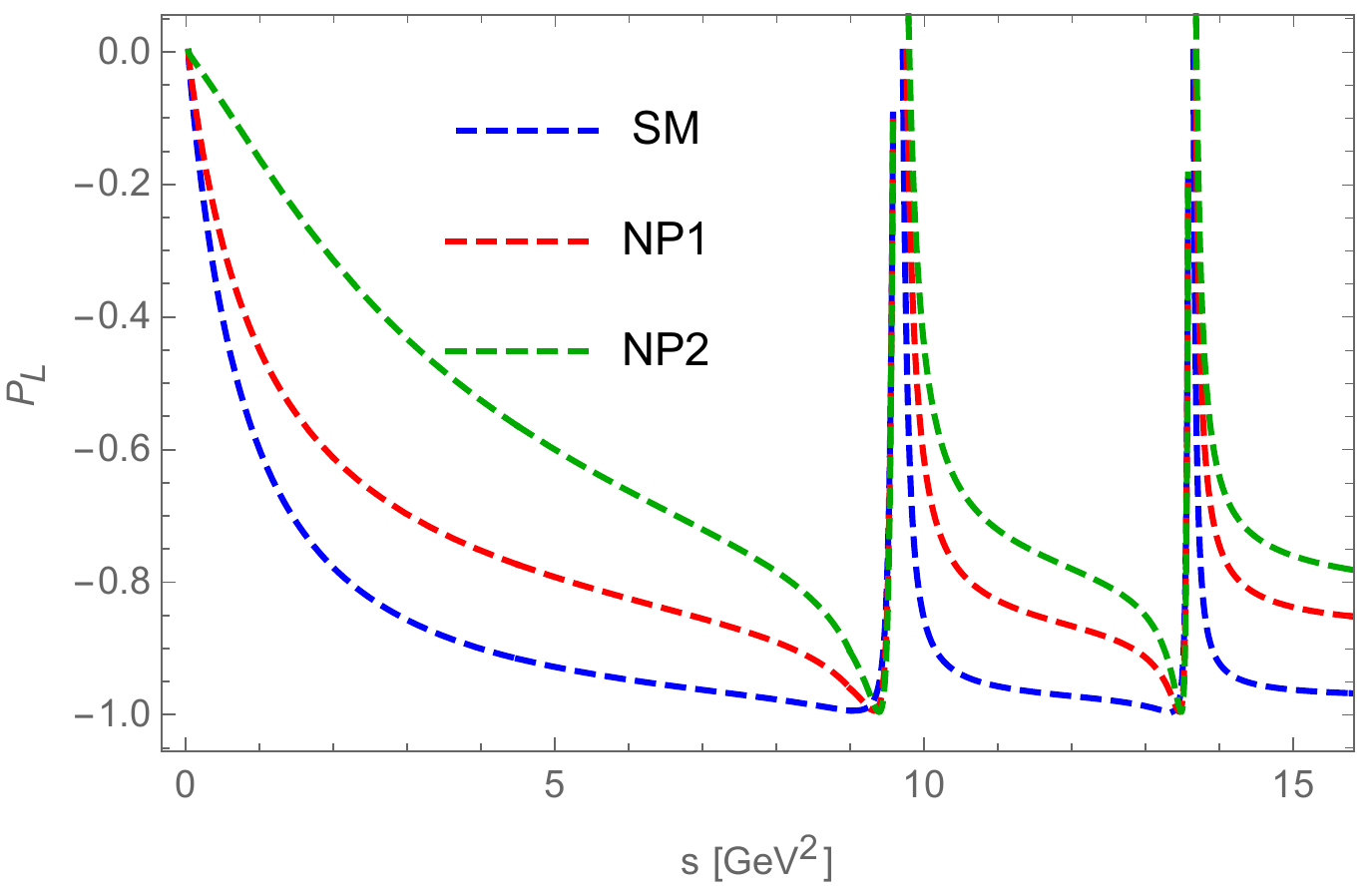}
\quad
\includegraphics[scale=0.5]{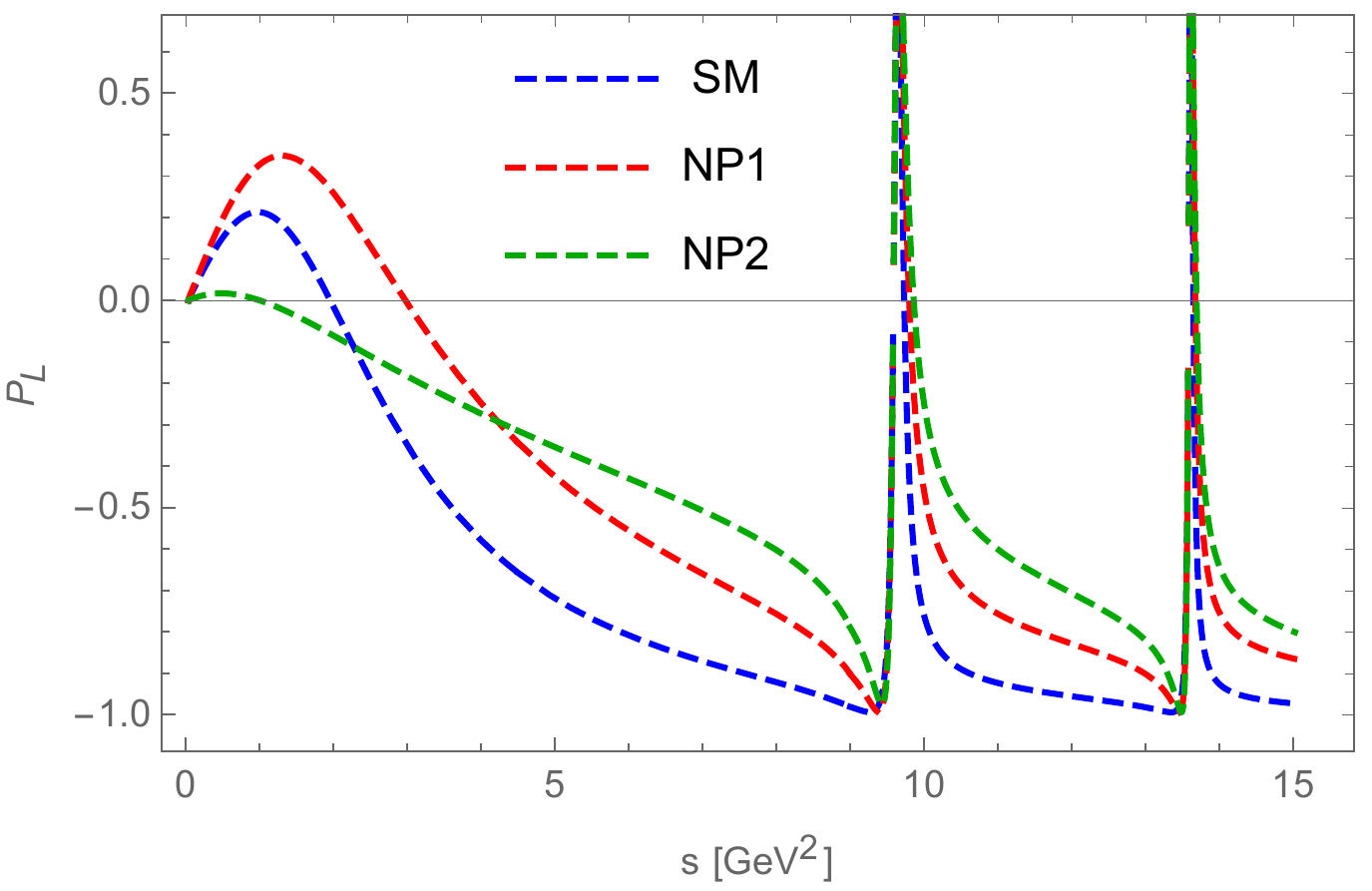}
\quad
\includegraphics[scale=0.5]{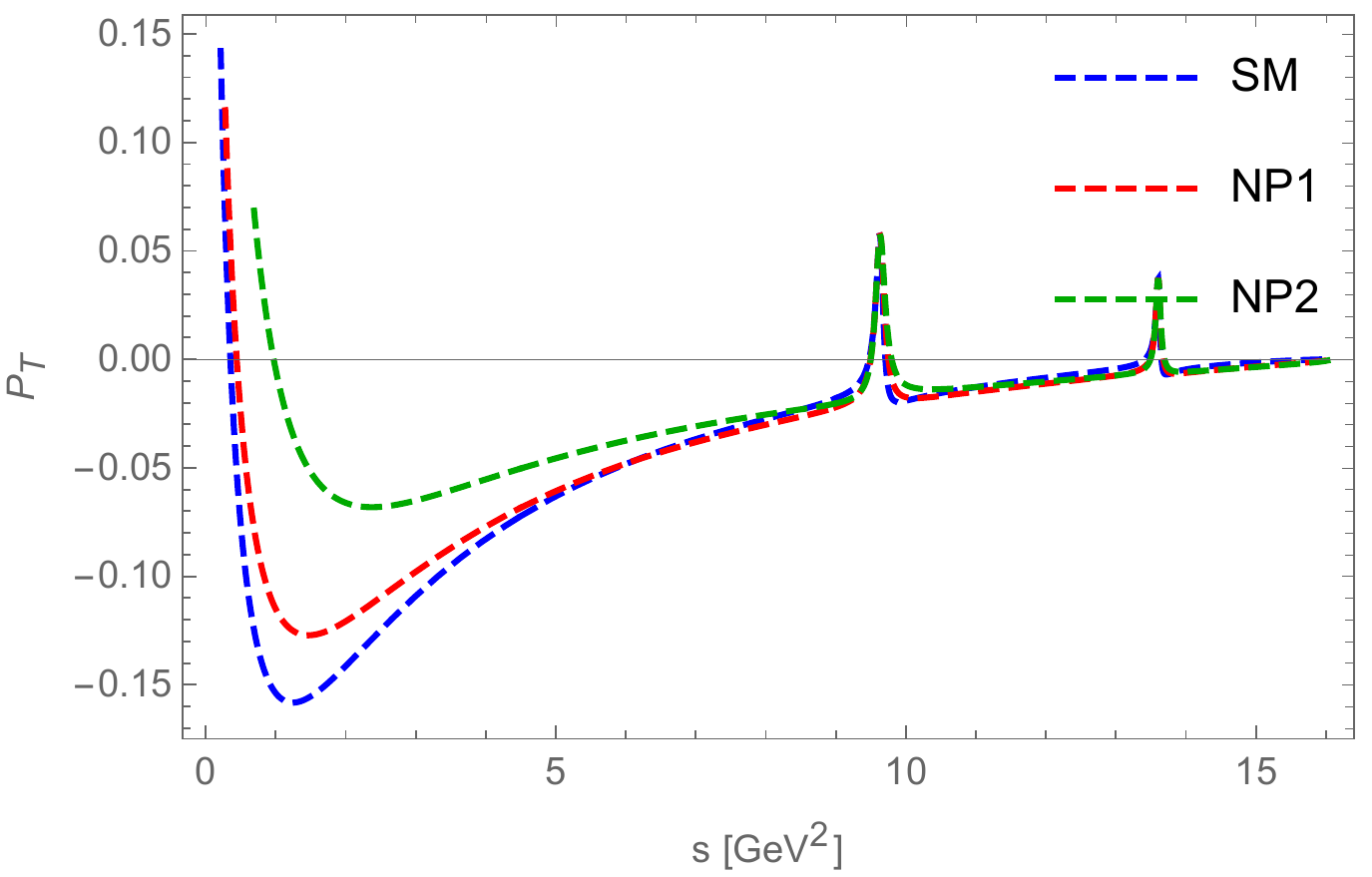}
\quad
\includegraphics[scale=0.5]{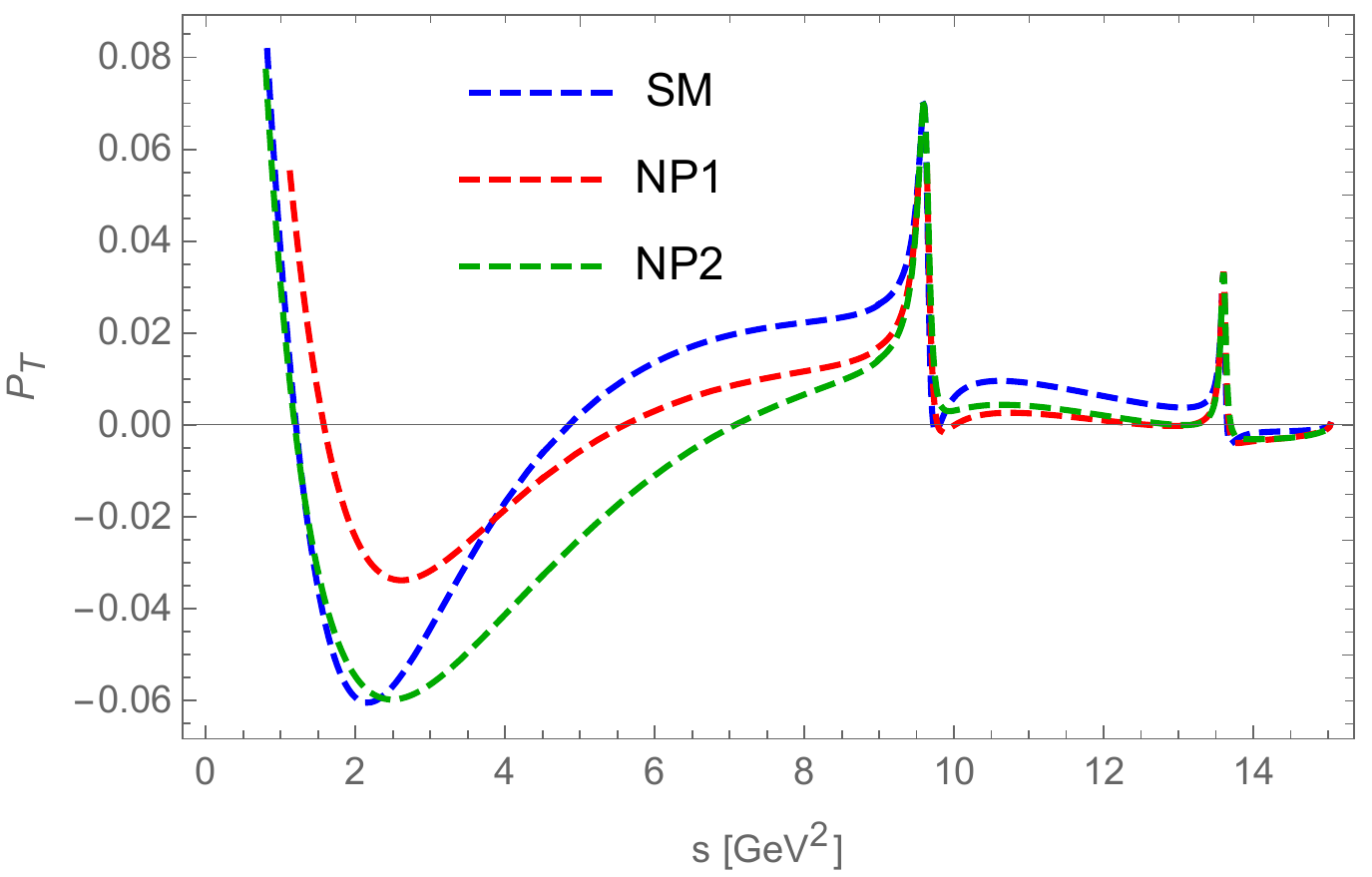}
\quad
\includegraphics[scale=0.5]{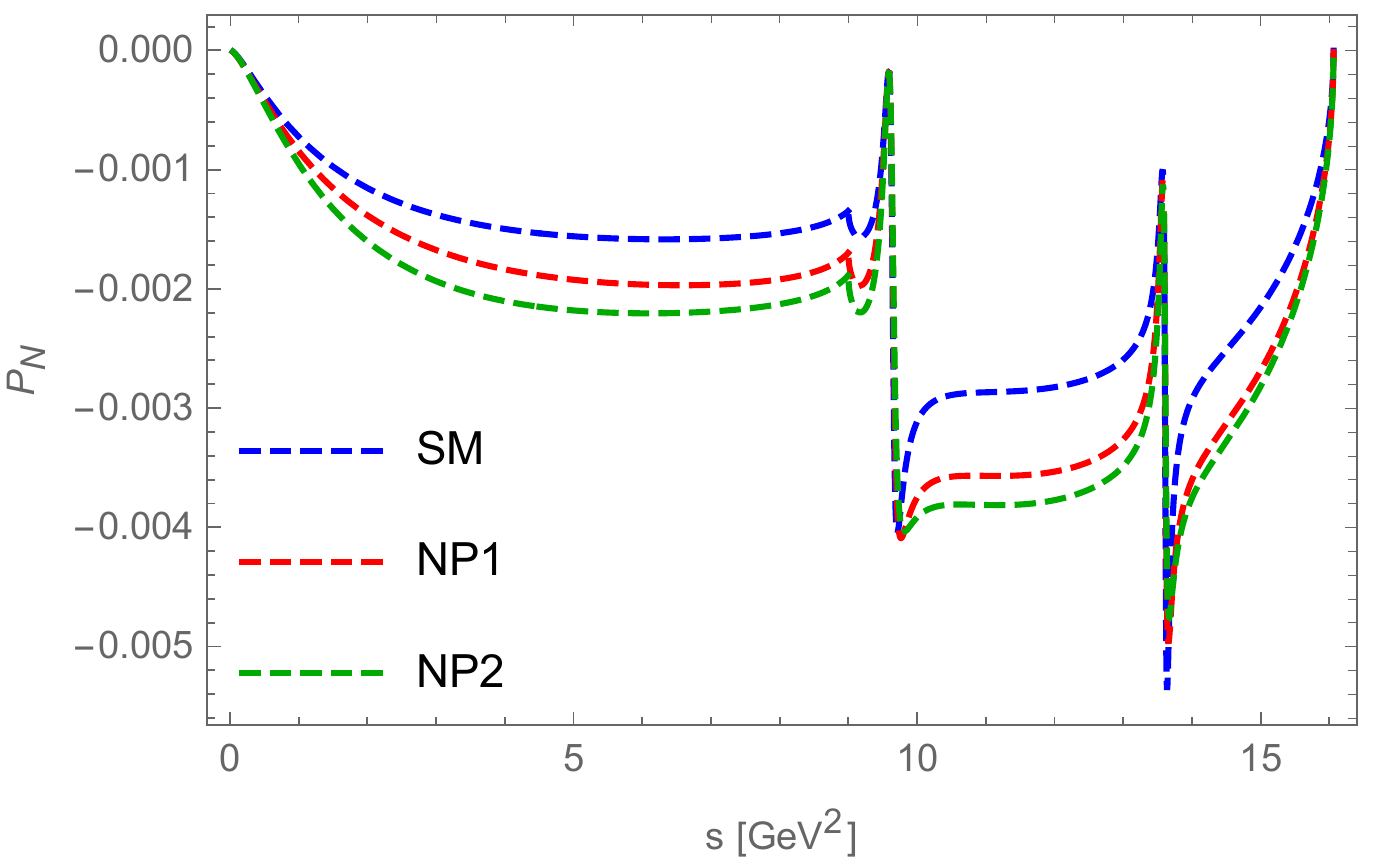}
\quad
\includegraphics[scale=0.5]{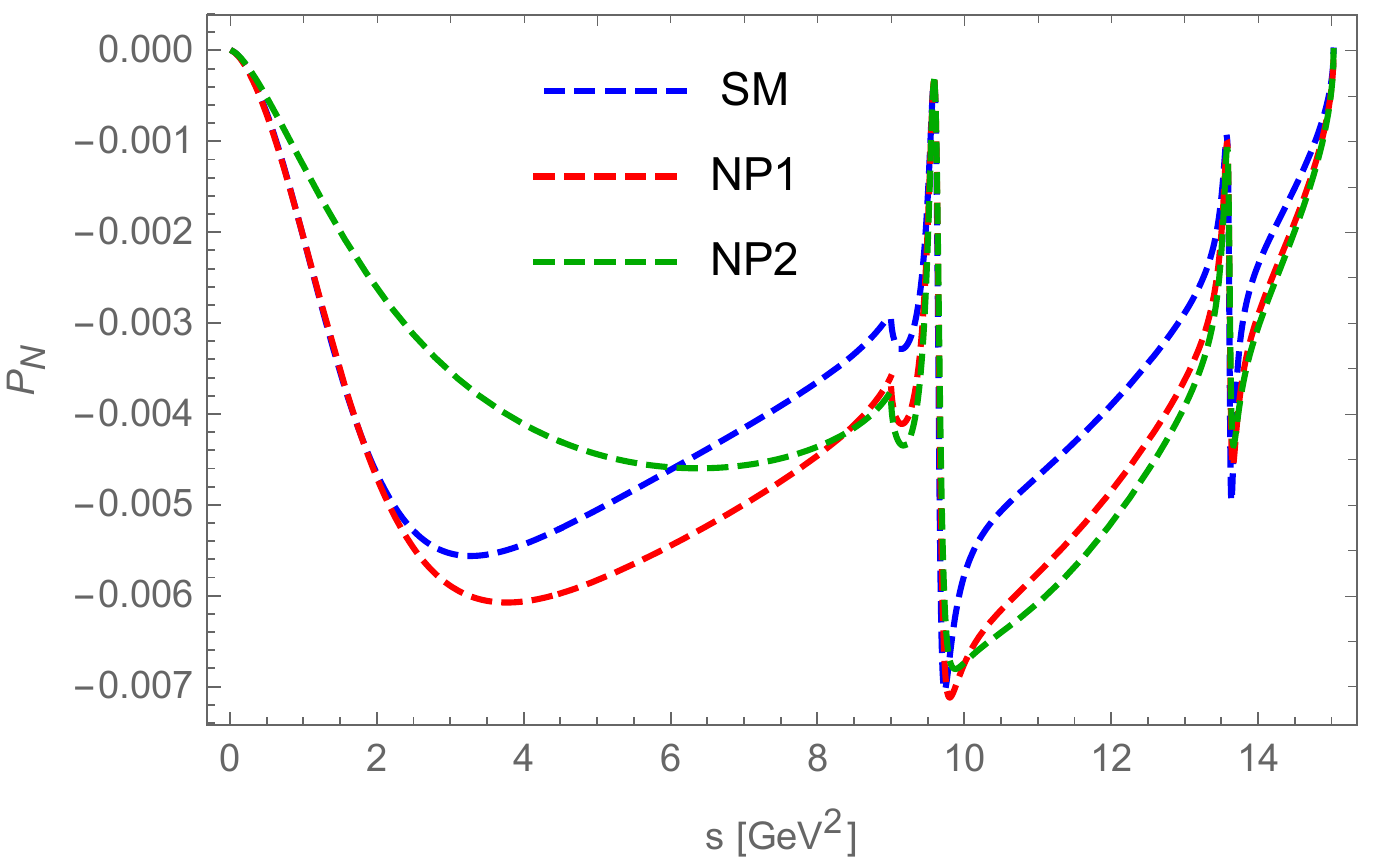}
\caption{Lepton polarization asymmetries are shown in the SM as well as in the NP  scenarios for $B \to K_1(1270) \mu^+ \mu^-$ (left panel) and $B \to K_1(1400) \mu^+ \mu^-$ (right panel) processes.}\label{Fig:Pol}
\end{figure}

\begin{figure}
\includegraphics[scale=0.5]{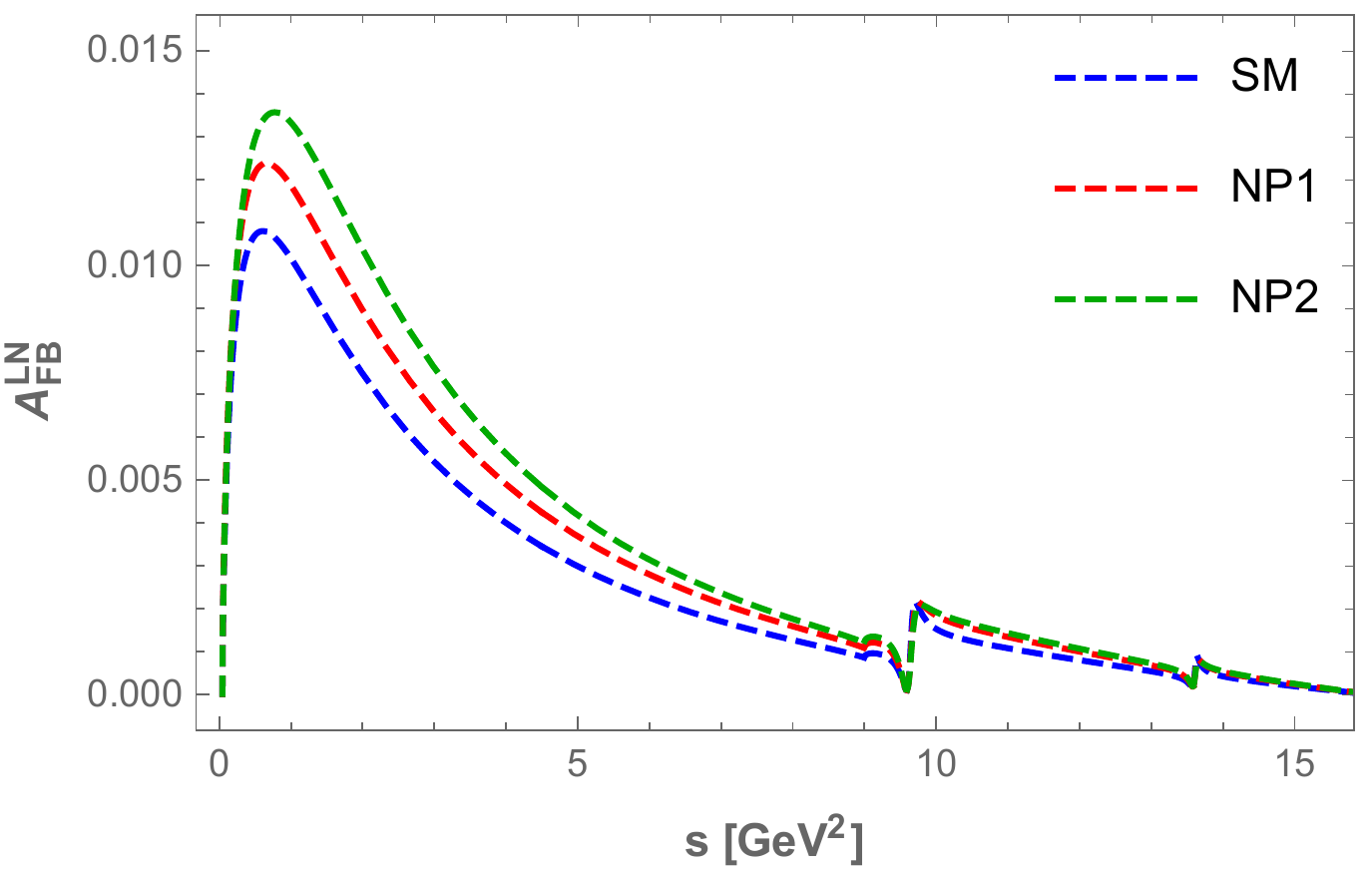}
\quad
\includegraphics[scale=0.5]{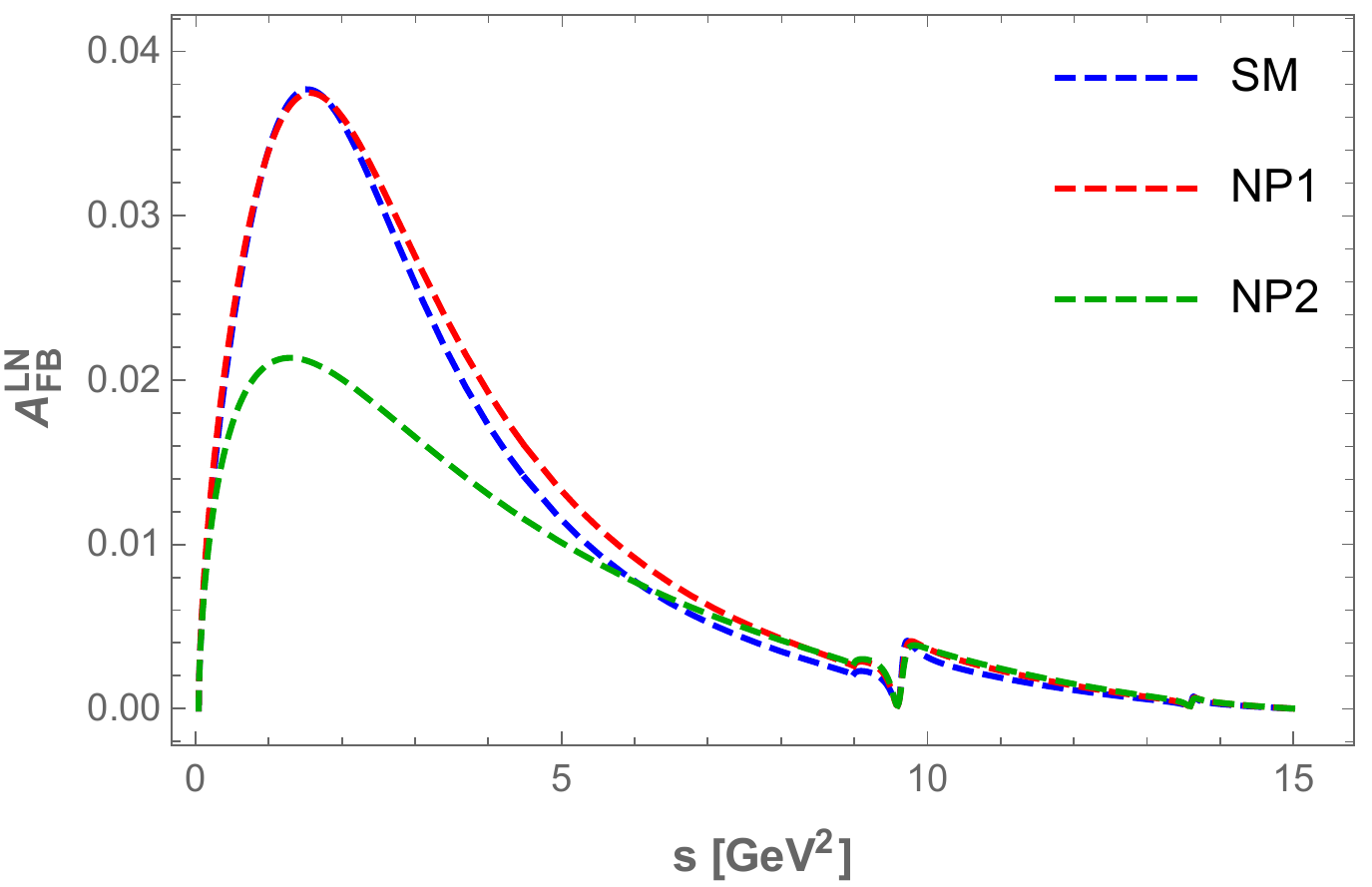}
\quad
\includegraphics[scale=0.5]{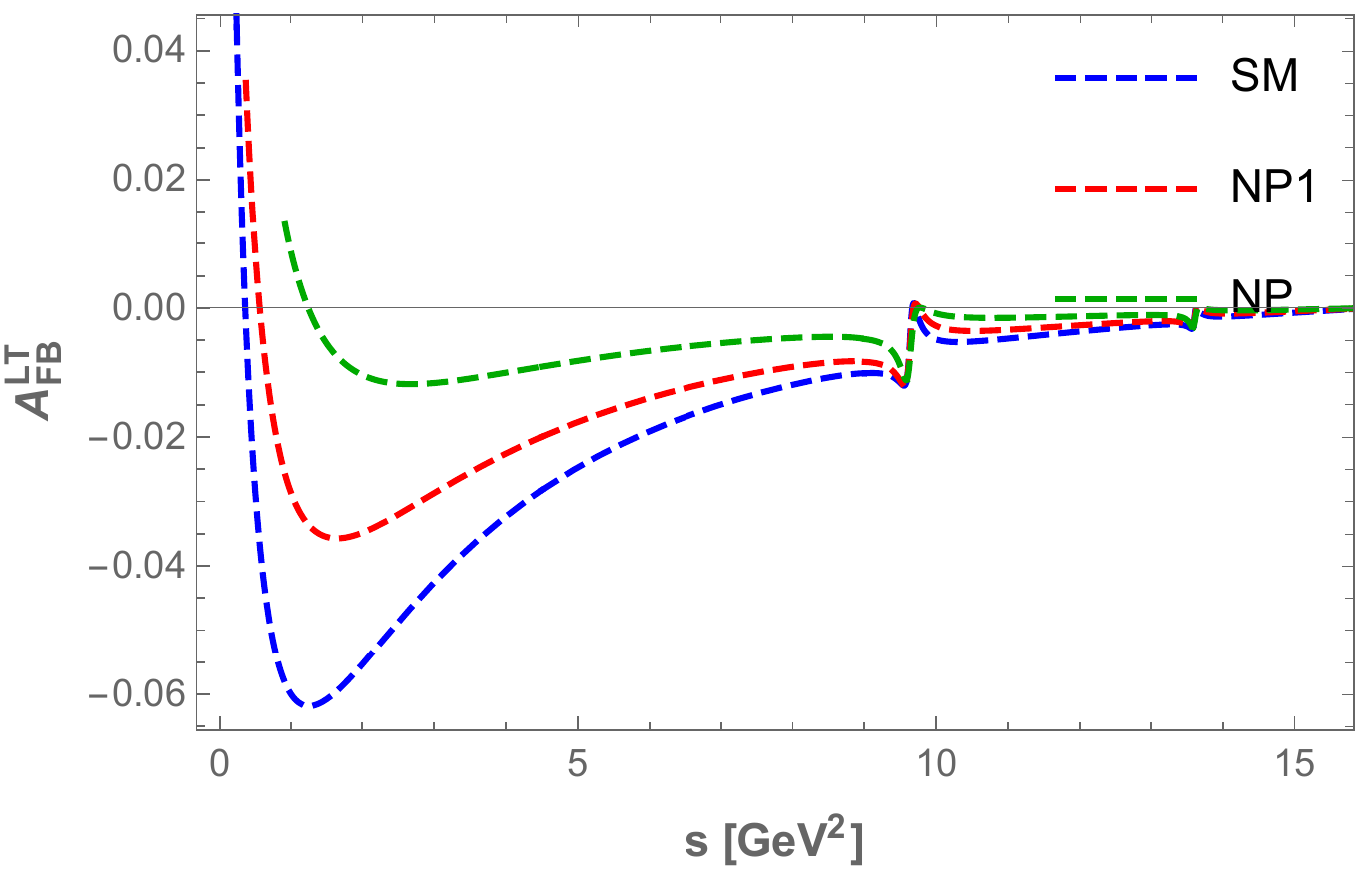}
\quad
\includegraphics[scale=0.5]{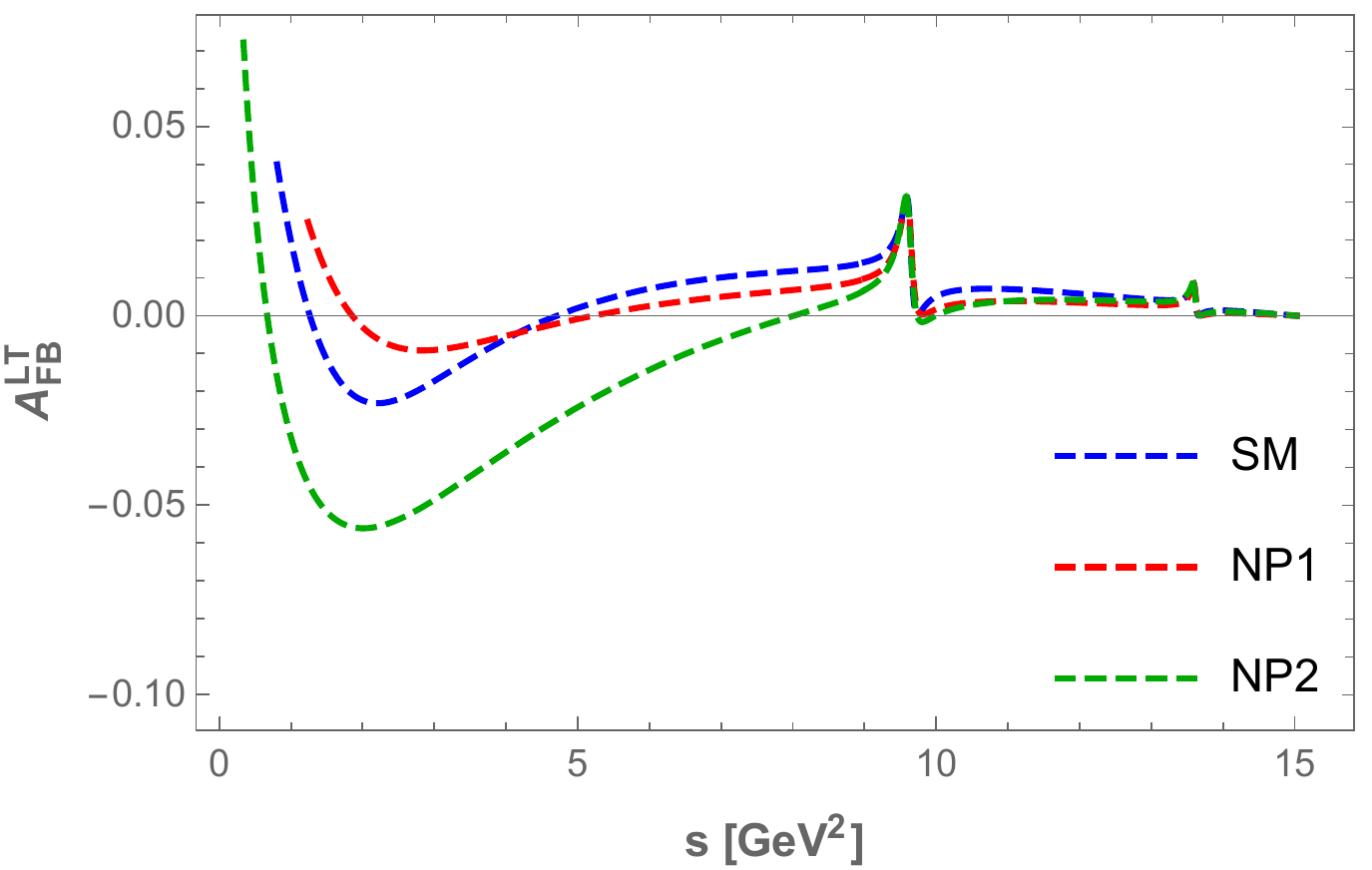}
\quad
\includegraphics[scale=0.5]{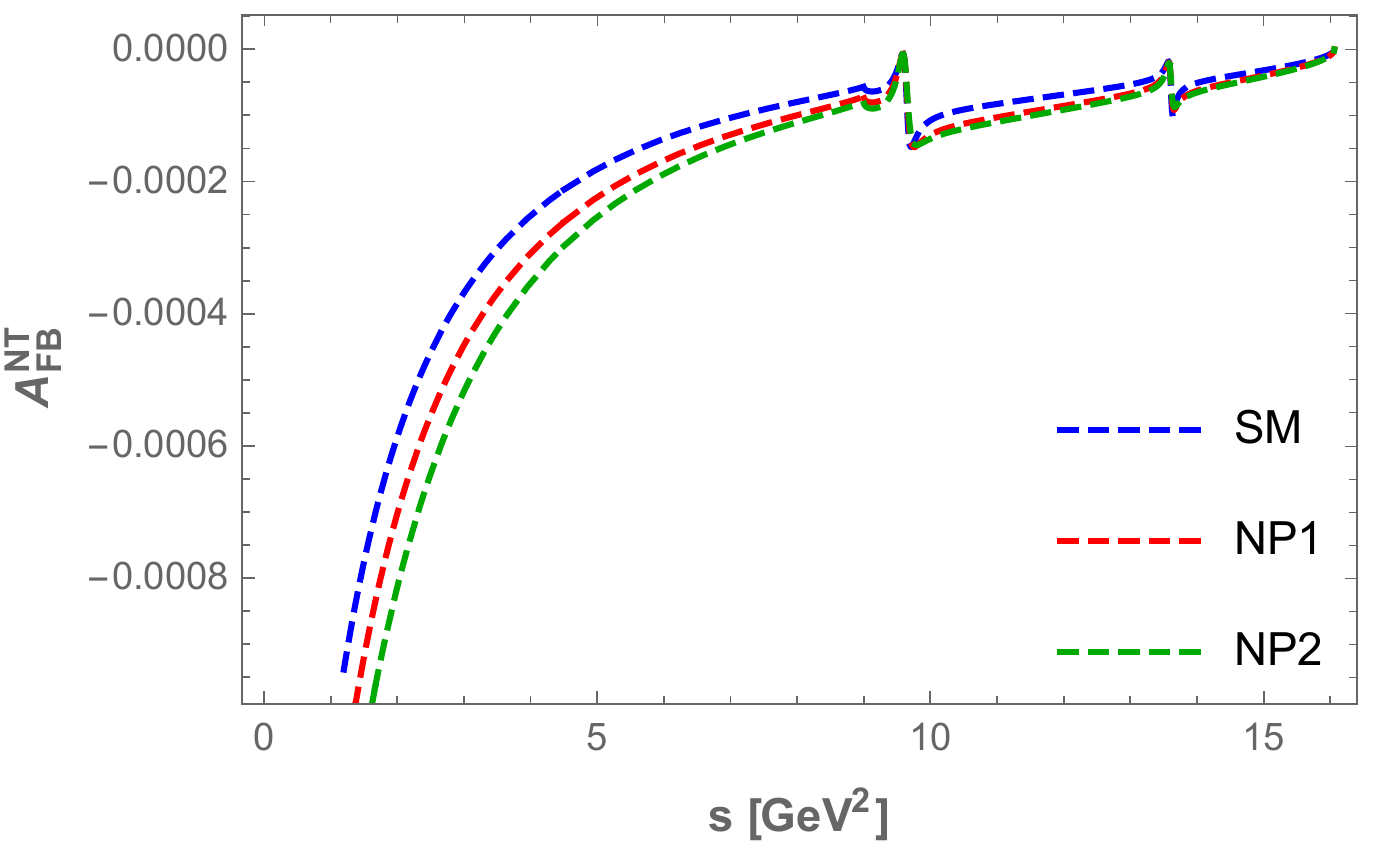}
\quad
\includegraphics[scale=0.5]{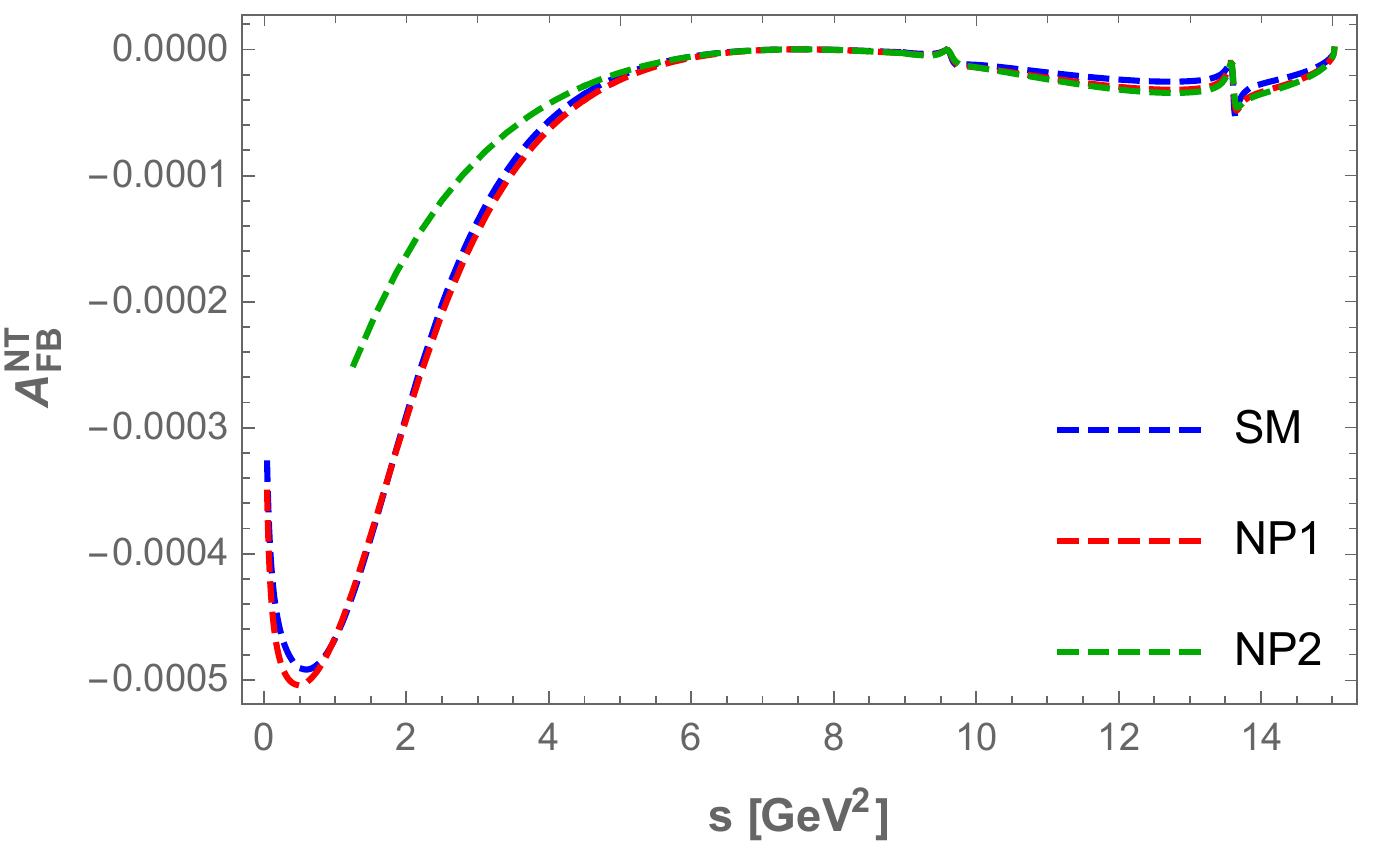}
\caption{Variation of polarized forward-backward asymmetry for $B \to K_1(1270) \mu^+ \mu^-$ (left panel and $B \to K_1(1400) \mu^+ \mu^-$ (right panel) processes. }\label{Fig:polarized-FB}
\end{figure}

\begin{figure}
\includegraphics[scale=0.5]{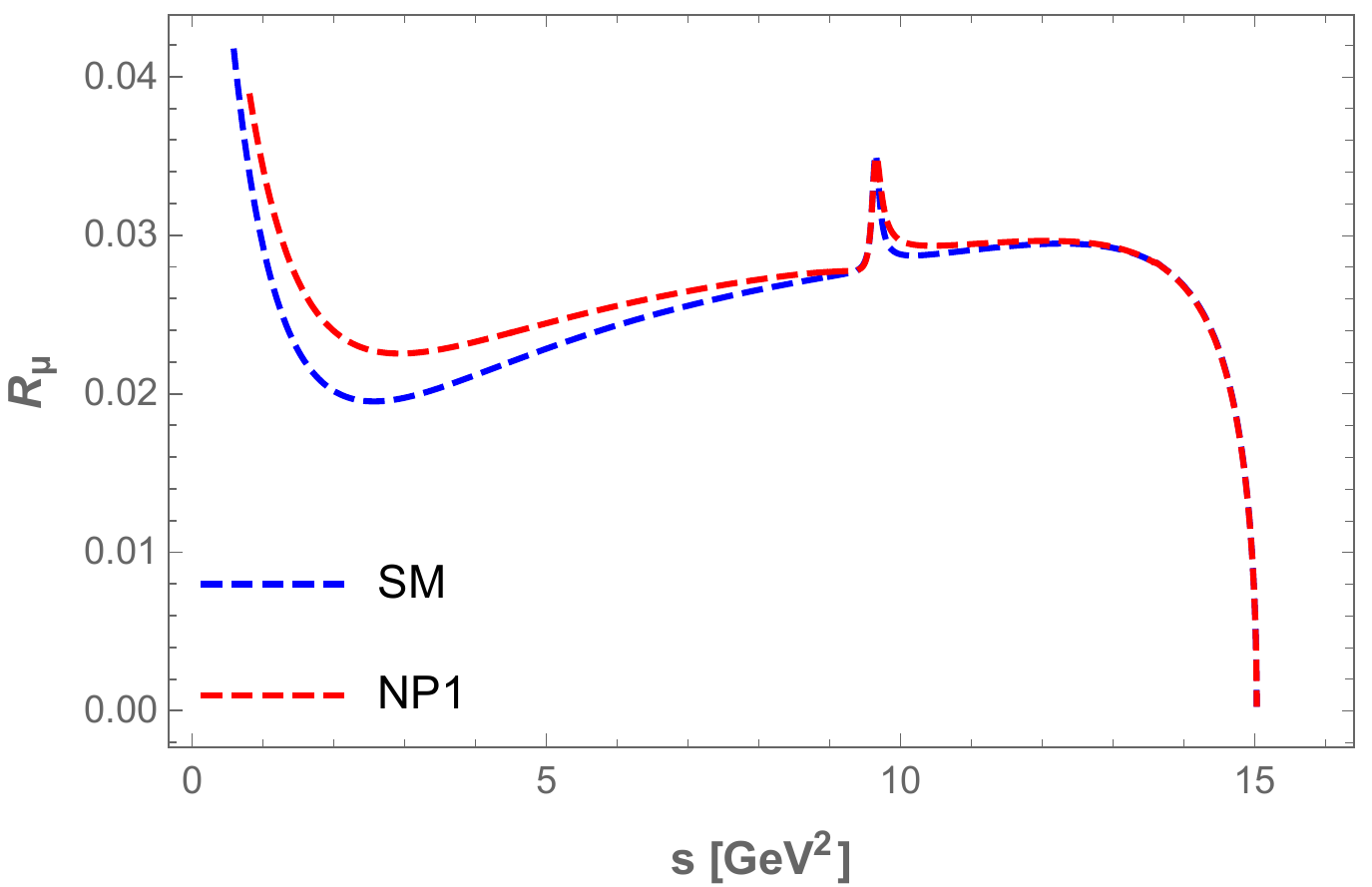}
\quad
\includegraphics[scale=0.5]{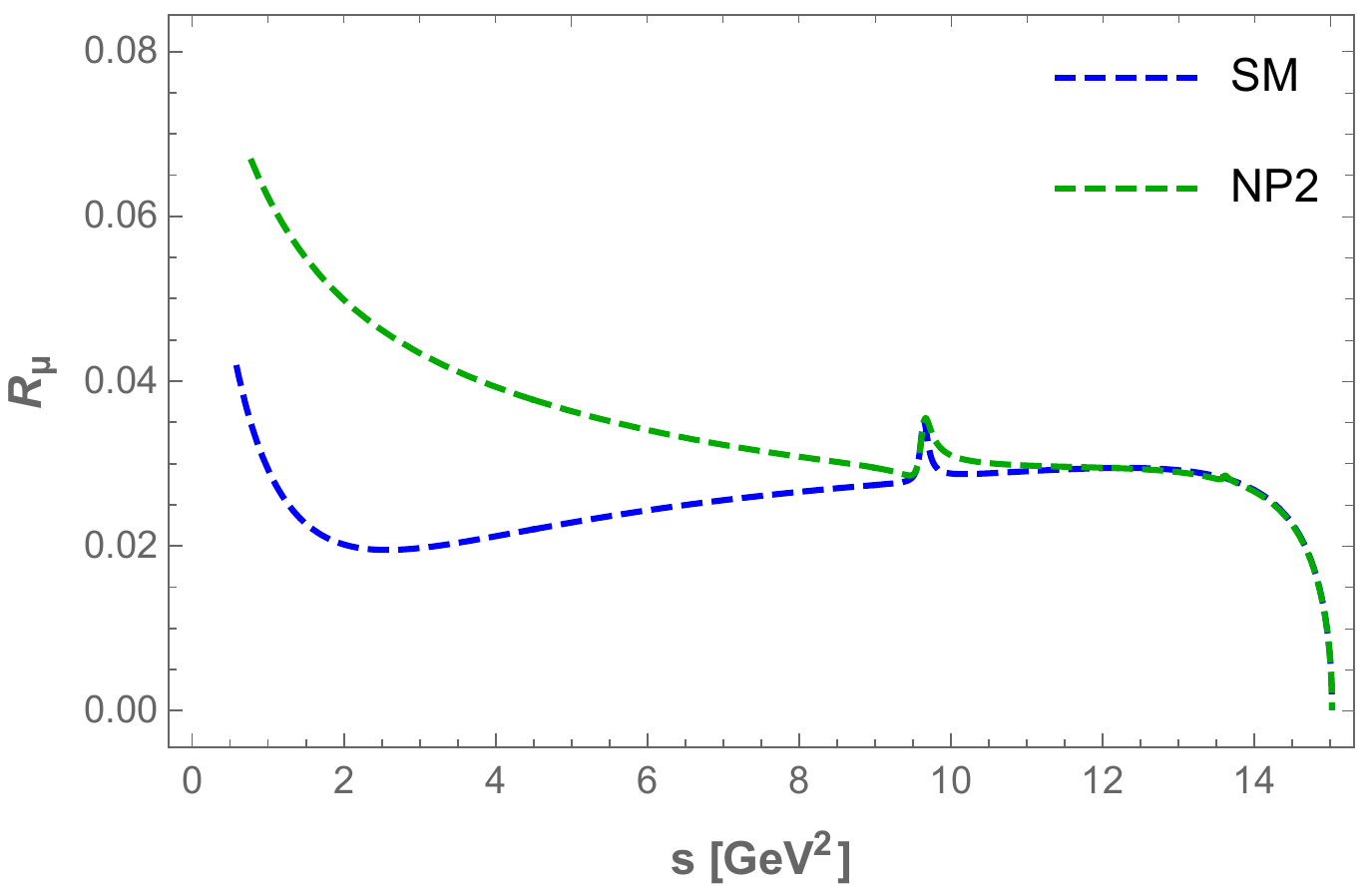}
\caption{The left (right) panel displays the variation of $R_\mu$ parameter with $s$ for SM and new physics scenario-I (scenario-II). }\label{Fig:Rmu}
\end{figure}

Next, we would like to see the impact of new physics on various observables, for which we have fixed the value of the mixing angle at its central value $\theta= -34^\circ$. We consider three specific new physics scenarios, in the first case we consider the NP contributions only in operators which are non-zero in the SM,
and the values of the NP coefficients as $C_7^{\rm NP}=0.013$, $C_9^{\rm NP}=-1.03$, and $C_{10}^{\rm NP}=0.08$ \cite{Bhom:2020lmk}.  For the second scenario, we consider the case $C_9^{\rm NP}= -0.829$ and $C_9^{'\rm NP}=-0.462$,  and for the third case we use $C_9^{\rm NP}= -0.526$ and $C_{10}^{\rm NP}=0.573$, which are obtained from the current data on $b \to s \mu^+ \mu^-$ anomalies,  that are relatively free from hadronic uncertainties. In Fig. \ref{Fig:BR-NP}, we show the $q^2$ variation of branching fraction, the lepton non-universality observable $R_{K_1}$ and the forward-backward asymmetry for $B \to K_1(1270) \mu^+ \mu^-~(B \to K_1(1400) \mu^+ \mu^-)$ process in the left (right) panel, both in the SM and the two NP1 and NP2 scenarios. The plots for NP3 scenario are very similar and close to those of  NP1, so we have not shown them explicitly. The branching fractions are shown in the top panel of the figure, where the dashed lines are due to the central values of the input parameters whereas the bands are due to the $1\sigma$ uncertainties. From the figure, it can be noticed that  for $B \to K_1(1270) \mu \mu$ process, the branching fractions are lower than the SM values for both types of NP scenarios, whereas for $B \to K_1(1400) \mu^+ \mu^-$, the branching ratio of NP scenario II (NP2) is higher than the SM while for scenario-I (NP1), it is lower than the SM prediction. In the middle panel the lepton flavour non universality ratio is displayed,
which is lower than the SM predicted value for both the NP scenarios for $B \to K_1(1270) \mu^+\mu^-$, while for $B\to K_1(1400) \mu^+ \mu^-$, while it is lower than the SM for NP1 and higher than the SM value in the lower $q^2$ bin for NP2. The measurement of this observable in both the decay modes will help to distinguish between these two NP scenarios.    The behaviour of forward-backward asymmetry is shown in the lower panel and it is found that the zero crossing points in the both types of NP scenarios differ from its SM value and shift towards higher value of $q^2$.
In Fig. \ref{Fig:Pol}, the lepton polarization asymmetries are displayed. From the plots it is found that the behaviour of lepton polarization asymmetries in NP-II scenario is quite different from SM as well as NP-I cases. It is also inferred that the longitudinal polarization asymmetry receives the dominant contributions for both the decay modes. The polarized forward-backward asymmetries are presented in Fig \ref{Fig:polarized-FB}. In this case also the effect of NP2 is significantly different from SM as well as NP1 scenario, though its effect is more prominent in $B \to K_1(1400) \mu^+ \mu^-$ process. Finally in Fig. \ref{Fig:Rmu}, we show the $q^2$ variation of $R_\mu$ parameter for the case of NP1 (left panel) and NP2 (right panel) and it is found in the low $q^2$ regime, the impact of NP-II is relatively significant. The integrated values of the branching ratios in the low-$q^2$ bin well below the charmonium resonance region  ($q^2 \in [1,6]~{\rm GeV^2}$) are presented in Table \ref{Br-lowq} both for the SM and the NP scenarios and the numerical values of all other observables are presented in Table -\ref{Num-lowq}. The theoretical uncertainties arising from the hadronic form factors, CKM matrix elements and other input parameters are provided only for those observables for which SM predictions are more than a percent level.
The value of $R_\mu$ ratio in the low-$q^2$ region ([1,6]~${\rm GeV}^2$) is found to be 0.02 in the SM and 0.024/0.043/0.021 in the NP scenarios-1/2/3.  
\begin{table}[h]
\begin{center}
\caption{The predicted  values of the branching ratios in the low $q^2$ bin $q^2 \in [1,6]~{\rm GeV}^2$  for the $B^0 \rightarrow K_1^0(1270) \mu^+ \mu^-$  and $B^0 \rightarrow K_1^0(1400) \mu^+ \mu^-$ processes, both in the SM and NP scenarios.  }\label{Br-lowq}
\begin{tabular}{|c | c | c| }
\hline
 ~Various in different scenarios ~ & ~ ${\rm Br}( B^0 \rightarrow K_1(1270) \mu^+ \mu^-)$~ &~ ${\rm Br}(B^0 \rightarrow K_1(1400) \mu^+\mu^-)$~~\\ 
\hline 
\hline
Standard Model & ~$(4.257\pm 0.851) \times 10^{-7}$ ~& ~$(8.548\pm1.71) \times10^{-9}$  \\
\hline
NP scenario-I & ~ $(3.433\pm 0.687) \times 10^{-7}$~& ~ $(8.409\pm 1.682)\times 10^{-9}$\\
\hline
NP scenario-II  &~$ (3.057 \pm 0.611) \times 10^{-7}$~
 &~ $(1.307\pm 0.261)\times 10^{-8}$ ~\\ 
 \hline
NP scenario-III  &~$ (3.192 \pm 0.638 ) \times 10^{-7}$~
 &~ $(6.622\pm 1.324)\times 10^{-9}$ ~\\  
 
 \hline
\end{tabular}
\end{center}
\end{table}
\begin{table}[h]
\begin{center}
\caption{The predicted  values of the  the lepton nonuniversality ratio $R_{K_1}$, forward-backward asymmetry and lepton polarisation asymmetries  in the low $q^2$ bin $q^2 \in [1,6]~{\rm GeV}^2$  for the $B^0 \rightarrow K_1^0(1270) \mu^+ \mu^-$  and $B^0 \rightarrow K_1^0(1400) \mu^+ \mu^-$ processes in the  SM as well as in NP scenarios. }\label{Num-lowq}
\begin{tabular}{|c | c | c| c | c | c|}
\hline
 Observables &$ B^0 \rightarrow K_1(1270) \mu \mu$ &$B^0 \rightarrow K_1(1400) \mu\mu$ & Observables &$ B^0 \rightarrow K_1(1270) \mu \mu$ &$B^0 \rightarrow K_1(1400) \mu\mu$ \\
 \hline
 \hline
 ~ $R_{K_1}^{\rm SM}$ & $0.995\pm 0.05$  &$0.987\pm  049$&  $\langle A_{FB}\rangle ^{\rm SM} $ & $0.081\pm0.004$ & $0.149\pm 0.007$\\
 
 ~ $R_{K_1}^{\rm NP1}$ & $0.803 \pm 0.04$&$ 0.971 \pm 0.048$ &~ $\langle A_{FB}\rangle ^{\rm NP1} $ & $0.026\pm 0.001$ & $-(0.059\pm 0.003)$\\ 
~ $R_{K_1}^{\rm NP2}$ & $0.715 \pm 0.036$ &$1.51\pm 0.075$&~ $\langle A_{FB}\rangle ^{\rm NP2} $ &$-(0.007\pm .0003)$ &$-(0.107\pm 0.005)$ \\ 
~ $R_{K_1}^{\rm NP3}$ & $0.746\pm 0.037$ &$0.765\pm 0.038$&~ $\langle A_{FB}\rangle ^{\rm NP3} $ &$0.053\pm 0.003$ &$0.025\pm 0.001$ \\ 
 \hline
    $\langle P_{L}\rangle ^{\rm SM}$ &$ -(0.8625\pm 0.043)$ & $-(0.488\pm 0.024)$ & $\langle P_{T}\rangle ^{\rm SM}$ & $-(0.095\pm 0.005)$ &$-(0.019\pm 0.001)$  \\ 
     $\langle P_{L}\rangle ^{\rm NP1}$ & $-(0.713\pm 0.036)$& $-(0.137 \pm 0.007)$ & $\langle P_{T}\rangle ^{\rm NP1} $ & $-(0.085\pm 0.004) $&$-(0.009\pm .0004) $ \\ 
     $\langle P_{L}\rangle ^{\rm NP2} $ & $-(0. 478\pm 0.024)$&$-(0.219\pm 0.011)$ &$\langle P_{T}\rangle ^{\rm NP2}$ & $-(0.052 \pm 0.003)$ & $-(0.032\pm 0.002)$\\ 
     
  $\langle P_{L}\rangle ^{\rm NP3} $ & $-(0. 825\pm 0.041) $&$-(0.317\pm 0.016)$ &$\langle P_{T}\rangle ^{\rm NP3}$ & $-(0.094 \pm .0005)$ & $-(0.012\pm .0006)$\\ 
   \hline
 $\langle P_{N}\rangle ^{\rm SM} $ & $-1.39\times 10^{-3}$ &$-5.35 \times  10^{-3}$& $\langle { A_{FB}^{LN}}\rangle ^{\rm SM} $ & $4.86 \times 10^{-3}$ &$0.023$\\ 
  $\langle P_{N}\rangle ^{\rm NP1}$ & $-1.69 \times 10^{-3}$&$-5.30 \times 10^{-3}$&   $\langle { A_{FB}^{LN}}\rangle ^{\rm NP1}$ & $5.92 \times 10^{-3}$&$0.023$\\  
    $\langle P_{N}\rangle ^{\rm NP2}$ & $-1.93 \times 10^{-3}$ &$-3.51 \times10^{-3}$& $ \langle { A_{FB}^{LN}}\rangle ^{\rm NP2}$ & $6.77  \times 10^{-3}$ &$0.015$\\ 
     $\langle P_{N}\rangle ^{\rm NP3}$ & $-1.61 \times 10^{-3}$ &$-6.03 \times 10^{-3}$& $ \langle { A_{FB}^{LN}}\rangle ^{\rm NP3}$ & $5.63 \times 10^{-3}$ &$0.026$\\ 
   \hline 
 
$\langle{ A_{FB}^{LT}}\rangle ^{\rm SM}$ & $-0.037$ & $-6.06\times 10^{-3}$& $\langle { A_{FB}^{NT}}\rangle ^{\rm SM} $ & $-0.357 \times 10^{-3}$ &$-0.141\times 10^{-3}$\\ 
  $\langle { A_{FB}^{LT}}\rangle ^{\rm NP1} $ & $-0.024$&$0.101\times 10^{-3}$ &   $\langle { A_{FB}^{NT}}\rangle ^{\rm NP1}$ & $-0.435 \times 10^{-3}$&
  $-0.140 \times 10^{-3}$\\ 
   $\langle { A_{FB}^{LT}}\rangle ^{\rm NP2}$ &$ -0.008 $ & $-0.038 $& $\langle{ A_{FB}^{NT}}\rangle ^{\rm NP2}$ & $-0.497 \times 10^{-3}$ &
   $-0.92  \times 10^{-4}$\\ 
   
    $\langle { A_{FB}^{LT}}\rangle ^{\rm NP3}$ &$ -0.037 $ & $-3.35 \times 10^{-3}$& $\langle{ A_{FB}^{NT}}\rangle ^{\rm NP3}$ & $-0.413 \times 10^{-3}$ &
   $-0.159  \times 10^{-3}$\\ 
   \hline 
 \hline
\end{tabular}
\end{center}
\end{table}

We now proceed to calculate the branching fractions  for $B \to K_1 \mu^+ \mu^-$ processes in the whole $q^2$ region, for which it is necessary to eliminate the backgrounds coming from the resonance regions. This can be done by using the following veto windows so that backgrounds coming from the dominant resonances $B \to K_1 J/\psi (\psi')$ with $J/\psi (\psi') \to \mu^+ \mu^-$ can be eliminated, 
\bea
&&(m_{J/\psi}-0.02)^2~{\rm GeV}^2 \leq s \leq (m_{J/\psi}+0.02)^2 ~{\rm GeV}^2, ~~~~~{\rm and}\nn\\
&&(m_{\psi'}-0.02)^2 ~{\rm GeV}^2\leq s \leq (m_{\psi'}+0.02)^2 ~{\rm GeV}^2,
\eea
which basically corresponds to the invariant mass of the muon pair to be within 20 MeV of  the  $J/\psi(\psi')$ mass. Using the above mentioned cuts, the predicted 
 branching fractions for the whole $q^2$ region  are presented  in Table-\ref{whole-range}.

\begin{table}[h]
\begin{center}
\caption{The predicted   values of the branching fractions in the whole $q^2$ range  for the $B^0 \rightarrow K_1^0(1270) \mu^+ \mu^-$  and $B^0 \rightarrow K_1^0(1400) \mu^+ \mu^-$ processes, in the SM and in NP scenarios.  }\label{whole-range}
\begin{tabular}{|c | c | c| }
\hline
 ~Various in different scenarios ~ & ~ ${\rm Br}( B^0 \rightarrow K_1(1270) \mu^+ \mu^-)$~ &~ ${\rm Br}(B^0 \rightarrow K_1(1400) \mu^+\mu^-)$~~\\ 
\hline 
\hline
Standard Model & ~$(1.477\pm 0.295) \times 10^{-6}$ ~& ~$(4.084 \pm 0.817) \times 10^{-8}$  \\
\hline
NP scenario-I & ~ $(1.184 \pm 0.237)\times 10^{-6}$~& ~ $(3.473\pm 0.695))\times 10^{-8}$\\
\hline
NP scenario-II  &~$(1.103\pm 0.221)\times 10^{-6}$~
 &~ $(4.158\pm 0.832 )\times 10^{-8}$ ~\\ 
 \hline
NP scenario-III  &~$ (1.122 \pm 0.224 ) \times 10^{-7}$~
 &~ $(3.236\pm 0.647)\times 10^{-8}$ ~\\  
 
 \hline
\end{tabular}
\end{center}
\end{table}

\section{Summary and outlook}

The recent results from LHCb  experiment, show some level of discrepancies in the FCNC mediated transitions $b \to s \ell \ell$, e.g., the branching fractions of $B \to K^{(*)} \mu\mu$ and  $B_s \to \phi \mu \mu$, angular observables of $B \to K^* \mu \mu$ process, such as   $P'_{4,5}$,   as well as the  LFU violating ratios $R_{K^{(*)}}$    in $B \to K^{(*)} \ell \ell$ processes. All these discrepancies  are generally  attributed  to the possible interplay  of some kind of new physics in  $b \to s \mu^+ \mu^-$ channels. 
Hence,  considerable interest has been paid to these decay processes in all possible ways to establish or  rule out the role of NP. The general presumption is that, if indeed NP is responsible for the observed deviations in  $b \to s\mu^+ \mu^-$ processes, it must also show up in other modes having the same quark level transition.  In this context, we have studied various observables of $B \to (K_1(1270)/K_1(1400)) \mu^+ \mu^-$ processes in depth. The main objective of our work is to understand  the behaviour of these observables under the influence of new physics, associated with $b \to s \ell^+ \ell^- $ anomalies.  
It should be emphasized that the existing $b \to s \mu \mu$ anomalies  can be realized in a model independent approach, as  new augmentation to the Wilson coefficient $C_{9 \mu}$,  along with some room for  other Wilson coefficients. 
Even though such a contribution to $C_{9 \mu}$  appears to be a reasonable way of elucidating a large set of discrepancies, theory predictions for some $b \to s \mu \mu$  observables may have better consistency with data, once additional contributions are incorporated in other WCs (such as $C_{9\mu}$ or $C_{10 \mu })$. Recently a global fit has been performed in \cite{Bhom:2020lmk} considering  the   recent $b \to s\mu \mu$ data  and it has been shown that, all the anomalies can be elucidated with the following set values for the NP  Wilson coefficients: $(C_7^{\rm NP}, C_9^{\rm NP}, C_{10}^{\rm NP})=(0.013,-1.03,0.08)$.

In this work, we have considered a two-dimensional hypothesis with three specific scenarios for real NP  Wilson coefficients: $(C_9^{\rm NP}-C_9^{'\rm NP})$, $(C_{10}^{\rm NP}-C_{10}^{'\rm NP})$ and $(C_9^{\rm NP}-C_{10}^{\rm NP})$, and extracted the values of these new coefficients from the existing data on $b \to s \mu \mu$  anomalies, that are relatively free from hadronic uncertainties. We found that the combination $(C_9^{\rm NP}-C_9^{'\rm NP})=(-0.829,-0.463)$  and  $(C_9^{\rm NP}-C_{10}^{\rm NP})=(-0.526, 0.573)$  explain the anomalies preferably well.

We then studied the implications of these new physics scenarios on the semileptonic decay $B \to (K_1(1270)/K_1(1400)) \mu \mu$.  The axial vector mesons $K_1(1270)$ and $K_1(1400)$ are admixture states of the $1^1P_1$ and  $1^3P_1$  states with mixing angle $\theta$, which is not yet known precisely. Its value extracted  from the radiative decays $B \to K_1^* \gamma$ is $\theta = -(34 \pm 13)^\circ$. To see the impact of the mixing angle on various observables, we first looked into the SM branching ratio, forward-backward asymmetry and the longitudinal  lepton polarization asymmetry of $B \to K_1(1270)/K_1(1400) \mu^+ \mu^-$ processes for three different  values of $\theta= -34^\circ, -21^\circ,-47^\circ$ and found that the observables of $B \to K_1(1400) \mu^+ \mu^-$ processes are quite sensitive to the mixing angle as the contributions from $B \to K_{1A}$ and $B \to K_{1B}$ come with a relative minus sign, whereas those associated with $B \to K_1(1270) \mu^+ \mu^-$ process depend very mildly on the mixing angle. Next we analysed these decay modes considering these new physics scenarios. In the first case, we considered the structure of the NP which  includes new contributions only in operators which are non-zero in the SM and  the values of these  new Wilson coefficients $C_{7,9,10}^{\rm NP}$ are extracted  from the currently available data on $b \to s \mu \mu$ anomalies \cite{Bhom:2020lmk}. In the second case we considered the NP contributions in terms of two new Wilson coefficients $(C_9^{\rm NP},C_9^{'\rm NP})$, i.e., in addition to the standard left-handed quark currents, we have also taken into account the right-handed current and in the third case the new physics contributions are considered in terms of $(C_9^{\rm NP}, C_{10}^{\rm NP})$ coefficients. Since the effect due to the NP3 coefficients are  similar to NP1 case, we have not shown explicitly the corresponding results in the  plots and provided only the corresponding numerical results. We found that  in the second category of NP scenario, various  observables deviate significantly  from their corresponding SM predictions whereas for NP scenarios 1 and 3, there are only marginal deviations from SM results. It should be emphasized that lepton flavour universal violating ratio $R_{K_1}$ deviates significantly for all the three types of new physics scenarios. 
The measurement of these observables  would be highly instrumental in exploiting the full potential of $b \to s \mu \mu $ decays to look for new physics signal and ultimately uncover its true nature.
To conclude, these decay processes  offer an  alternative probe to scrutinize  the role of NP associated with the current $B$ anomalies in semileptonic transitions and could be accessible with the currently running LHCb and Belle II experiments.

 \acknowledgements
AB would like to acknowledge DST INSPIRE program for financial support.
RM would like to thank Science and Engineering Research Board (SERB), Govt. of India for financial support. The computational work done at CMSD, University of Hyderabad is duly acknowledged.

\medskip

\bibliography{BL}

\end{document}